\documentclass[aps,prd,reprint,nofootinbib,longbibliography]{revtex4-1}

\usepackage{amsmath}
\usepackage{mathtools}
\usepackage[dvipdfmx]{graphicx}
\usepackage[dvipdfmx]{color}
\usepackage[normalem]{ulem}
\usepackage[com pat=1.1.0]{tikz-feynman}
\usepackage{tikz}
\usepackage{aas_macros}
\usetikzlibrary{external}
\usepackage{comment}
\usetikzlibrary{arrows}
\usepackage[colorlinks=true, allcolors=blue]{hyperref}

\begin{document}

\title{Neutron star kick driven by asymmetric fast-neutrino flavor conversion}

\author{Hiroki Nagakura}
\email{hiroki.nagakura@nao.ac.jp}
\affiliation{Division of Science, National Astronomical Observatory of Japan, 2-21-1 Osawa, Mitaka, Tokyo 181-8588, Japan}

\author{Kohsuke Sumiyoshi}
\email{sumi@numazu-ct.ac.jp}
\affiliation{National Institute of Technology, Numazu College, Ooka 3600, Numazu, Shizuoka 410-8501, Japan}

\begin{abstract}
Multi-dimensional nature of core-collapse supernova (CCSN) leads to asymmetric matter ejection and neutrino emission, that potentially accounts for the origin of neutron star (NS) kick. Asymmetric neutrino radiation fields are, in general, accompanied by large-scale inhomogeneous fluid distributions, in particular for electron-fraction ($Y_e$) distributions. Recently, it has also been revealed that lower $Y_e$ environments in proto-neutron star envelope can offer preferable conditions for collective neutrino oscillations. In this paper, we show that a dipole asymmetry of fast neutrino-flavor conversion (FFC), one of the collective neutrino oscillation modes, can power a NS kick, and that it would generate a characteristic correlation between asymmetric distributions of heavy elements in the ejecta and the direction of NS kick. We strengthen our argument for the FFC-driven NS kick mechanism by performing axisymmetric neutrino transport simulations with full Boltzmann neutrino transport. We show that this mechanism can generate linear momentum of neutrinos to account for typical proper motions of NS. Although more detailed studies are necessary, the present study opens a new channel to give a natal NS kick.
\end{abstract}
\maketitle

\section{Introduction}
Main sequence stars with having masses more than $\sim 10$ times solar masses are destined to undergo gravitational collapse of the central core, and then form neutron stars (NS) or stellar-mass black holes. The released gravitational energy is mostly taken away by neutrinos, and the rest of the energy ($\lesssim 1 \%$) powers ejection of materials with electromagnetic emission, generating core-collapse supernova (CCSN). The theory of CCSN suggests that explosions are, in general, asymmetric regardless of the details of mechanism, which is consistent with polarization observations \cite{2008ARA&A..46..433W}, distributions of heavy elements in CCSN ejecta \cite{2017ApJ...842L..24A}, and high proper velocities of NS \cite{2005MNRAS.360..974H} (see also references therein). Detailed insights on the inner dynamics of CCSNe can be brought by comparing theoretical models and these observations.

Observations of neutron stars embedded in supernova remnants (SNR) exhibit that the typical proper velocity of NS is a few hundreds of km/s but some NSs have even more than a thousand km/s \cite{1994Natur.369..127L,1998ApJ...505..315C,2002ApJ...568..289A,2005MNRAS.360..974H,2006ApJ...643..332F,2005ApJ...630L..61C,2007ApJ...670..635W,2017ApJ...844...84H,2018ApJ...856...18K,2021A&A...651A..40M,2023arXiv231019879H}. Popular scenarios to explain such high NS velocities are that the linear momentum is imparted to NS by asymmetric matter ejection \cite{2010ApJ...725L.106W,2010PhRvD..82j3016N,2012MNRAS.423.1805N,2013A&A...552A.126W,2017ApJ...837...84J,2023arXiv231112109B} or asymmetric neutrino emission \cite{1987IAUS..125..255W,1993A&AT....3..287B,2005ApJ...632..531S,2006ApJS..163..335F,2019ApJ...880L..28N,2022MNRAS.517.3938C} or both \cite{2004ApJ...601L.175F} during the development of CCSN explosion. It should be mentioned that, when we sum up all the absolute values of neutrino momentum, it reaches $\sim 10^{43} {\rm g~cm/s}$, while the required linear momentum for the typical NS proper motion is $\sim 10^{41} {\rm g~cm/s}$, implying that a percent anisotropy of neutrino emission is high enough to generate NS kick. This consideration also exhibits a requirement of high precision in numerical simulations for modeling NS kick. Considerable care must be paid for the total momentum conservation in CCSN simulations; in fact, small errors can easily lead to numerical artifacts \cite{2019ApJ...878..160N}. Because of the delicate problem, the self-consistent multi-dimensional CCSN simulations would be a unique way to quantify NS kick, and the long-term simulations may also be important, since interaction to asymmetric fallback material onto NS may also affect the proper motion (and spin) \cite{2022ApJ...926....9J,2023MNRAS.526.2880M}. Finally it should also be mentioned that the NS proper motions are affected by long-term asymmetric electro-magnetic emission, known as {\it post natal} processes \cite{1975ApJ...201..447H,2001ApJ...549.1111L,2019MNRAS.488.4161P,2021MNRAS.508.3345I}, which may need to be taken into account for comparing to observations.

In this paper, we present a new possibility that fast neutrino-flavor conversion (FFC) plays an important role on NS kick. FFC is associated with one of flavor instabilities in collective neutrino oscillation \cite{2005PhRvD..72d5003S}, in which flavor correlation (or coherence) grows exponentially due to refractive effects of neutrino self-interactions (see reviews, e.g., \cite{2021ARNPS..71..165T,2022Univ....8...94C,2022arXiv220703561R,2023arXiv230111814V,2023arXiv230803962F}). Recent studies suggest that the flavor instability occurs for wide mass range of progenitors by various mechanisms \cite{2021PhRvD.103f3033A,2021PhRvD.104h3025N} and the stellar rotation would facilitate the occurrence of FFC \cite{2022ApJ...924..109H}. Since the flavor conversion can occur in much shorter timescale than dynamical timescale of the system, neutrino radiation field is instantaneously changed by flavor conversions, once the instability switches on. In this paper, we show that occurrences of FFC in asymmetric neutrino radiation field can further enhance the asymmetry, which potentially generates high velocities of NS natal kick

This paper is organized as follows. In Sec.~\ref{sec:basicpic}, we first describe our basic picture of how FFC can give a linear momentum to NS. We then strengthen our arguments by performing Boltzmann neutrino transport simulations under CCSN fluid profiles, in which we incorporate effects of FFCs by a phenomenological prescription. The numerical methods and models are summarized in Sec.~\ref{sec:methodandmodel} and the results are summarized in Sec.~\ref{sec:numresults}. We conclude our work in Sec.~\ref{sec:summary}.

\section{FFC-driven neutron star kick}\label{sec:basicpic}

\begin{figure*}[ht]
\begin{minipage}{1.0\textwidth}
\centering
\includegraphics[width=0.9\linewidth]{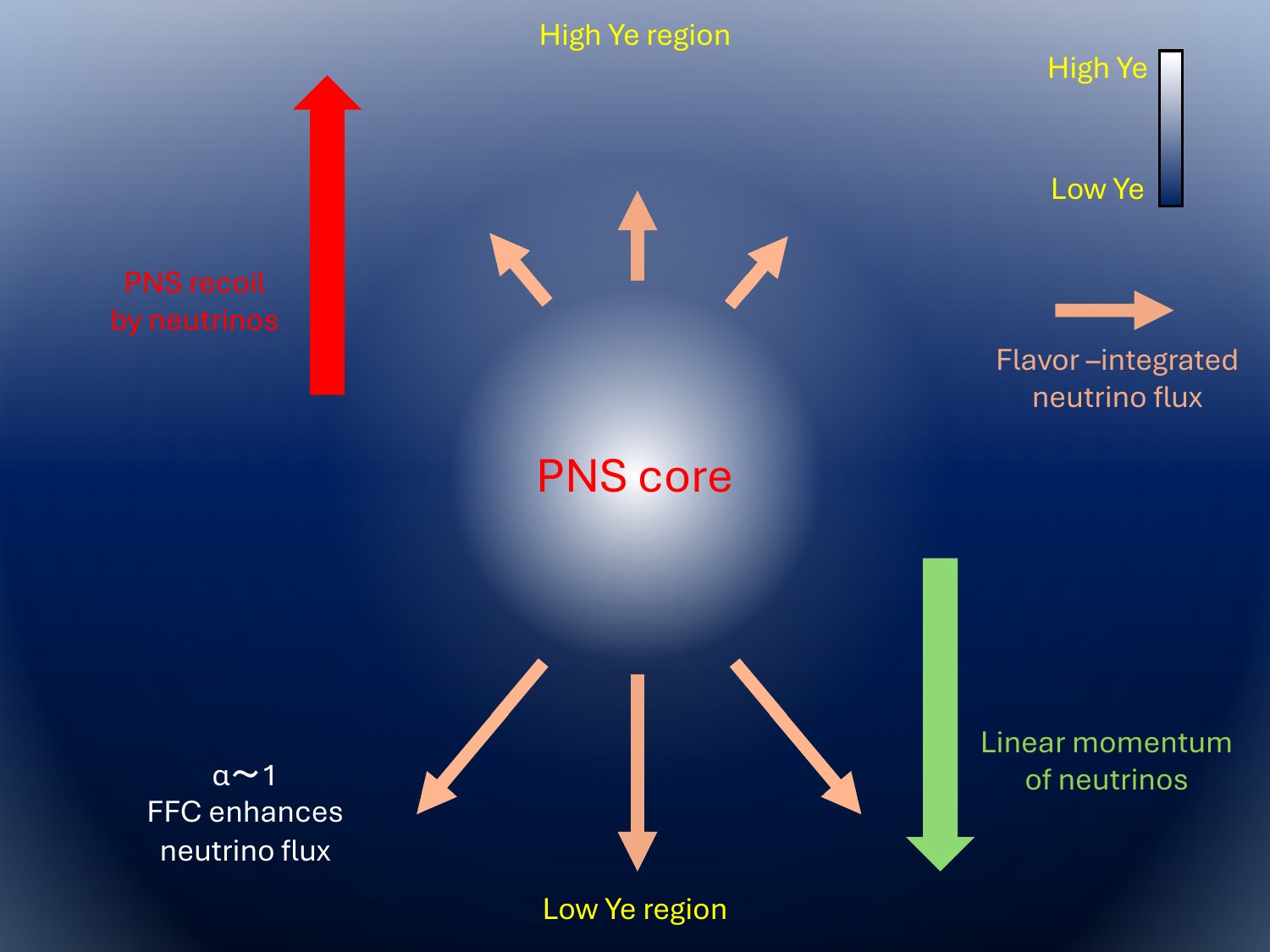}
\end{minipage}
\caption{Schematic picture of FFC-driven NS kick. Background color represents $Y_e$; white and black regions correspond to high- and low $Y_e$ regions, respectively. FFC occurs in low $Y_e$ region, which leads to high flavor-integrated neutrino fluxes (see the text for more details). This generates a linear momentum of neutrinos in the low $Y_e$ direction (green arrow), and consequently the NS obtains the same amount of linear momentum in the opposite direction to the neutrino linear momentum (red arrow).}
\label{fig_schematic}
\end{figure*}

Assuming that muon-type ($\nu_{\mu}$) and tau-type neutrinos ($\nu_{\tau}$), which are collectively denoted as $\nu_x$, and their antipartners ($\bar{\nu}_x$) are identical, the onset of FFC is dictated by the disparity of angular distributions of electron-type neutrinos ($\nu_e$) and their antipartners ($\bar{\nu}_e$) (but see also \cite{2021PhRvD.103f3013C} when we take into account differences between $\nu_x$ and $\bar{\nu}_x$). More specifically, electron-type lepton number ($\nu_e$ - $\bar{\nu}_e$ or ELN) angular crossings marks the onset of the flavor instability \cite{2022PhRvD.105j1301M}. It has been suggested that the crossing tends to occur in regions where the number densities of $\nu_e$ and $\bar{\nu}_e$ ($n_{\nu_e}$ and $n_{\bar{\nu}_e}$) are close to each other \cite{2019PhRvD.100d3004A,2019ApJ...886..139N}. 
Following the convention in this field, we use $\alpha$ to measure the asymmetry of number densities of $\nu_e$ ($n_{\nu_e}$) and $\bar{\nu}_e$ ($n_{\bar{\nu}_e}$), which is defined as
\begin{eqnarray}
\alpha \equiv \frac{n_{\bar{\nu}_e}}{n_{\nu_e}}.
\label{eqn:defalpha}
\end{eqnarray}

It should be mentioned that FFCs do not occur inside proto-neutron star (PNS) ($\rho \gtrsim 10^{14} {\rm g/cm^3}$, where $\rho$ denotes the baryon mass density), since $n_{\nu_e}$ is much larger than $n_{\bar{\nu}_e}$, i.e., $\alpha \ll 1$ due to strong degeneracy of $\nu_e$\footnote{We note that FFC may occur in the high density region in the late post bounce (or PNS cooling) phase, since the $\nu_e$ degeneracy becomes mild due to deleptonization.}. In the PNS envelope ($10^{11} {\rm g/cm^3} \lesssim \rho \lesssim 10^{14} {\rm g/cm^3}$), on the other hand, the chemical potential of $\nu_e$ decreases and matter temperature increases with radius, resulting in a rapid decrease of $\nu_e$ degeneracy. Spherically symmetric CCSN models have showed, however, that the chemical potential of $\nu_e$ is still high enough to suppress ELN angular crossings, implying that the neutrino distributions are stable to FFC \cite{2017ApJ...839..132T,2020PhRvR...2a2046M}. However, the situation can be qualitatively changed in multi-dimensional models. For instances, PNS convections facilitate deleptonization of CCSN core \cite{2020MNRAS.492.5764N}, that results in decreasing electron-fraction ($Y_e$), and consequently the degeneracy of $\nu_e$ is lower than spherically symmetric models, and it could be even negative. This implies the region with $\alpha \sim 1$ appears in the convective layer. Since the anisotropy of $\bar{\nu}_e$ in momentum space is relatively higher than $\nu_e$, ELN angular crossings emerge in these regions \cite{2020PhRvD.101f3001G,2020PhRvD.101b3018D}.

Another representative FFC in multi-dimensional CCSN models is triggered by large-scale coherent asymmetric neutrino emission. The asymmetric neutrino emission is accompanied by a radiation-hydrodynamical instability, namely lepton-emission self-sustained asymmetry (LESA \cite{2014ApJ...792...96T,2019ApJ...881...36G,2019MNRAS.487.1178P}) or coherent asymmetric $Y_e$ distributions associated with PNS kick \cite{2019ApJ...880L..28N}. Due to the anti-correlation of asymmetric neutrino emission between $\nu_e$ and $\bar{\nu}_e$, $\alpha$ becomes close to unity in the region where $Y_e$ is low and $\bar{\nu}_e$ ($\nu_e$) is stronger (weaker), generating ELN angular crossings \cite{2019ApJ...886..139N,2021PhRvD.104h3025N}. As we shall show below, the NS kick can be accompanied by this type of FFC associated with asymmetric neutrino emission.

One may wonder why FFC can impart a linear momentum to NS. In fact, no linear momentum are generated only by FFCs, because the flavor conversion changes flavor states but the flavor-integrated energy and momentum are conserved. The key player is neutrino-matter interactions such as neutrino emission, absorption, and scatterings. Since they depend on neutrino flavors, the exchange of momentum between neutrinos and matter is affected by FFCs.

Our NS kick scenario was inspired by results from previous studies. It has been shown in numerical simulations that FFCs can enhance neutrino cooling if they occur in optically thick or semi-transparent regions. This trend has  been observed rather commonly regardless of different approaches, e.g., global neutrino-radiation-hydrodynamic simulations of CCSNe and BNSMs with phenomenological approaches of FFC \cite{2021PhRvL.126y1101L,2022PhRvD.105h3024J,2022PhRvD.106j3003F,2023PhRvD.107j3034E,2023PhRvL.131f1401E} and direct quantum kinetic simulations \cite{2023PhRvL.130u1401N,2023PhRvD.108l3003N}.
The physical mechanism can be understood as follows. In general, electron-type neutrinos are more abundant than heavy-leptonic ones, implying that FFC works to reduce the number of electron-type neutrinos. On the other hand, heavy-leptonic neutrinos have lower opacities than electron-type ones due to the lack of charged-current reactions\footnote{We note that on-shell muons may appear in the envelop of PNS \cite{2017PhRvL.119x2702B,2020PhRvD.102l3001F,2020PhRvD.102b3037G}. However, the muon number density is much lower than that of electrons, indicating that the trend is qualitatively similar in this case.}, implying that heavy-leptonic neutrinos can easily escape from the region. This indicates that the increase of heavy-leptonic neutrinos by FFCs results in increasing neutrino flux. Due to the large neutrino flux of heavy-leptonic neutrinos, FFCs keep converting electron-type neutrinos to heavy-leptonic ones, while electron-type neutrinos are produced by charged-current reactions. As a result, neutrinos can extract energies from matter more efficiently by FFCs. This leads to the acceleration of neutrino cooling. We note that neutrinos can carry not only the energy but also momentum, indicating that momentum loss by neutrinos are also enhanced by FFC. This results in generating a linear momentum.

Let us summarize the FFC-driven NS kick scenario (see also Fig.~\ref{fig_schematic}). In PNS envelope, large-scale asymmetric matter distributions are created for some reason (e.g., LESA), and $Y_e$ in some regions can be low enough to generate ELN angular crossings, leading to occurrences of FFCs. Because electron-type neutrinos are more populated than other species, the flavor conversion increases the number density of heavy-leptonic one, while heavy-leptonic neutrinos are more transparent than electron-type neutrinos, leading to the increase of neutrino flux. The flux reduces the number density of heavy-leptonic neutrinos, which sustains the flavor conversion from electron-type to heavy-leptonic neutrinos, while electron-type neutrinos can be efficiently produced by charged-current reactions. This implies that neutrinos and matter in the region share the momentum each other more efficiently than in other regions, breaking the global momentum balance in the system, which gives a linear momentum to NS.

This mechanism suggests that the direction of NS kick should be in the direction with higher-$Y_e$ environment, implying that this process generate a correlation between ejecta composition and NS kick direction (see also \cite{2023MNRAS.519.2623F}). The X-ray observations for young SNRs, which have the ability to measure
spatial distributions of heavy elements (see, e.g., \cite{2020ApJ...889..144H}),
would be very useful to place a constraint of the mechanism. We also note that more detailed information may be given near future by XRISM mission \cite{2020arXiv200304962X}.

To strengthen our proposed scenario, we demonstrate in the following sections that FFCs can induce linear momentum of neutrinos by performing axisymmetric Boltzmann neutrino transport simulations. It should be noted that these simulations are meant as a proof-of-principle, and more detailed studies are needed to assess whether the FFC-driven NS kick mechanism can be responsible for observed velocities of NS proper motions. Nevertheless, we demonstrate that the FFC has the ability to change a linear momentum of neutrinos by $\sim$ a few $\times 10^{40} {\rm g~cm/s}$. This represents a possibility that the FFC-driven mechanism is a new channel to contribute NS natal kick.

\section{Numerical method and model}\label{sec:methodandmodel}
In this section, we describe some essential information of our numerical method and model in our neutrino transport simulations. In Sec.~\ref{subsec:fluid}, we first describe the background fluid profile. In Sec.~\ref{subsec:NeutTransFFC}, we summarize our neutrino transport code and also describe an approximate neutrino-mixing scheme to include effects of FFCs into classical neutrino transport.

\subsection{Fluid distribution}\label{subsec:fluid}

In this study, we refer a fluid profile from a spherically symmetric CCSN model, which was developed by a numerical code for a neutrino-radiation hydrodynamics
with full Boltzmann neutrino transport \cite{2018ApJ...854..136N}. In the simulation, the neutrino radiation field is determined
by solving the Boltzmann equation. We adopt the fluid profile at 300~ms after the core bounce obtained by following the time evolution from the Fe core of the 15$M_{\odot}$ star by \cite{2002RvMP...74.1015W}. The equation of state by the variational method \cite{2017JPhG...44i4001F} was adopted in the simulation and the same equation of state is used in the current study.   

As mentioned in Sec.~\ref{sec:basicpic}, spherically symmetric CCSN models are unlikely to generate ELN angular crossings in the post shock region, and we confirmed that there are no ELN angular crossings in our neutrino data. This is mainly due to the high $Y_e$ distributions compared to multi-dimensional models (see \cite{2020MNRAS.492.5764N}), since higher $Y_e$ leads to stronger degeneracy of $\nu_e$. $Y_e$ needs to be, hence, lower in order to generate ELN angular crossings (or FFCs). Another important condition for the FFC-driven NS kick scenario is that $Y_e$ distributions need to be globally asymmetric; more specifically, it should have a dipole asymmetry. Based on the above considerations, we assume that $Y_e$ has a dipole deformation, which is given as
\begin{eqnarray}
Y_e(r,\theta) = Y^{1D}_e(r) (1+\epsilon \cos \theta)
\label{eq:Yedeform}
\end{eqnarray}
where $\theta$ is the polar angle measured from the $z$-axis, and $Y^{1D}_e$ denotes the $Y_e$ profile in a spherically symmetric CCSN model. In the expression $\epsilon$ represents a deformation parameter, and we set $\epsilon=0.15$. As we shall show in Sec.~\ref{subsec:baselinemodel}, ELN angular crossings appear in the southern hemisphere, i.e., lower $Y_e$ region in the classical neutrino transport simulation (baseline model; see below for more details). We note that the matter distributions are frozen during the neutrino transport simulations.

Let us remark a few caveats about fluid background. First, one has to keep in mind that the deformed $Y_e$ profile given by Eq.~\ref{eq:Yedeform} is not realistic but rather a toy model. The artificial change of fluid distributions would lead to unphysical equilibrium state between neutrino and matter. In fact, high (low) $Y_e$ region has stronger $\nu_e$ ($\bar{\nu}_e$) emission, making neutrino distributions approache toward a different chemical equilibrium state from the original one. More self-consistent treatment of neutrino-radiation fields is necessary for more quantitative arguments. We defer such simulations to future work. 

As the other important remark, the asymmetric degree of $Y_e$ in the present study ($\epsilon=0.15$) is high compared to more realistic CCSN simulations. Again this setting is meant to construct a model to test our hypothesis. We note that the required asymmetry in $Y_e$ distribution for occurrences of FFC (or ELN angular crossings) strongly depends on the angle-averaged $Y_e$. In the present study, we adopt a spherically symmetric CCSN model, which has a systematically high $Y_e$ than that in multi-dimensional CCSN models due to suppression of PNS convection \cite{2020MNRAS.492.5764N}. In fact, $\epsilon \lesssim 0.1$ is large enough for occurrences of FFCs in more realistic multi-dimensional models \cite{2019ApJ...886..139N,2021PhRvD.104h3025N}. In the present study, we need to keep in mind these caveats to interpret our results.

\begin{figure*}[ht]
\begin{minipage}{1.0\textwidth}
\centering
\includegraphics[width=0.9\linewidth]{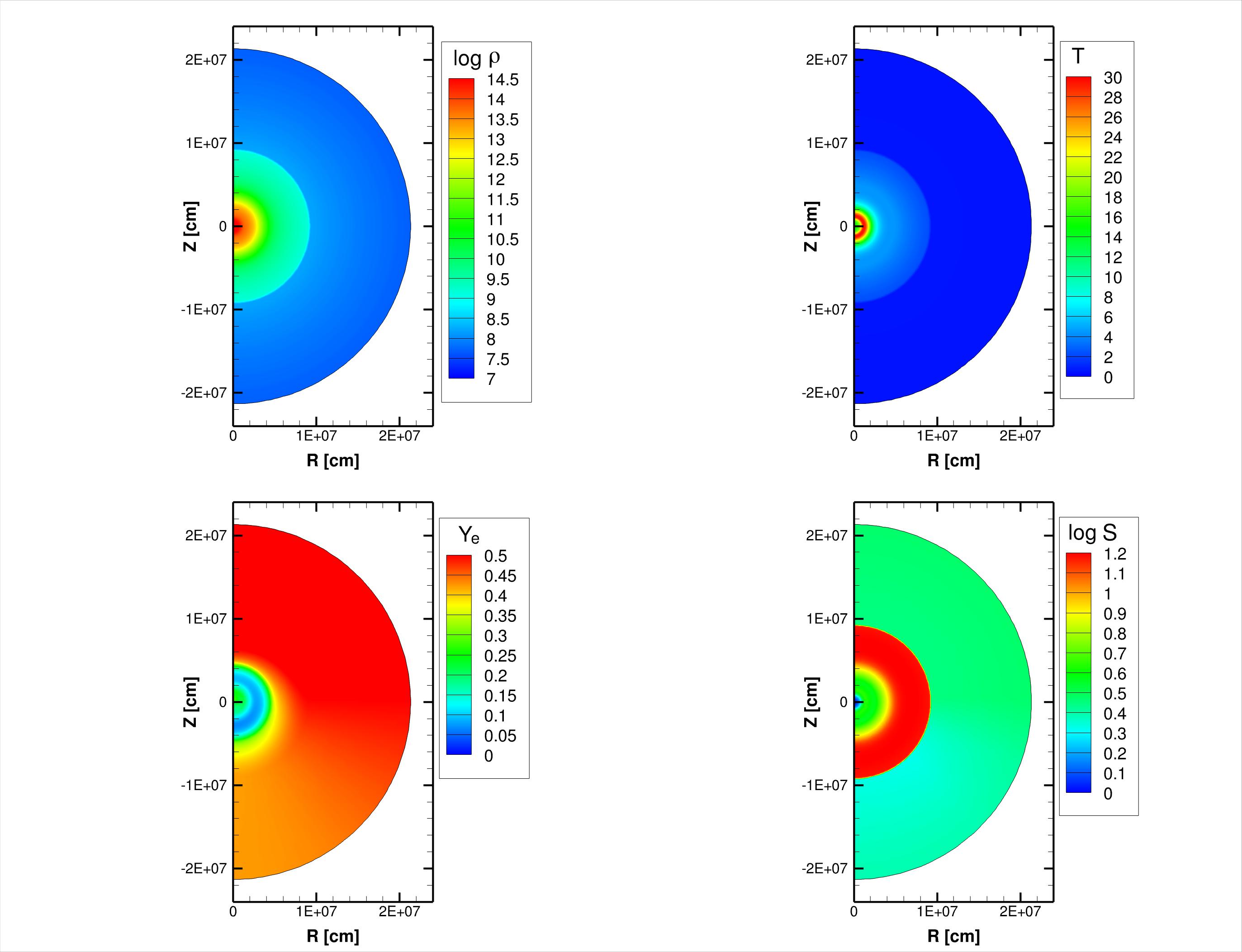}
\end{minipage}
\caption{Profiles of matter distributions employed in this study.
Upper left, upper right, lower left, and lower right panels show density (gcm$^{-3}$), temperature (MeV), $Y_e$, and entropy per baryon ($k_B$), respectively. We note that density and temperature profiles are assumed to be in spherically symmetric, whereas $Y_e$ is deformed. This leads to a dipole deformation in entropy distribution. See text for more details. }
\label{fig_hydro_profile}
\end{figure*}
We show in Fig. \ref{fig_hydro_profile} the fluid profile employed in this study. As in the original spherically symmetric CCSN model, the distributions of density and temperature remain spherical. The outer radius of PNS envelope, which is defined at the density of $10^{11}$~g cm$^{-3}$, is $\sim$40 km. At this time snapshot, the shock wave is located at $\sim$90 km, where a large entropy jump is remarkable. Note that the distribution of entropy is not spherical due to the deformed distribution of $Y_e$. The $Y_e$ is higher on the north side than at the equator and lower on the south side due to the given deformation (see Eq.~\ref{eq:Yedeform}). This deformed distribution in $Y_e$ leads to the different neutrino distributions for $\nu_e$ and $\bar{\nu}_e$ accompanied by their anti-correlated asymmetric emission as we examine in the following.

\subsection{Classical neutrino transfer with FFC}
\label{subsec:NeutTransFFC}
The neutrino radiation field is determined by utilizing the numerical code to directly solve the Boltzmann equation \cite{2012ApJS..199...17S}. This neutrino transport code has been used to model neutrino radiation field in static fluid backgrounds (see, e.g., \cite{2015ApJS..216....5S,2021ApJ...907...92S}). It should be mentioned that we employ the Boltzmann transport code rather than the radiation-hydrodynamic one which has been used to develop multi-dimensional CCSN models \cite{2018ApJ...854..136N,2019ApJS..240...38N,2019ApJ...880L..28N,2020ApJ...902..150H,2020ApJ...903...82I}. As we shall describe in Sec.~\ref{subsec:NeutTransFFC}, the Boltzmann transport module needs to be extended so as to incorporate effects of FFCs in the present study; for instances, the number of neutrino species is increased from three to four, and a new module for neutrino-mixing also needs to be incorporated. These extensions are complicated for radiation-hydrodynamic code, because there are many other modules interwined to the transport one. On the other hand, the code structure in \cite{2012ApJS..199...17S} is simpler than the radiation-hydrodynamic one; hence, we employ the latter code in the present study. We note that the purpose of the simulations is to provide evidence that FFC can change a linear momentum in neutrino radiation field along with the scenario described in Sec.~\ref{sec:basicpic}, indicating that transport simulations in static fluid background satisfy the purpose of the present study.

Before describing the numerical setup, we make a few remarks about differences between the two codes. In our radiation-hydro code, we use a two-energy-grid technique \cite{2014ApJS..214...16N}, which was developed so as to to take into account fluid-velocity dependence in neutrino transport. On the other hand, such a technique is not used in \cite{2012ApJS..199...17S}, and we do not distinguish laboratory- and fluiid-rest frames. It should also be mentioned that some updates of neutrino-matter interactions have been made in neutrino-radiation-hydro code: electron(positron)-scatterings, electron-capture by heavy nuclei, and weak interactions of light nuclei (see \cite{2017ApJ...847..133R} for more details). For the sake of completeness, we checked the descrepancy in the results between two codes under spherically symmetric model, and found that $\nu_e$ and $\bar{\nu}_e$ fluxes in \cite{2012ApJS..199...17S} are higher and lower, respectively, than those obtained in \cite{2019ApJS..240...38N}.

Although these discrepancies between the two codes prevent us from ascertaining quantitative impacts of FFC on NS kick, the transport code in \cite{2012ApJS..199...17S} still has the ability to capture essential features for the FFC-induced NS kick scenario. In fact, our simulations successfully demonstrate that FFC increases energy fluxes of heavy-leptonic neutrinos, which corresponds to the most important ingredient in the mechanism.

Boltzmann equations are solved by the S$_n$ method. Under spatial axisymmetric condition, we determine the time evolution of the neutrino distributions as functions of radius ($r$), zenith angle ($\theta$), two angles ($\theta_{\nu}$ and $\phi_{\nu}$) and energy ($\varepsilon_{\nu}$) in momentum space. The radial and polar angle coordinates cover the range of $0\sim210$~km and $0\sim\pi$ by 256 and 128 grids, respectively. The neutrino energy is divided into 14 bins. The neutrino angle distributions, $\theta_{\nu}$ and $\phi_{\nu}$, are described by 10 and 6 bins, respectively.

\begin{figure*}[ht]
\begin{minipage}{1.0\textwidth}
\centering
\includegraphics[width=0.45\linewidth]{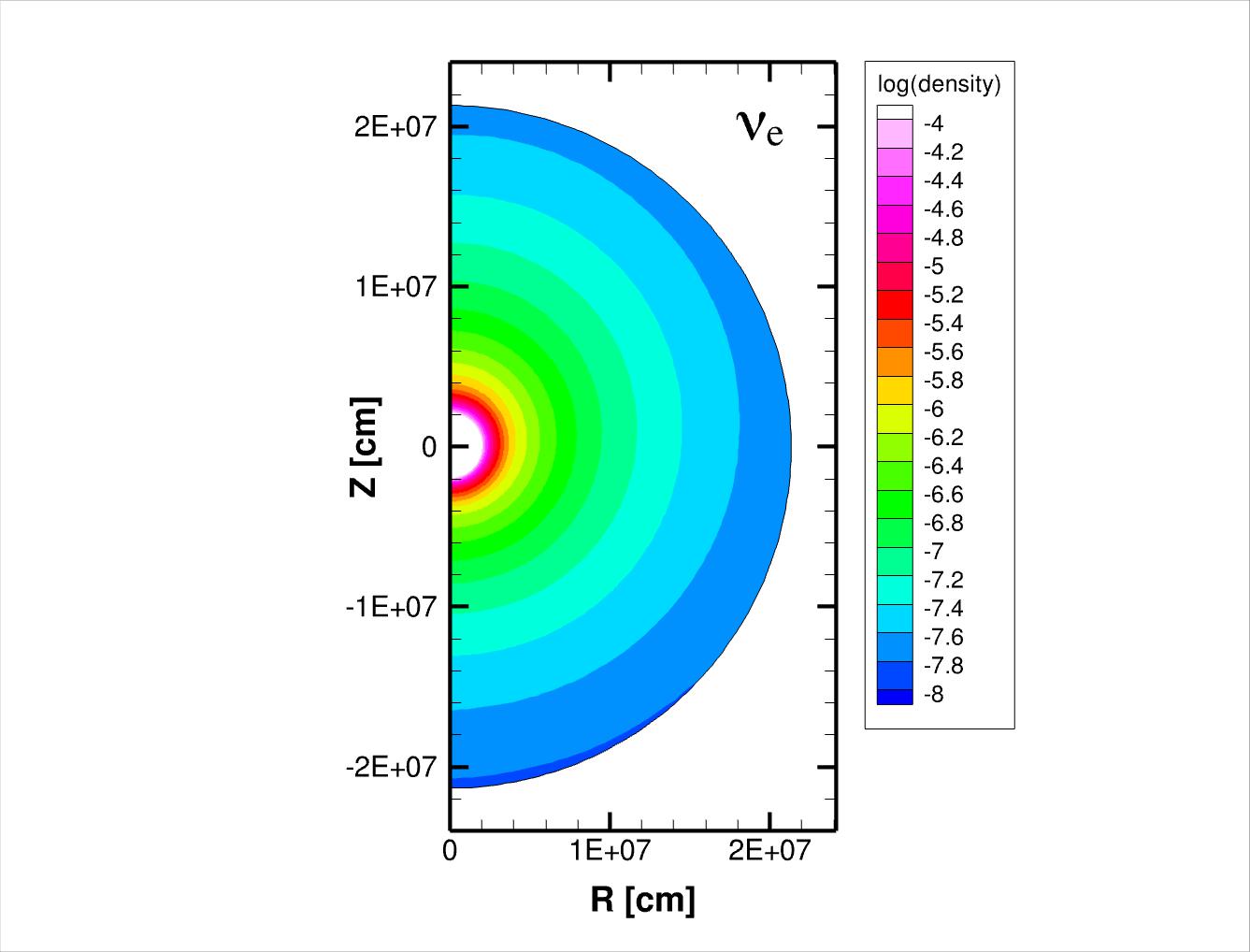}
\includegraphics[width=0.45\linewidth]{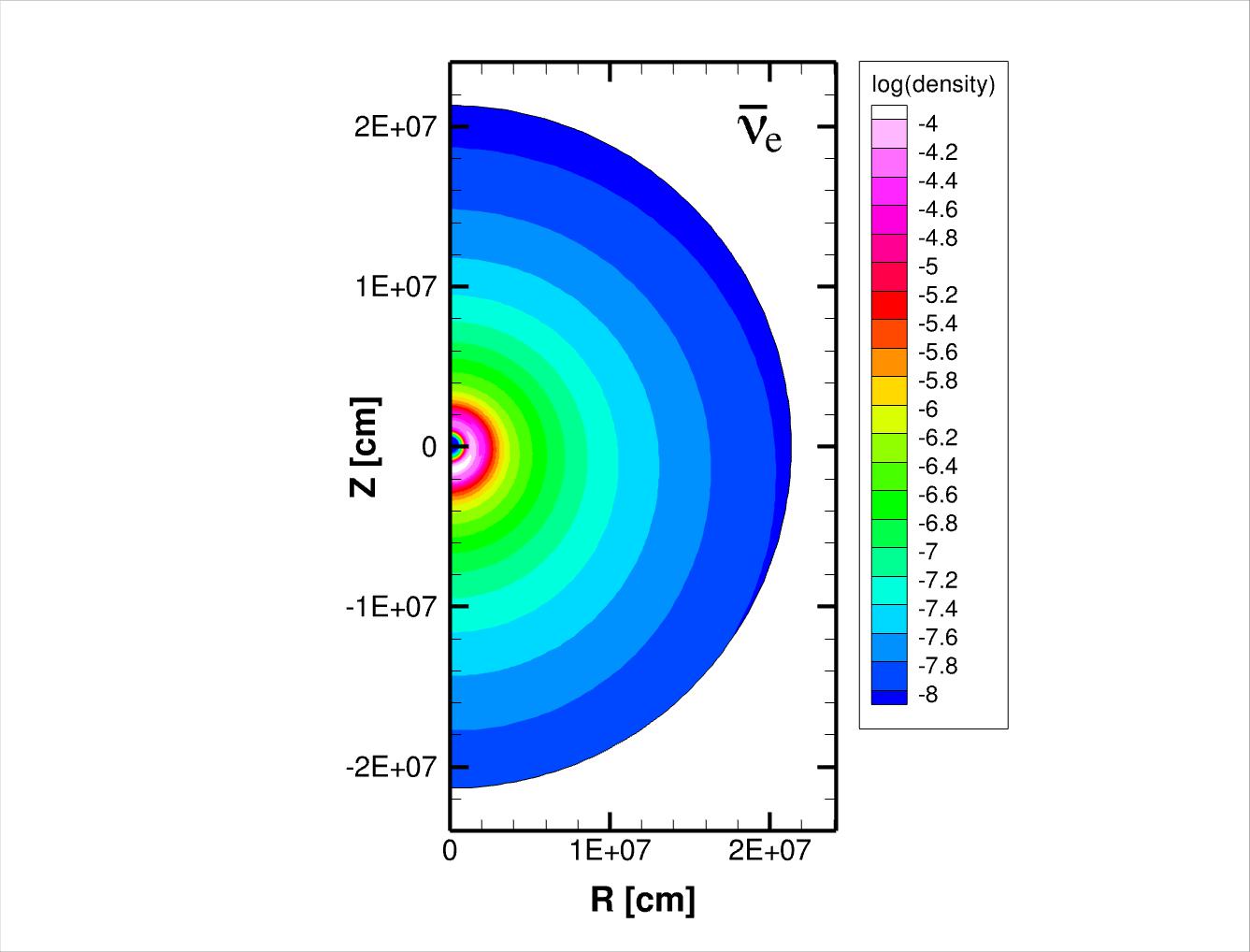}
\includegraphics[width=0.45\linewidth]{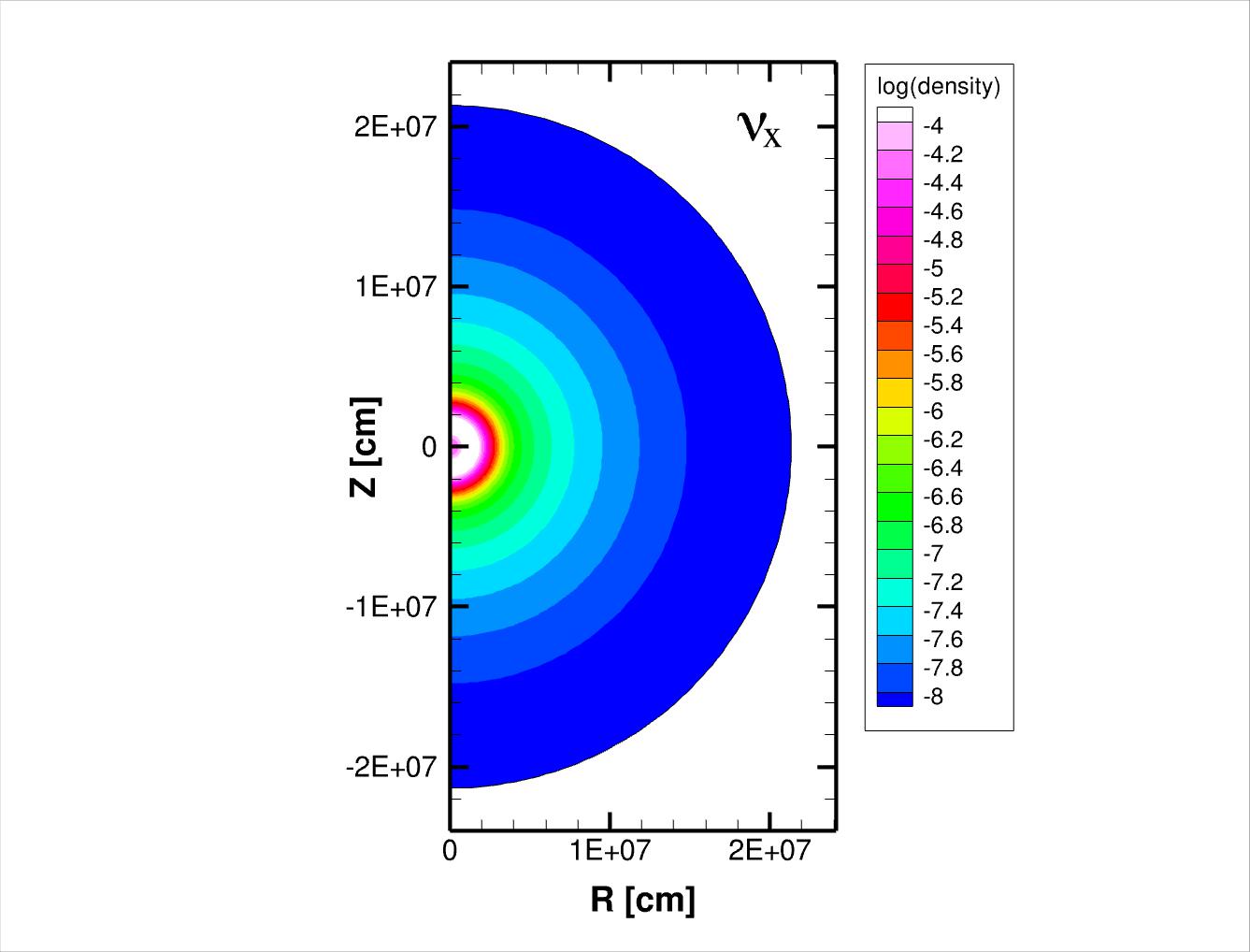}
\includegraphics[width=0.45\linewidth]{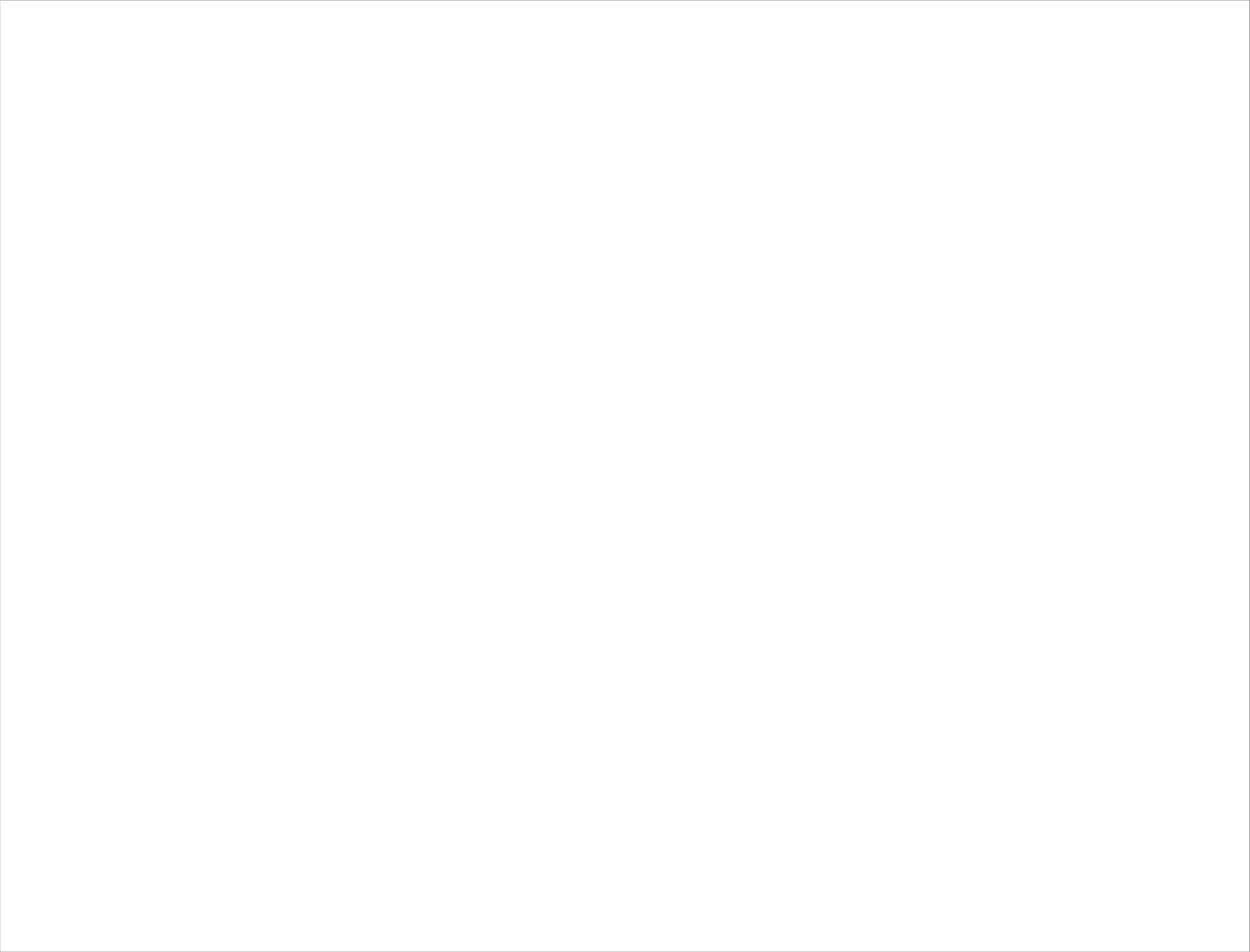}
\end{minipage}
\caption{Neutrino number densities obtained in baseline model
are shown on the meridian slice by color contour map.  
The number densities in units of fm$^{-3}$ for $\nu_e$, $\bar{\nu}_e$, and $\nu_{\mu}$ are shown in upper-left, upper-right, and lower-left panels, respectively.}
\label{fig_3nu_density_flux}
\end{figure*}

We use the equation of state by the variational method with the mixture of nuclei under the nuclear statistical equilibrium \cite{2017JPhG...44i4001F} to obtain the thermodynamical properties and composition of hot and dense matter.  
The basic set of weak interaction (emission, absorption, pair production and annhilation) is implemented in the collision term of the Boltzmann equation with angle- and energy-dependent expressions \cite{2012ApJS..199...17S}.

Given a static fluid background (Sec.~\ref{subsec:fluid}), we follow the time evolution of neutrino distributions until the radiation field settles into a steady state. In this study, we run two simulations: basedline model and FFC model. The basedline model correspond to a purely classical transport case. In FFC model, we incorporate effects of FFCs on classical neutrino transport (see below).

To incorporate effects of flavor conversions, two updates in our Boltzmann solver are required. First, $\nu_x$ and $\bar{\nu}_x$ should be treated independently, since flavor conversions differentiate their distributions. Since our original Boltzmann solver has treated $\nu_x$ and $\bar{\nu}_x$ collectively (i.e., 3 species neutrino transfer), we updated the Boltzmann code to handle 4 species ($\nu_e$, $\bar{\nu}_e$, $\nu_x$, $\bar{\nu}_x$).

The other necessary update is neutrino-mixing scheme. At each time step, we check if neutrino distributions have ELN-XLN (XLN denotes heavy-leptonic neutrino number) angular crossings at each spatial mesh. If they are detected, we shuffle neutrinos instantaneously as the following manner. The neutrino mixing is treated by introducing survival probability of $\nu_e$ ($p$) and $\bar{\nu}_e$ ($\bar{p}$) as
\begin{eqnarray}
&f_{\nu_e} = p \, f^{0}_{\nu_e} + \left(1- p \, \right) f^{0}_{\nu_x} , &\label{eq:flavconv_nue} \\
&f_{\bar{\nu}_e} = \bar{p} \, f^{0}_{\bar{\nu}_e} + \left(1- \bar{p} \, \right) f^{0}_{\bar{\nu}_x} ,& \label{eq:flavconv_nueb} 
\end{eqnarray}
for electron-type (anti-)neutrinos and 
\begin{eqnarray}
&f_{\nu_x} = \frac{1}{2}  \left(1- p \, \right) f^{0}_{\nu_e} 
+ \frac{1}{2} \left(1+ p \, \right) f^{0}_{\nu_x} ,& \label{eq:flavconv_nux} \\
&f_{\bar{\nu}_x} = \frac{1}{2}  \left(1- \bar{p} \, \right) f^{0}_{\bar{\nu}_e} 
+ \frac{1}{2}  \left(1+ \bar{p} \, \right) f^{0}_{\bar{\nu}_x} ,& \label{eq:flavconv_nuxb}
\end{eqnarray}
for $\mu$ and $\tau$-types (anti-)neutrinos \cite{2021MNRAS.500..319N}.
In the expression, $f^{0}$ denotes the distribution function of neutrinos before the mixing.
Just for simplicity, we set $p=\bar{p}=\frac{1}{3}$ for all energies and angles, corresponding to flavor equipartition, in this study.

Let us make a few remarks about our neutrino mixing scheme. First, as pointed out by \cite{2024arXiv240115247J}, the instantaneous mixing prescription suffers from a self-consistency issue. This issue can be resolved by more appropriate prescriptions such as miscidynamics \cite{2023arXiv230614982J} and Bhatnagar–Gross–Krook (BGK) subgrid model \cite{2023arXiv231216285N}. It should be noted, however, that a recent study in \cite{2024arXiv240317269X} showed that classical neutrino transport with an instantaneous mixing scheme agreed reasonably well with results of quantum kinetic one. This would be attributed to the fact that FFCs occur in optically thick or semi-transparent regions where neutrino self-interactions dominate over neutrino-matter interactions. Since we have in mind a scenario that FFC occurs in these regions, the instantaneous mixing prescription would be a reasonable approximation.

One may also speculate that shallow ELN angular crossings, which have been observed in multi-dimensional CCSN models \cite{2019ApJ...886..139N,2021PhRvD.104h3025N}, do not have abilities to lead flavor equipartition. According to detailed studies of asymptotic states of FFC \cite{2023PhRvD.107f3033N,2023PhRvD.107j3022Z}, however, the shallow and narrow ELN angular crossings can lead to large flavor conversions and even flavor equipartition if $\nu_e$ and $\bar{\nu}_e$ number densities are nearly equal to each other \cite{2021PhRvD.104j3003W}. In fact, we observed flavor equiparitions in CCSN models with quantum kinetic neutrino transport \cite{2023PhRvL.130u1401N,2023PhRvD.108l3003N,2024arXiv240219252X}. It should be noted, however, that the flavor equipartition corresponds to an extreme case and the actual flavor conversion may be less vigorous, indicating that the present study may overestimate the impact of FFC on NS kick. This exhibits that higher fidelity scheme in determination of asymptotic states of FFC (see, e.g., \cite{2023PhRvD.107j3022Z,2023PhRvD.107l3021Z,2023PhRvD.108f3003X}) is necessary for more quantitative discussion, but such an improvement is beyond the scope of this paper.


\section{Numerical results}\label{sec:numresults}

\subsection{Baseline model (no flavor conversions)}
\label{subsec:baselinemodel}

Let us first highlight some important features of neutrino radiation field in basedline model, that is helpful to understand the result of FFC model. In Fig.~\ref{fig_3nu_density_flux},
we display neutrino number densities for $\nu_e$, $\bar{\nu}_e$, and $\nu_{x}$. The distribution of $\nu_e$ is almost spherical but shifted to the north side due to the higher $Y_e$. Contrary to $\nu_e$, the distribution of $\bar{\nu}_e$ is shifted to the south side due to the lower $Y_e$. Hence, there is north-south anti-symmetry between $\nu_e$ and $\bar{\nu}_e$ radiation fields. The distribution of $\nu_{x}$ is spherical at the center, reflecting the spherical distribution of density and temperature.  
This is because $\nu_{x}$ is thermally produced by pair-processes, and these processes are less sensitive to $Y_e$.

\begin{figure}[ht]
\begin{minipage}{0.5\textwidth}
\centering
\includegraphics[width=1.0\linewidth]{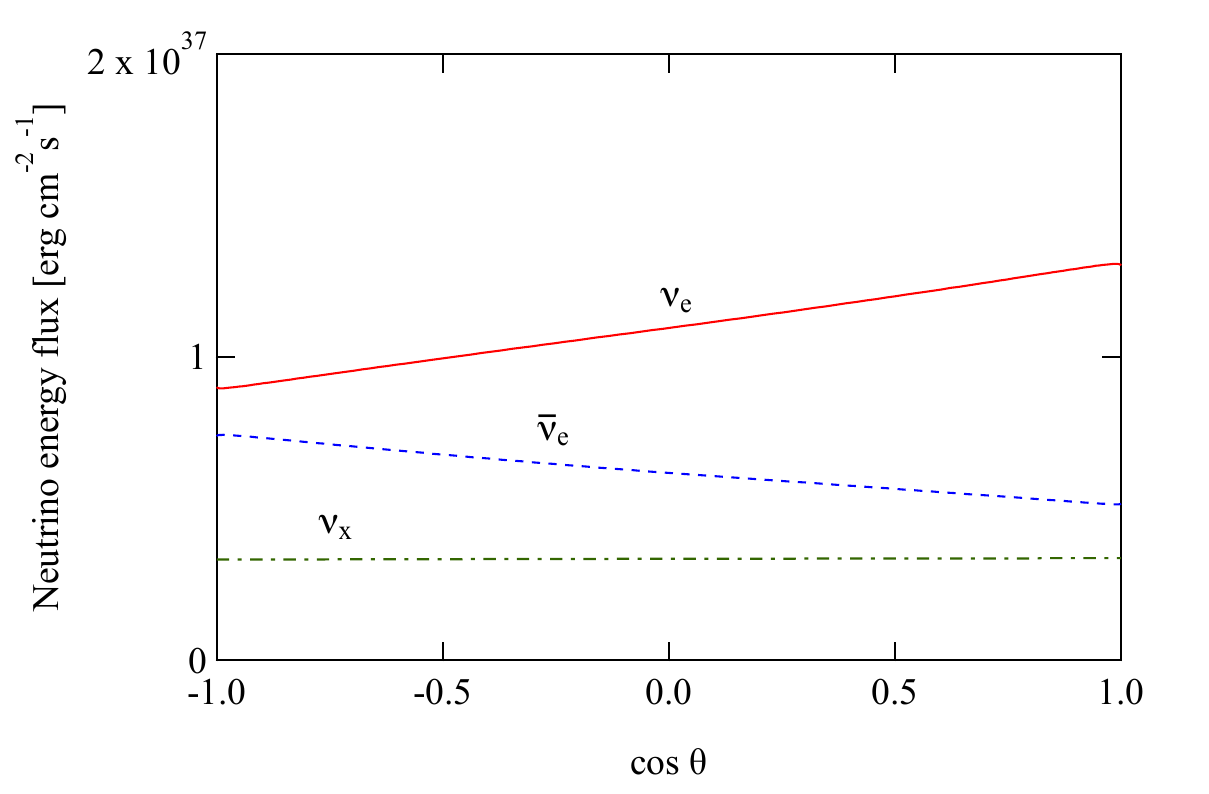}
\end{minipage}
\caption{  
Radial component of energy fluxes for each species of neutrinos at the outer boundary ($210 {\rm km}$) for baseline model.
Sold, dashed, and dot-dashed lines denote the energy flux of $\nu_e$, $\bar{\nu}_e$, and $\nu_{\mu}$, respectively.}
\label{fig_3nu_momentum_rate_z}
\end{figure}

It would also be interesting to quantify the asymmetry of neutrino energy fluxes for different neutrino species at the outer boundary of computational domain ($r = 210 {\rm km}$), which is shown in Fig. \ref{fig_3nu_momentum_rate_z}. The fluxes along the radial coordinate clearly exhibit that there are north-south asymmetries for $\nu_e$ and $\bar{\nu}_e$, in which the fluxes are larger in the north direction for the former and in the south direction for the latter, respectively. On the other hand, the energy flux of $\nu_x$ is much less angular dependent than $\nu_e$ and $\bar{\nu}_e$. These trends are consistent with the spatial distributions of neutrino number densities as shown in Fig.~\ref{fig_3nu_density_flux}.
 
There are two important remarks here. First, $\nu_e$ energy flux is remarkably higher than $\bar{\nu}_e$ in baseline model. This is partially due to the discrepancy between the two Boltzmann codes (see Sec.~\ref{subsec:fluid}). It should also be mentioned that we adopt the angular averaged $Y_e$ (or $Y^{1D}_e$ in Eq.~\ref{eq:Yedeform}) from a result of spherically symmetric CCSN model. As mentioned already, $Y_e$ tends to be higher around PNS envelope in spherically symmetric model due to lack of PNS convection \cite{2020MNRAS.492.5764N}. The disparity between $\nu_e$ and $\bar{\nu}_e$ should be, hence, relaxed in more self-consistent multi-dimensional models. Second, the energy flux of $\nu_x$ is the lowest, but it does not mean that $\nu_x$ is subdominant to carry energies. Importantly, the $\nu_x$ represents a single species of heavy-leptonic neutrinos, indicating that their total fluxes is four times higher than the value displayed in Fig. \ref{fig_3nu_momentum_rate_z}. This also exhibits that the small asymmetry of $\nu_x$ can contribute the NS kick (see also \cite{2019ApJ...880L..28N}).

\begin{figure}[ht]
\begin{minipage}{0.5\textwidth}
\centering
\includegraphics[width=1.0\linewidth]{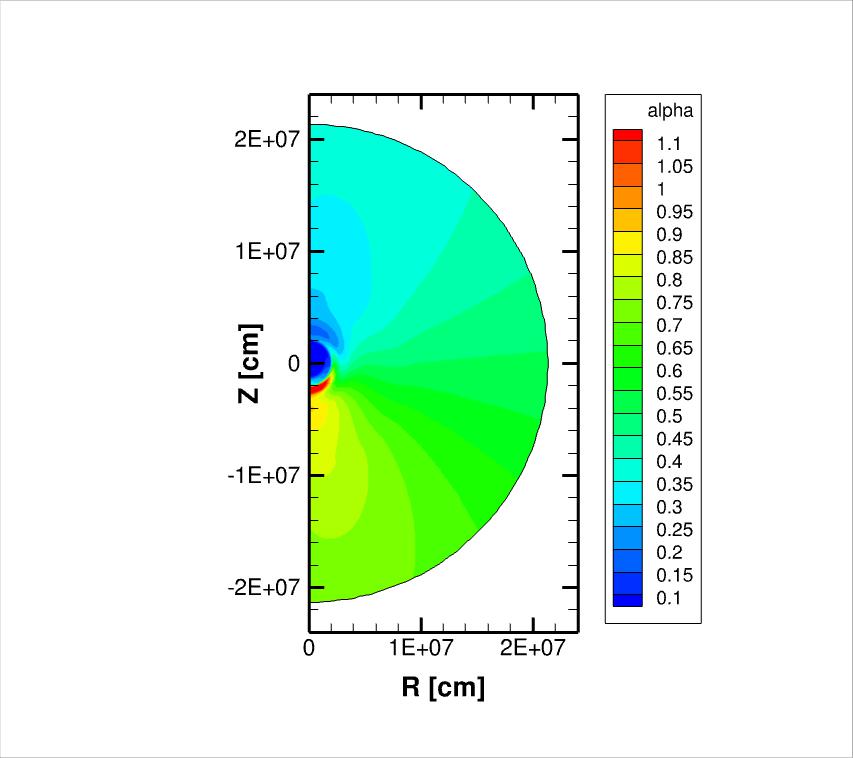}
\end{minipage}
\caption{Ratio of the number density of $\bar{\nu}_e$ to the number density of $\nu_e$ for baseline model is shown on the meridian slice by color map.}
\label{fig_3nu_density_ratio}
\end{figure}

\begin{figure}[ht]
\begin{minipage}{0.5\textwidth}
\centering
\includegraphics[width=1.0\linewidth]{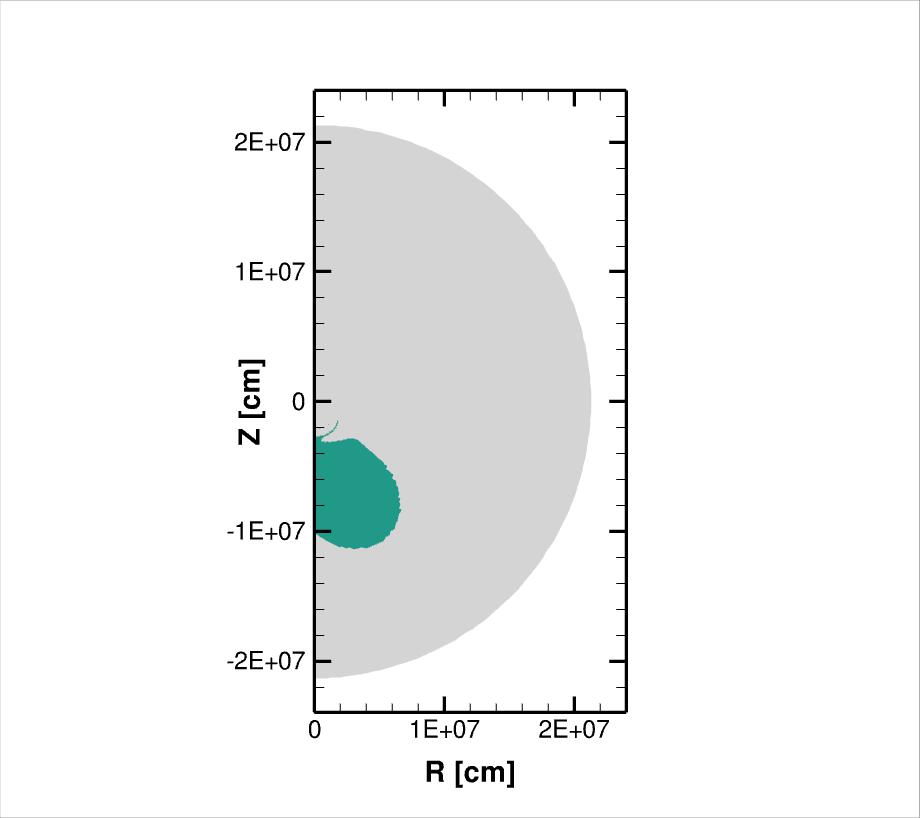}
\end{minipage}
\caption{2D color map highlighting the region of 
the ELN angular crossing in baseline model.
}
\label{fig_3nu_ELN_region}
\end{figure}

Figure~\ref{fig_3nu_density_ratio} displays a 2D color map of $\alpha$ (ratio of the densities of $\bar{\nu}_e$ and $\nu_e$; see Eq.~\ref{eqn:defalpha}) for baseline model. We find that a region with $\alpha \sim 1$ appears in the southern hemisphere. This offers a preferable condition for occurrences of FFC, which is portrayed in Fig.~\ref{fig_3nu_ELN_region}. In the figure, we highlight regions having ELN angular crossings by green color, where the ELN angular distribution is defined as
\begin{eqnarray}
G_{v_{\nu_e}}=\sqrt{2} \, \frac{G_F}{\hbar c}\int_{0}^{\infty}\frac{\varepsilon_{\nu}^2 d\varepsilon_{\nu}}{(hc)^3} \, [f_{\nu_e}(p_{\nu}) - f_{\bar{\nu}_e}(p_{\nu})].
\label{eqn:eln}
\end{eqnarray}
In the expression, $G_F$, $\hbar$, and $c$ denote Fermi constant, reduced Planck constant, and the speed of light, respectively. We note that XLN is always zero in the baseline model, implying that the occurrences of FFC can be assessed only by ELN angular distributions. The occurrence of ELN angular crossing corresponds to the case that the sign of $G_{v_{\nu_e}}$ changes in the angular distribution. We find that ELN angular crossings occur in the region with $\alpha \sim 1$ (southern hemisphere), which corresponds to lower $Y_e$ region where $\bar{\nu}_e$ ($\nu_e$) emission is also stronger (weaker). This implies that FFC must occur in the region, which leads to a different steady radiation field in FFC model; the detail is discussed in the next subsection.

\subsection{FFC model}\label{subsec:FFCmodel}

\begin{figure*}[ht]
\begin{minipage}{1.0\textwidth}
\centering
\includegraphics[width=0.45\linewidth]{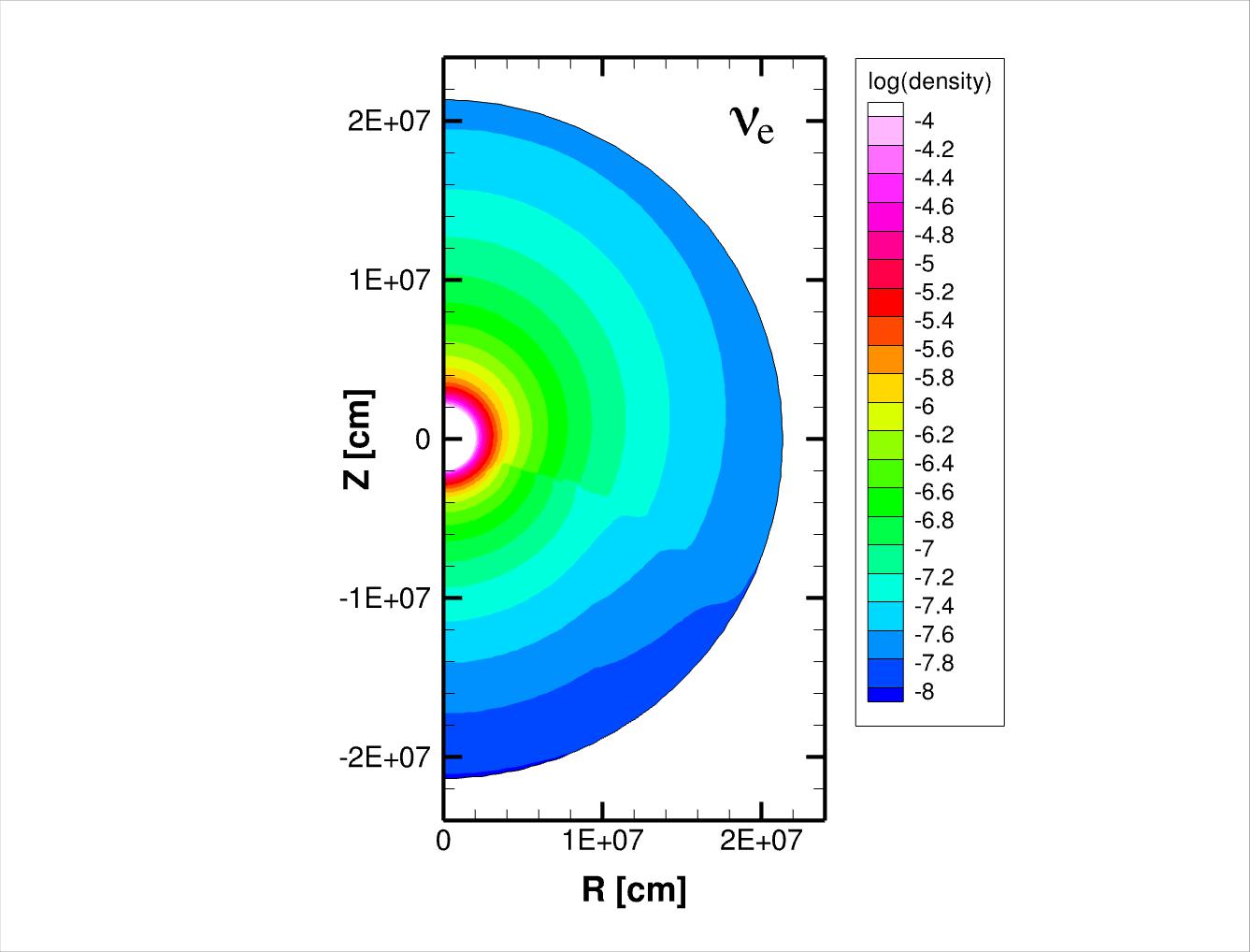}
\includegraphics[width=0.45\linewidth]{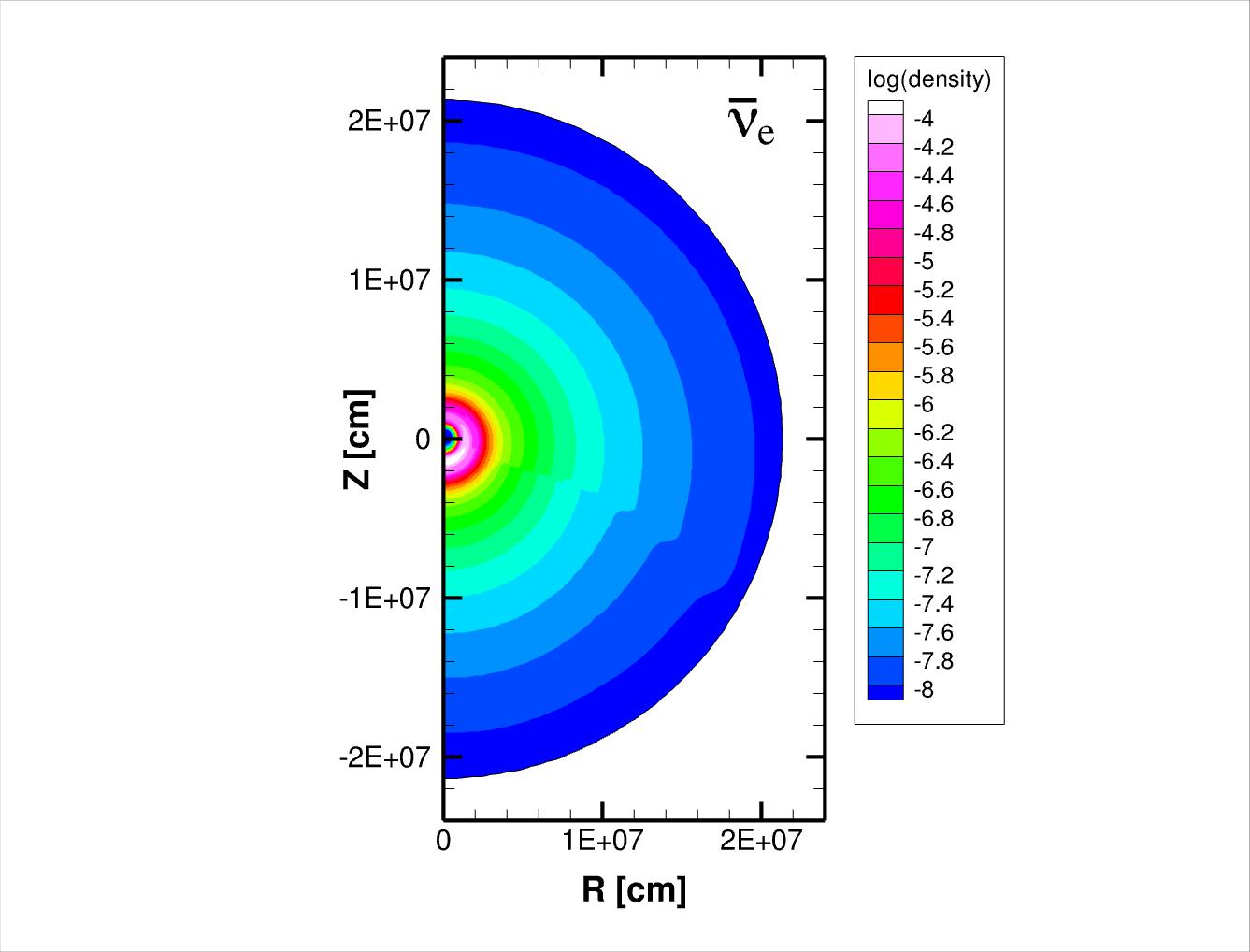}
\includegraphics[width=0.45\linewidth]{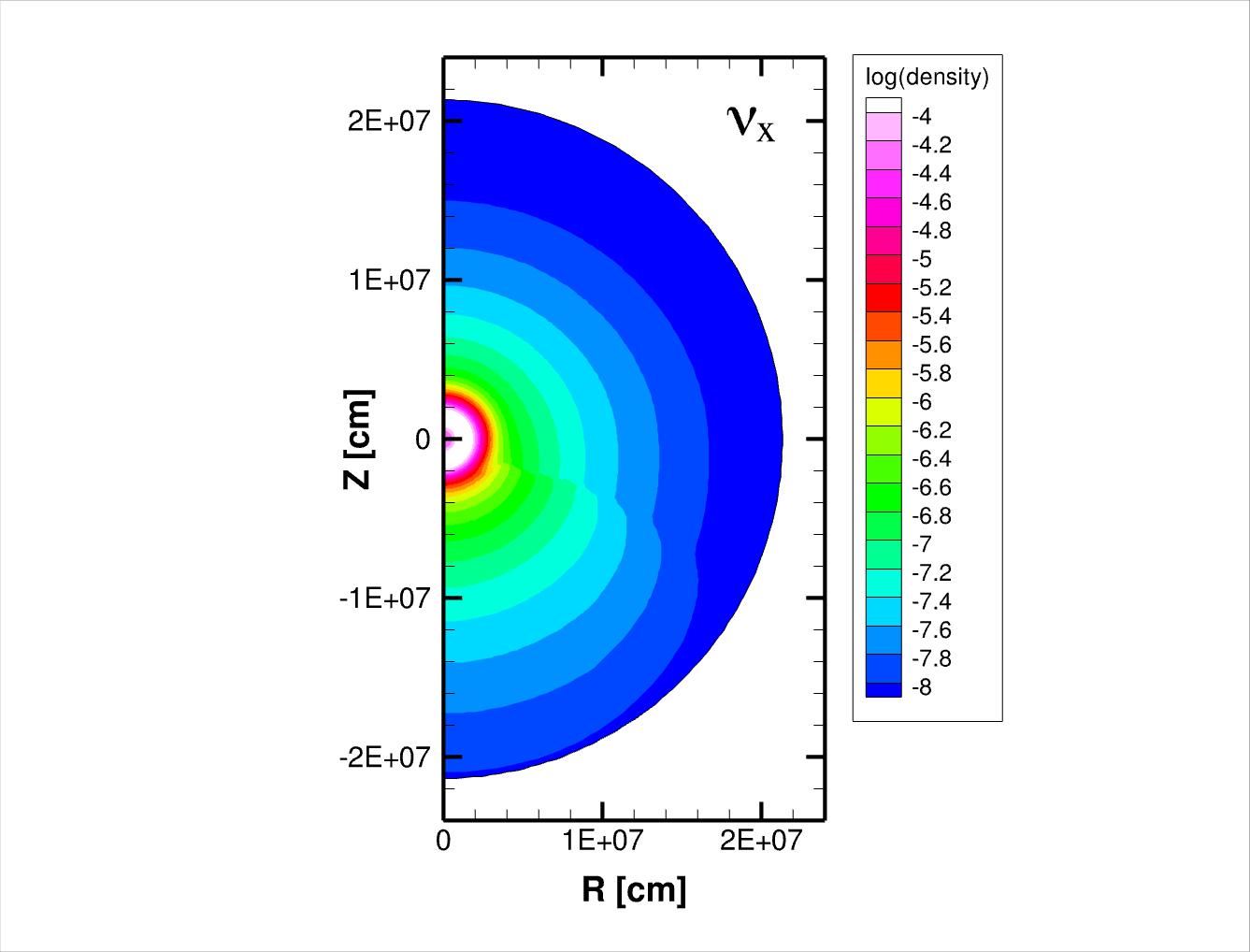}
\includegraphics[width=0.45\linewidth]{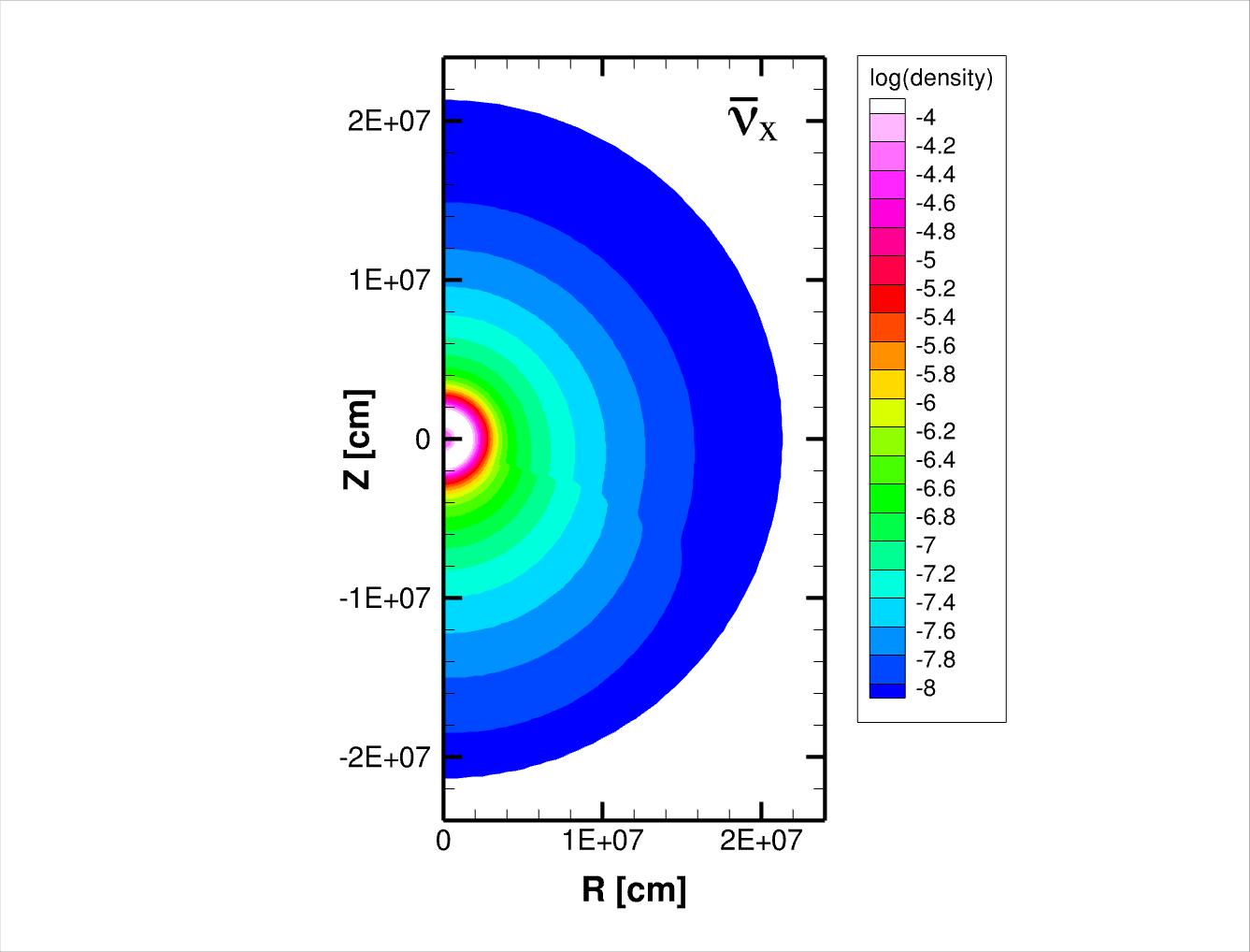}
\end{minipage}
\caption{Same as Fig.~\ref{fig_3nu_density_flux} but for FFC model. Since $\nu_x$ and $\bar{\nu}_x$ are no longer identical, they are displayed in different panels: $\nu_e$ (upper-left), $\nu_x$ (lower-left), $\bar{\nu}_e$ (upper-right), and  $\bar{\nu}_x$ (lower-right).
}
\label{fig_4nu_density_flux}
\end{figure*}

We show in Fig. \ref{fig_4nu_density_flux} the neutrino number densities for $\nu_e$, $\bar{\nu}_e$, $\nu_{x}$, and $\bar{\nu}_{x}$.
Let us remind the reader that $\nu_{x}$ and $\bar{\nu}_{x}$ are no longer identical in FFC model. We find that distributions of all species of neutrinos are very different from those in baseline model (see also Fig.~\ref{fig_3nu_density_flux}). The $\nu_e$ distribution is shifted to the north side and has a rapid decline on the south side. This deficit arises from the conversion of $\nu_e$ to $\nu_{x}$ due to FFCs. The $\bar{\nu}_e$ distribution has a similar deformation in the opposite direction. It should be noted, however, that $\bar{\nu}_e$ in the southern region is somewhat diminished due to the conversion from $\bar{\nu}_e$ to $\bar{\nu}_{x}$, indicating that the asymmetry of $\bar{\nu}_e$ becomes mild compared to the baseline model. For $\nu_{x}$ and $\bar{\nu}_{x}$, they are enhanced in the southern region, which is due to flavor conversions from $\nu_e$ and $\bar{\nu}_e$, respectively.

\begin{figure}[ht]
\begin{minipage}{0.5\textwidth}
\centering
\includegraphics[width=1.0\linewidth]{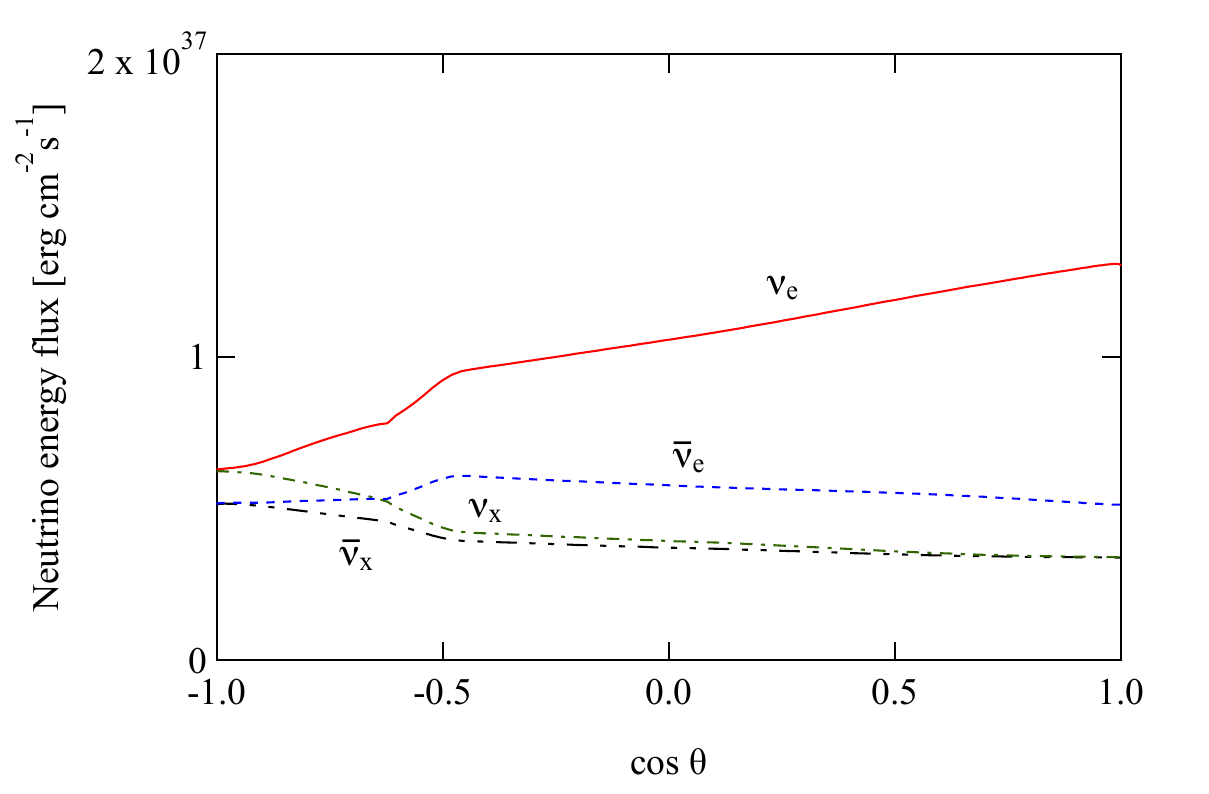}
\end{minipage}
\caption{
Same as Fig.~\ref{fig_3nu_momentum_rate_z} but for FFC model. Solid (red), dashed (blue), dot-dashed (green), and dot-dot-dashed (black) lines denote the energy flux of $\nu_e$, $\bar{\nu}_e$, $\nu_{x}$, and $\bar{\nu}_{x}$, respectively.
}
\label{fig_4nu_momentum_rate_z}
\end{figure}

In Fig.~\ref{fig_4nu_momentum_rate_z}, we display the energy fluxes for all species of neutrinos, measured at the outer boundary ($r = 220 {\rm km}$). This corresponds to the counterpart of Fig.~\ref{fig_3nu_momentum_rate_z} that displays the result of baseline model. It is worthy of note that $\nu_e$ and $\nu_x$ have nearly the same flux to each other at $\cos{\theta} = -1$ (or the south pole), while $\bar{\nu}_e$ and $\bar{\nu}_x$ also reach nearly flavor equipartition there. We note that the same trend has also been observed in quantum kinetic neutrino transport simulations \cite{2023PhRvL.130u1401N,2023PhRvD.108l3003N}, in which neutrinos and antineutrinos have different flavor equipartition states. This exhibits that the qualitative trend of non-linear evolutions of FFCs is captured by our mixing scheme (see Eqs.~\ref{eq:flavconv_nue}-\ref{eq:flavconv_nuxb}) with a condition of flavor equipartition.

As displayed in Figs.~\ref{fig_4nu_density_flux}~and~\ref{fig_4nu_momentum_rate_z}, it is noteworthy that the asymmetry of $\nu_{x}$ is higher than $\bar{\nu}_{x}$. This trend can be understood from the result of baseline model. As shown in Fig.~\ref{fig_3nu_density_ratio}, $\nu_e$ number density in the southern region is still higher than $\bar{\nu}_e$ except for the inner region of PNS envelop ($20 {\rm km} \lesssim r \lesssim 25 {\rm km}$). Relevant to this trend, $\nu_e$ flux at the outer boundary is also higher than $\bar{\nu}_e$ in the southern region (see the region of $\cos \theta <0$ in Fig.~\ref{fig_3nu_momentum_rate_z}). In such an environment, neutrinos undergo more flavor conversions than antineutrinos \cite{2023PhRvL.130u1401N,2023PhRvD.108l3003N}. It should be mentioned that this is not only due to FFC but also neutrino-matter interactions. In general, FFC is a pairwise conversion, indicating that the number of neutrinos and antineutrinos that experience flavor conversions should be the same. However, the collision term, in particular emission and absorption processes, can change the number of neutrinos and antineutrinos, which is responsible for the difference between $\nu_x$ and $\bar{\nu}_x$ in FFC model\footnote{It should be noted that our approximate mixing scheme does not guarantee the pairwise conversion.}. The dominance of $\nu_e$ indicates that the charged-current reaction of $\nu_e$ is stronger than $\bar{\nu}_e$. As a result, both the number density and flux of $\nu_x$ tend to be larger than those in $\bar{\nu}_x$.

However, there is a caveat to keep in mind that this trend ($n_{\nu_x} > n_{\bar{\nu}_x}$) may disappear in more realistic CCSN models. As already pointed out in Sec.~\ref{subsec:baselinemodel}, the angular-averaged $Y_e$ profile in PNS envelop is higher in our model (because we adopt the $Y_e$ profile from a spherically symmetric CCSN simulation), that results in stronger $\nu_e$ emission in the entire direction. This would be an artifact and the trend is at least reduced in more realistic situations. Nevertheless, our numerical simulations lend confidence our claim that both number densities and fluxes for $\nu_x$ and $\bar{\nu}_x$ become higher in the southern region, enhancing a linear momentum in neutrino radiation field. Below, we quantify the linear momentum of neutrinos for both baseline and FFC models, and we show that the result is in line with our FFC-driven NS kick scenario as described in Sec.~\ref{sec:basicpic}.

\subsection{Linear momentum carried by neutrinos}\label{subsec:anamomtrans}
We estimate linear momentum of neutrinos by following \cite{2019ApJ...878..160N}. Assuming steady state of neutrino radiation field, the total momentum balance of neutrinos in z-direction can be written as,
\begin{eqnarray}
\sum_{i}
\frac{1}{r^2 \sin \theta} \, \partial_{\alpha}(r^2 \sin \theta \, T_{\nu_i}^{\alpha z}) = \sum_{i} G_{\nu_i}^{z},
\label{eqn:momentum_conservation}
\end{eqnarray}
where $T_{\nu_i}^{\alpha z}$ and $G_{\nu_i}^{z}$ are the z-projection of the energy-momentum tensor of neutrinos and the momentum gain/loss by neutrino-matter interactions, respectively. In the expression, the index $i$ specifies the neutrino species: $\nu_i=\nu_e$, $\bar{\nu}_e$, $\nu_{\mu}$, $\bar{\nu}_{\mu}$, $\nu_{\tau}$, $\bar{\nu}_{\tau}$. The energy-momentum tensor of neutrinos can be computed as
\begin{eqnarray}
T_{\nu_i}^{\alpha \beta} = \int \frac{d\varepsilon \, \varepsilon^2}{(2 \pi)^3} \int d\Omega 
\, \varepsilon \, n^{\alpha} \, n^{\beta} f_{\nu_i}(\varepsilon,\Omega)
\label{eq:defTalphabeta}
\end{eqnarray}
where $f_{\nu_i}$ is the distribution function of neutrino, and $n^{\alpha}$ is the unit vector specifying neutrino flight directions in four-dimensional spacetime. Note that the energy-momentum tensor depends on space, although we omit to show them in Eq.~\ref{eq:defTalphabeta} just for simplicity.

The linear momentum carried by neutrinos per unit time at the surface of $r$ ($P^z_{\nu_i}$ which has a dimension of $[{\rm g cm/s^2}]$)) can be computed as
\begin{eqnarray}
P_{\nu_i}^z(r) = 2\pi r^2 \int_{0}^{\pi} T_{\nu_i}^{rz}(r, \theta) \sin\theta d\theta ,
\label{eqn:momentum_transfer_rate}
\end{eqnarray}
where the $z-$projection of the radial component of the energy-momentum tensor, $T_{\nu_i}^{rz}$, is given by,
\begin{eqnarray}
T_{\nu_i}^{rz} = T_{\nu_i}^{rr} \cos \theta - T_{\nu_i}^{r\theta} \sin \theta .
\label{eqn:momentum_transfer_rate_component}
\end{eqnarray}
We note that the flavor-integrated $P^z_{\nu_i}$ can also be evaluated from the volume integral of $G^z_{\nu_i}$ as (see Eq.~\ref{eqn:momentum_conservation}),
\begin{eqnarray}
\sum_{i} P_{\nu_i}^z(r) = \sum_{i} 2\pi \int_{0}^{r} \int_{0}^{\pi} r^2 \sin\theta \, G_{\nu_i}^{z}(r', \theta) \, d\theta dr',
\label{eq:linmomneutrinoperunittime}
\end{eqnarray}
which represents the linear momentum transfer from matter to neutrinos. Because of the conservation of law of total energy and momentum in the system, the fluid needs to gain the linear momentum of the opposite sign of Eq.~\ref{eq:linmomneutrinoperunittime}, representing the recoil by asymmetric neutrino emission. This leads to a NS natal kick.

\begin{figure}[ht]
\begin{minipage}{0.5\textwidth}
\centering
\includegraphics[width=1.0\linewidth]{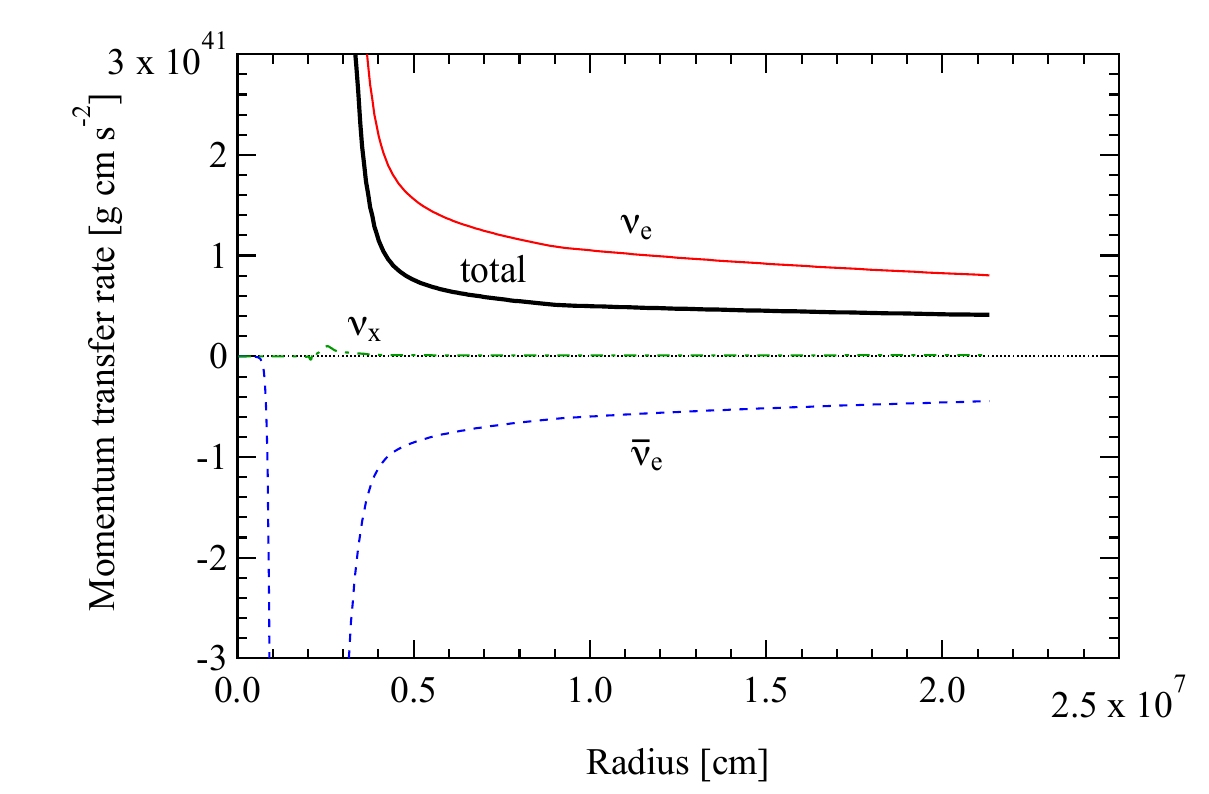}
\includegraphics[width=1.0\linewidth]{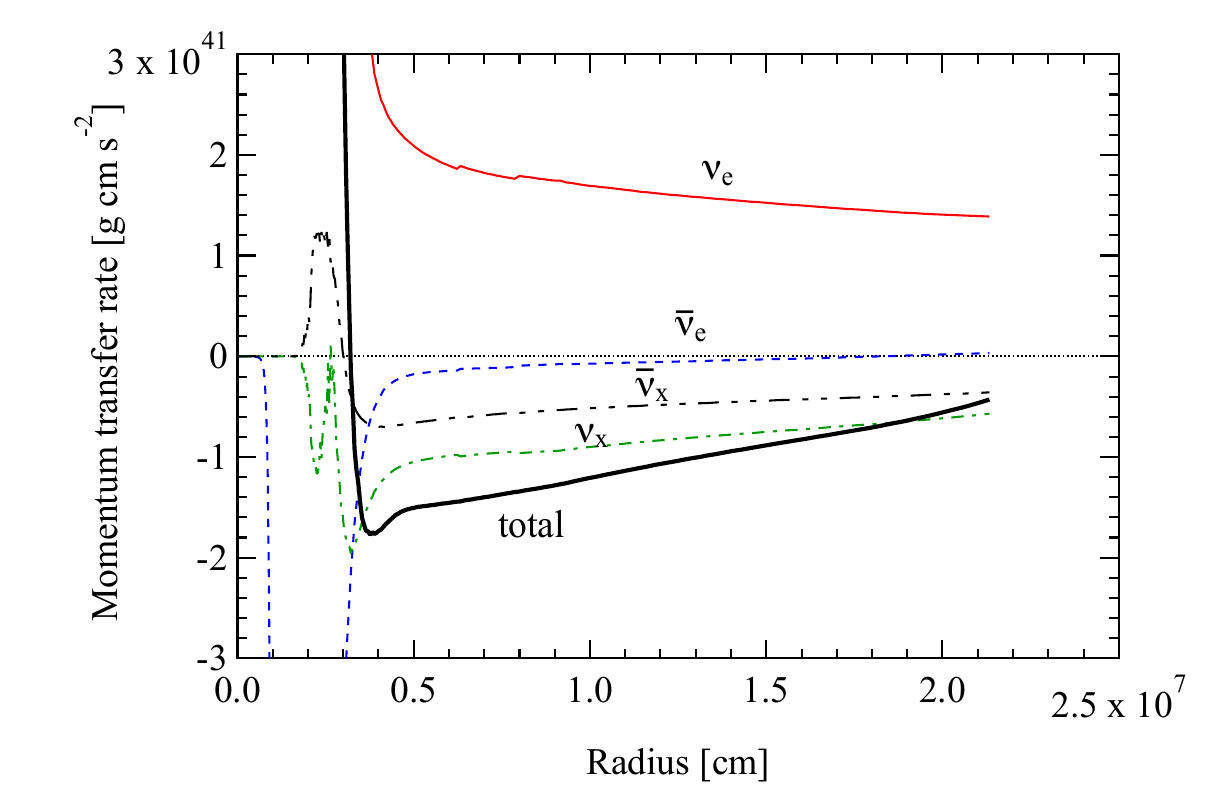}
\end{minipage}
\caption{Rates of momentum transfer in the z-direction are shown for baseline and FFC models as functions of radius in upper and lower panels, respectively.  
Solid (red), dashed (blue), dot-dashed (green), and dot-dot-dashed (black) lines denote the rate for $\nu_e$, $\bar{\nu}_e$, $\nu_{x}$, and $\bar{\nu}_{x}$, respectively.
Thick lines denote the total rate.}
\label{fig:momtransfercompare}
\end{figure}

In Fig.~\ref{fig:momtransfercompare}, we compare the species dependent $P_{\nu_i}^z$ as a function of radius for both baseline (upper panel) and FFC (lower panel) models. In baseline model (upper panel), $\nu_x$ is nearly spherical, and hence their linear momentum is negligible. On the other hand, $\nu_e$ has higher emission than $\bar{\nu}_e$, while the asymmetric degree is roughly the same to each other. As a result, the total linear momentum is in the direction of $\nu_e$, i.e., the northern direction  (or high $Y_e$ hemisphere) in the baseline model. As shown in the lower panel, however, FFCs substantially change the linear momentum of neutrinos. Interestingly, the linear momentum carried by the sum of $\nu_e$ and $\bar{\nu}_e$ is in the northern direction, and the magnitude is even larger than the baseline model. This is attributed to the fact that the global asymmetries in $\nu_e$ ($\bar{\nu}_e$) emission are enhanced (reduced) by FFCs (see Sec.~\ref{subsec:FFCmodel}). Nevertheless, the total flavor-integrated linear momentum is flipped and pointed in the southern direction (or low $Y_e$ hemisphere), indicating that the linear momentum carried by $\nu_x$ and $\bar{\nu}_x$ overwhelm $\nu_e$ and $\bar{\nu}_e$. This is exactly what we expected in our FFC-driven NS kick scenario (see Sec.~\ref{sec:basicpic}). The total neutrino emission is enhanced in the region where FFC occurs. We also find that the increase of $\nu_x$ asymmetry is remarkable, representing many $\nu_e$s in the southern direction undergo flavor conversions to $\nu_x$.

As shown in Fig.~\ref{fig:momtransfercompare}, the total linear momentum carried by neutrinos per unit time is $\sim 5 \times 10^{40} {\rm g cm/s^2}$ and $\sim - 5 \times 10^{40} {\rm g cm/s^2}$ at the outer boundary for baseline and FFC models, respectively, which exhibits that FFCs change the linear momentum of neutrinos by $\sim 10^{41} {\rm g cm/s^2}$. One thing we do emphasize here is that this is a rough estimation and detailed inspections are necessary for more quantative arguments. In fact, the total linear momentum does not reach the asymptotic value at the outer boundary. The change of total momentum at the outer radii is mainly due to the coherent scattering by heavy nuclei, which corresponds to the dominant opacity source for neutrinos in pre-shock region. In the hemisphere of higher luminosity, neutrinos undergo those scattering more frequent than in the other hemisphere, leading to the decrease of total linear momentum. This potentially reduces the impact of FFCs on NS kick, suggesting that the feedback should be taken into account for more accurate estimation.

The lifetime of asymmetric neutrino emission depends on CCSN dynamics, but it would be typically a few hundred of milliseconds \cite{2014ApJ...792...96T,2019ApJ...881...36G,2019MNRAS.487.1178P,2019ApJ...886..139N,2019ApJ...880L..28N}, suggesting that the impact of FFC in total linear momentum could be $\sim$ a few $\times 10^{40} {\rm g cm/s}$. This indicates FFC potentially has the ability to change the linear monentum of neutrinos by this order, i.e., generating $\sim$ a few $\times 0.1 \%$ asymmetry with respect to the total neutrino emission. It should be emphasized, however, that the present study is meant as a proof-of-principle, and our model is too simple to draw robust conclusion whether the mechanism can account for the observed velocity distributions of NS proper motions. Nevertheless, this demonstration offers a possibility that globally asymmetric FFCs can induce NS natal kick.

\section{Summary}\label{sec:summary}
In this paper, we propose a new channel to generate a neutron star (NS) natal kick during developments of CCSN explosions, upon which fast neutrino-flavor conversion (FFC), one of the quantum kinetic features of neutrinos, plays an important role. FFC tends to occur in the low $Y_e$ environments, in which the $\nu_e$ degeneracy becomes mild and consequently ELN angular crossings can occur. The large-scale asymmetric $Y_e$ distributions have been observed in recent multi-dimensional CCSN models such as LESA \cite{2014ApJ...792...96T,2019ApJ...881...36G,2019MNRAS.487.1178P} and a feedback mechanism between asymmetric neutrino emission and NS kick \cite{2019ApJ...880L..28N}. As shown in \cite{2019ApJ...886..139N,2021PhRvD.104h3025N}, FFC can occur in such asymmetric neutrino radiation fields, and we make a statement in this paper that the linear momentum of neutrinos are enhanced by FFCs in the hemisphere of low $Y_e$ environment.

One thing we do notice here is that this scenario is inspired by previous studies, and it has been reached a consensus that the enhancement of neutrino cooling is one of the characteristic features of FFC. This is attributed to the fact that FFC can increase the number of heavy-leptonic neutrinos, while they are more optically thin than electron-type neutrinos due to the lack of charged-current reactions. This indicates that the opacity for flavor-integrated neutrinos are reduced by FFCs, leading to the enhancement of neutrino cooling. Since the neutrinos carry not only energy but also momentum, a linear momentum can also be generated (see Sec.~\ref{sec:basicpic} for more details of the FFC-driven NS kick mechanism).

We also perform axisymmetric neutrino transport simulations, which validates our proposed scenario. In our models, we deform the fluid background, in particular $Y_e$ distribution, as a dipole shape, which can generate linear momentum of neutrinos in z-direction. In the present study, we perform two simulations, one of which corresponds to a purely classical model (baseline model) and the other incorporates effects of FFC in a phenomenological way (FFC model). We show that FFC enhances a neutrino momentum in the direction of low $Y_e$ environment and also provide physical processes how to generate the linear momentum. We also note that FFC-driven NS kick mechanism makes the NS to accelerate in the high $Y_e$ direction, indicating that the asymmetric distributions of synthesized heavy elements in the ejecta would be correlated to the NS kick direction. This will be checked by future observations of SNR.

To compare these observations, we need more detailed study for FFC-driven NS kick scenario based on three-dimensional CCSN simulations by systematically changing progenitor models. It should also be mentioned that more accurate prescriptions for FFC are also necessary for quantitative arguments. In fact, we assume that flavor equipartition achieves for all places where ELN-XLN angular crossings appear. However, this treatment is obviously a crude treatment, and it should be improved in future studies. A new subgrid model, namely BGK-model in \cite{2023arXiv231216285N}, will help us to carry out global CCSN simulations with better FFC prescriptions; the results will be reported in our forthcoming papers.

\section{Acknowledgments}
This work is supported by the HPCI System Research Project (Project ID: 220173, 220047, 220223, 230033, 230204, 230270), XC50 of CfCA at the National Astronomical Observatory of Japan (NAOJ), Yukawa-21 at Yukawa Institute for Theoretical Physics of Kyoto University, Research Center for Nuclear Physics, Osaka University, Information Technology Center, University of Tokyo, and the High Energy Accelerator Research Organization (KEK). For providing high performance computing resources, Computing Research Center, KEK, and JLDG on SINET of NII are acknowledged. HN is supported by Grant-inAid for Scientific Research (23K03468). KS is supported by Grant-in-Aid for Scientific Research (19K03837, 20H01905, 24K00632). This work is supported by MEXT as "Program for Promoting Researches on the Supercomputer Fugaku" 
(Structure and Evolution of the Universe Unraveled by Fusion of Simulation and AI, JPMXP1020230406) and the Particle, Nuclear and Astro Physics Simulation Program (Nos. 2021-004, 2022-003, 2023-003) of Institute of Particle and Nuclear Studies, High Energy Accelerator Research Organization (KEK).
\bibliography{bibfile}

\begin{thebibliography}{89}%
\makeatletter
\providecommand \@ifxundefined [1]{%
 \@ifx{#1\undefined}
}%
\providecommand \@ifnum [1]{%
 \ifnum #1\expandafter \@firstoftwo
 \else \expandafter \@secondoftwo
 \fi
}%
\providecommand \@ifx [1]{%
 \ifx #1\expandafter \@firstoftwo
 \else \expandafter \@secondoftwo
 \fi
}%
\providecommand \natexlab [1]{#1}%
\providecommand \enquote  [1]{``#1''}%
\providecommand \bibnamefont  [1]{#1}%
\providecommand \bibfnamefont [1]{#1}%
\providecommand \citenamefont [1]{#1}%
\providecommand \href@noop [0]{\@secondoftwo}%
\providecommand \href [0]{\begingroup \@sanitize@url \@href}%
\providecommand \@href[1]{\@@startlink{#1}\@@href}%
\providecommand \@@href[1]{\endgroup#1\@@endlink}%
\providecommand \@sanitize@url [0]{\catcode `\\12\catcode `\$12\catcode
  `\&12\catcode `\#12\catcode `\^12\catcode `\_12\catcode `\%12\relax}%
\providecommand \@@startlink[1]{}%
\providecommand \@@endlink[0]{}%
\providecommand \url  [0]{\begingroup\@sanitize@url \@url }%
\providecommand \@url [1]{\endgroup\@href {#1}{\urlprefix }}%
\providecommand \urlprefix  [0]{URL }%
\providecommand \Eprint [0]{\href }%
\providecommand \doibase [0]{http://dx.doi.org/}%
\providecommand \selectlanguage [0]{\@gobble}%
\providecommand \bibinfo  [0]{\@secondoftwo}%
\providecommand \bibfield  [0]{\@secondoftwo}%
\providecommand \translation [1]{[#1]}%
\providecommand \BibitemOpen [0]{}%
\providecommand \bibitemStop [0]{}%
\providecommand \bibitemNoStop [0]{.\EOS\space}%
\providecommand \EOS [0]{\spacefactor3000\relax}%
\providecommand \BibitemShut  [1]{\csname bibitem#1\endcsname}%
\let\auto@bib@innerbib\@empty
\bibitem [{\citenamefont {{Wang}}\ and\ \citenamefont
  {{Wheeler}}(2008)}]{2008ARA&A..46..433W}%
  \BibitemOpen
  \bibfield  {author} {\bibinfo {author} {\bibfnamefont {L.}~\bibnamefont
  {{Wang}}}\ and\ \bibinfo {author} {\bibfnamefont {J.~C.}\ \bibnamefont
  {{Wheeler}}},\ }\bibfield  {title} {\enquote {\bibinfo {title}
  {{Spectropolarimetry of supernovae.}}}\ }\href {\doibase
  10.1146/annurev.astro.46.060407.145139} {\bibfield  {journal} {\bibinfo
  {journal} {Annual Review of Astronomy and Astrophysics}\ }\textbf {\bibinfo
  {volume} {46}},\ \bibinfo {pages} {433--474} (\bibinfo {year} {2008})},\
  \Eprint {http://arxiv.org/abs/0811.1054} {arXiv:0811.1054 [astro-ph]}
  \BibitemShut {NoStop}%
\bibitem [{\citenamefont {{Abell{\'a}n}}\ \emph {et~al.}(2017)\citenamefont
  {{Abell{\'a}n}}, \citenamefont {{Indebetouw}}, \citenamefont {{Marcaide}},
  \citenamefont {{Gabler}}, \citenamefont {{Fransson}}, \citenamefont
  {{Spyromilio}}, \citenamefont {{Burrows}}, \citenamefont {{Chevalier}},
  \citenamefont {{Cigan}}, \citenamefont {{Gaensler}}, \citenamefont {{Gomez}},
  \citenamefont {{Janka}}, \citenamefont {{Kirshner}}, \citenamefont
  {{Larsson}}, \citenamefont {{Lundqvist}}, \citenamefont {{Matsuura}},
  \citenamefont {{McCray}}, \citenamefont {{Ng}}, \citenamefont {{Park}},
  \citenamefont {{Roche}}, \citenamefont {{Staveley-Smith}}, \citenamefont
  {{van Loon}}, \citenamefont {{Wheeler}},\ and\ \citenamefont
  {{Woosley}}}]{2017ApJ...842L..24A}%
  \BibitemOpen
  \bibfield  {author} {\bibinfo {author} {\bibfnamefont {F.~J.}\ \bibnamefont
  {{Abell{\'a}n}}}, \bibinfo {author} {\bibfnamefont {R.}~\bibnamefont
  {{Indebetouw}}}, \bibinfo {author} {\bibfnamefont {J.~M.}\ \bibnamefont
  {{Marcaide}}}, \bibinfo {author} {\bibfnamefont {M.}~\bibnamefont
  {{Gabler}}}, \bibinfo {author} {\bibfnamefont {C.}~\bibnamefont
  {{Fransson}}}, \bibinfo {author} {\bibfnamefont {J.}~\bibnamefont
  {{Spyromilio}}}, \bibinfo {author} {\bibfnamefont {D.~N.}\ \bibnamefont
  {{Burrows}}}, \bibinfo {author} {\bibfnamefont {R.}~\bibnamefont
  {{Chevalier}}}, \bibinfo {author} {\bibfnamefont {P.}~\bibnamefont
  {{Cigan}}}, \bibinfo {author} {\bibfnamefont {B.~M.}\ \bibnamefont
  {{Gaensler}}}, \bibinfo {author} {\bibfnamefont {H.~L.}\ \bibnamefont
  {{Gomez}}}, \bibinfo {author} {\bibfnamefont {H.~Th.}\ \bibnamefont
  {{Janka}}}, \bibinfo {author} {\bibfnamefont {R.}~\bibnamefont {{Kirshner}}},
  \bibinfo {author} {\bibfnamefont {J.}~\bibnamefont {{Larsson}}}, \bibinfo
  {author} {\bibfnamefont {P.}~\bibnamefont {{Lundqvist}}}, \bibinfo {author}
  {\bibfnamefont {M.}~\bibnamefont {{Matsuura}}}, \bibinfo {author}
  {\bibfnamefont {R.}~\bibnamefont {{McCray}}}, \bibinfo {author}
  {\bibfnamefont {C.~Y.}\ \bibnamefont {{Ng}}}, \bibinfo {author}
  {\bibfnamefont {S.}~\bibnamefont {{Park}}}, \bibinfo {author} {\bibfnamefont
  {P.}~\bibnamefont {{Roche}}}, \bibinfo {author} {\bibfnamefont
  {L.}~\bibnamefont {{Staveley-Smith}}}, \bibinfo {author} {\bibfnamefont
  {J.~Th.}\ \bibnamefont {{van Loon}}}, \bibinfo {author} {\bibfnamefont
  {J.~C.}\ \bibnamefont {{Wheeler}}}, \ and\ \bibinfo {author} {\bibfnamefont
  {S.~E.}\ \bibnamefont {{Woosley}}},\ }\bibfield  {title} {\enquote {\bibinfo
  {title} {{Very Deep inside the SN 1987A Core Ejecta: Molecular Structures
  Seen in 3D}},}\ }\href {\doibase 10.3847/2041-8213/aa784c} {\bibfield
  {journal} {\bibinfo  {journal} {\apjl}\ }\textbf {\bibinfo {volume} {842}},\
  \bibinfo {eid} {L24} (\bibinfo {year} {2017})},\ \Eprint
  {http://arxiv.org/abs/1706.04675} {arXiv:1706.04675 [astro-ph.SR]}
  \BibitemShut {NoStop}%
\bibitem [{\citenamefont {{Hobbs}}\ \emph {et~al.}(2005)\citenamefont
  {{Hobbs}}, \citenamefont {{Lorimer}}, \citenamefont {{Lyne}},\ and\
  \citenamefont {{Kramer}}}]{2005MNRAS.360..974H}%
  \BibitemOpen
  \bibfield  {author} {\bibinfo {author} {\bibfnamefont {G.}~\bibnamefont
  {{Hobbs}}}, \bibinfo {author} {\bibfnamefont {D.~R.}\ \bibnamefont
  {{Lorimer}}}, \bibinfo {author} {\bibfnamefont {A.~G.}\ \bibnamefont
  {{Lyne}}}, \ and\ \bibinfo {author} {\bibfnamefont {M.}~\bibnamefont
  {{Kramer}}},\ }\bibfield  {title} {\enquote {\bibinfo {title} {{A statistical
  study of 233 pulsar proper motions}},}\ }\href {\doibase
  10.1111/j.1365-2966.2005.09087.x} {\bibfield  {journal} {\bibinfo  {journal}
  {\mnras}\ }\textbf {\bibinfo {volume} {360}},\ \bibinfo {pages} {974--992}
  (\bibinfo {year} {2005})},\ \Eprint {http://arxiv.org/abs/astro-ph/0504584}
  {arXiv:astro-ph/0504584 [astro-ph]} \BibitemShut {NoStop}%
\bibitem [{\citenamefont {{Lyne}}\ and\ \citenamefont
  {{Lorimer}}(1994)}]{1994Natur.369..127L}%
  \BibitemOpen
  \bibfield  {author} {\bibinfo {author} {\bibfnamefont {A.~G.}\ \bibnamefont
  {{Lyne}}}\ and\ \bibinfo {author} {\bibfnamefont {D.~R.}\ \bibnamefont
  {{Lorimer}}},\ }\bibfield  {title} {\enquote {\bibinfo {title} {{High birth
  velocities of radio pulsars}},}\ }\href {\doibase 10.1038/369127a0}
  {\bibfield  {journal} {\bibinfo  {journal} {\nat}\ }\textbf {\bibinfo
  {volume} {369}},\ \bibinfo {pages} {127--129} (\bibinfo {year}
  {1994})}\BibitemShut {NoStop}%
\bibitem [{\citenamefont {{Cordes}}\ and\ \citenamefont
  {{Chernoff}}(1998)}]{1998ApJ...505..315C}%
  \BibitemOpen
  \bibfield  {author} {\bibinfo {author} {\bibfnamefont {J.~M.}\ \bibnamefont
  {{Cordes}}}\ and\ \bibinfo {author} {\bibfnamefont {David~F.}\ \bibnamefont
  {{Chernoff}}},\ }\bibfield  {title} {\enquote {\bibinfo {title} {{Neutron
  Star Population Dynamics. II. Three-dimensional Space Velocities of Young
  Pulsars}},}\ }\href {\doibase 10.1086/306138} {\bibfield  {journal} {\bibinfo
   {journal} {\apj}\ }\textbf {\bibinfo {volume} {505}},\ \bibinfo {pages}
  {315--338} (\bibinfo {year} {1998})},\ \Eprint
  {http://arxiv.org/abs/astro-ph/9707308} {arXiv:astro-ph/9707308 [astro-ph]}
  \BibitemShut {NoStop}%
\bibitem [{\citenamefont {{Arzoumanian}}\ \emph {et~al.}(2002)\citenamefont
  {{Arzoumanian}}, \citenamefont {{Chernoff}},\ and\ \citenamefont
  {{Cordes}}}]{2002ApJ...568..289A}%
  \BibitemOpen
  \bibfield  {author} {\bibinfo {author} {\bibfnamefont {Z.}~\bibnamefont
  {{Arzoumanian}}}, \bibinfo {author} {\bibfnamefont {D.~F.}\ \bibnamefont
  {{Chernoff}}}, \ and\ \bibinfo {author} {\bibfnamefont {J.~M.}\ \bibnamefont
  {{Cordes}}},\ }\bibfield  {title} {\enquote {\bibinfo {title} {{The Velocity
  Distribution of Isolated Radio Pulsars}},}\ }\href {\doibase 10.1086/338805}
  {\bibfield  {journal} {\bibinfo  {journal} {\apj}\ }\textbf {\bibinfo
  {volume} {568}},\ \bibinfo {pages} {289--301} (\bibinfo {year} {2002})},\
  \Eprint {http://arxiv.org/abs/astro-ph/0106159} {arXiv:astro-ph/0106159
  [astro-ph]} \BibitemShut {NoStop}%
\bibitem [{\citenamefont {{Faucher-Gigu{\`e}re}}\ and\ \citenamefont
  {{Kaspi}}(2006)}]{2006ApJ...643..332F}%
  \BibitemOpen
  \bibfield  {author} {\bibinfo {author} {\bibfnamefont {Claude-Andr{\'e}}\
  \bibnamefont {{Faucher-Gigu{\`e}re}}}\ and\ \bibinfo {author} {\bibfnamefont
  {Victoria~M.}\ \bibnamefont {{Kaspi}}},\ }\bibfield  {title} {\enquote
  {\bibinfo {title} {{Birth and Evolution of Isolated Radio Pulsars}},}\ }\href
  {\doibase 10.1086/501516} {\bibfield  {journal} {\bibinfo  {journal} {\apj}\
  }\textbf {\bibinfo {volume} {643}},\ \bibinfo {pages} {332--355} (\bibinfo
  {year} {2006})},\ \Eprint {http://arxiv.org/abs/astro-ph/0512585}
  {arXiv:astro-ph/0512585 [astro-ph]} \BibitemShut {NoStop}%
\bibitem [{\citenamefont {{Chatterjee}}\ \emph {et~al.}(2005)\citenamefont
  {{Chatterjee}}, \citenamefont {{Vlemmings}}, \citenamefont {{Brisken}},
  \citenamefont {{Lazio}}, \citenamefont {{Cordes}}, \citenamefont {{Goss}},
  \citenamefont {{Thorsett}}, \citenamefont {{Fomalont}}, \citenamefont
  {{Lyne}},\ and\ \citenamefont {{Kramer}}}]{2005ApJ...630L..61C}%
  \BibitemOpen
  \bibfield  {author} {\bibinfo {author} {\bibfnamefont {S.}~\bibnamefont
  {{Chatterjee}}}, \bibinfo {author} {\bibfnamefont {W.~H.~T.}\ \bibnamefont
  {{Vlemmings}}}, \bibinfo {author} {\bibfnamefont {W.~F.}\ \bibnamefont
  {{Brisken}}}, \bibinfo {author} {\bibfnamefont {T.~J.~W.}\ \bibnamefont
  {{Lazio}}}, \bibinfo {author} {\bibfnamefont {J.~M.}\ \bibnamefont
  {{Cordes}}}, \bibinfo {author} {\bibfnamefont {W.~M.}\ \bibnamefont
  {{Goss}}}, \bibinfo {author} {\bibfnamefont {S.~E.}\ \bibnamefont
  {{Thorsett}}}, \bibinfo {author} {\bibfnamefont {E.~B.}\ \bibnamefont
  {{Fomalont}}}, \bibinfo {author} {\bibfnamefont {A.~G.}\ \bibnamefont
  {{Lyne}}}, \ and\ \bibinfo {author} {\bibfnamefont {M.}~\bibnamefont
  {{Kramer}}},\ }\bibfield  {title} {\enquote {\bibinfo {title} {{Getting Its
  Kicks: A VLBA Parallax for the Hyperfast Pulsar B1508+55}},}\ }\href
  {\doibase 10.1086/491701} {\bibfield  {journal} {\bibinfo  {journal} {\apj}\
  }\textbf {\bibinfo {volume} {630}},\ \bibinfo {pages} {L61--L64} (\bibinfo
  {year} {2005})},\ \Eprint {http://arxiv.org/abs/astro-ph/0509031}
  {arXiv:astro-ph/0509031 [astro-ph]} \BibitemShut {NoStop}%
\bibitem [{\citenamefont {{Winkler}}\ and\ \citenamefont
  {{Petre}}(2007)}]{2007ApJ...670..635W}%
  \BibitemOpen
  \bibfield  {author} {\bibinfo {author} {\bibfnamefont {P.~Frank}\
  \bibnamefont {{Winkler}}}\ and\ \bibinfo {author} {\bibfnamefont {Robert}\
  \bibnamefont {{Petre}}},\ }\bibfield  {title} {\enquote {\bibinfo {title}
  {{Direct Measurement of Neutron Star Recoil in the Oxygen-rich Supernova
  Remnant Puppis A}},}\ }\href {\doibase 10.1086/522101} {\bibfield  {journal}
  {\bibinfo  {journal} {\apj}\ }\textbf {\bibinfo {volume} {670}},\ \bibinfo
  {pages} {635--642} (\bibinfo {year} {2007})},\ \Eprint
  {http://arxiv.org/abs/astro-ph/0608205} {arXiv:astro-ph/0608205 [astro-ph]}
  \BibitemShut {NoStop}%
\bibitem [{\citenamefont {{Holland-Ashford}}\ \emph {et~al.}(2017)\citenamefont
  {{Holland-Ashford}}, \citenamefont {{Lopez}}, \citenamefont {{Auchettl}},
  \citenamefont {{Temim}},\ and\ \citenamefont
  {{Ramirez-Ruiz}}}]{2017ApJ...844...84H}%
  \BibitemOpen
  \bibfield  {author} {\bibinfo {author} {\bibfnamefont {Tyler}\ \bibnamefont
  {{Holland-Ashford}}}, \bibinfo {author} {\bibfnamefont {Laura~A.}\
  \bibnamefont {{Lopez}}}, \bibinfo {author} {\bibfnamefont {Katie}\
  \bibnamefont {{Auchettl}}}, \bibinfo {author} {\bibfnamefont {Tea}\
  \bibnamefont {{Temim}}}, \ and\ \bibinfo {author} {\bibfnamefont {Enrico}\
  \bibnamefont {{Ramirez-Ruiz}}},\ }\bibfield  {title} {\enquote {\bibinfo
  {title} {{Comparing Neutron Star Kicks to Supernova Remnant Asymmetries}},}\
  }\href {\doibase 10.3847/1538-4357/aa7a5c} {\bibfield  {journal} {\bibinfo
  {journal} {\apj}\ }\textbf {\bibinfo {volume} {844}},\ \bibinfo {eid} {84}
  (\bibinfo {year} {2017})},\ \Eprint {http://arxiv.org/abs/1705.08454}
  {arXiv:1705.08454 [astro-ph.HE]} \BibitemShut {NoStop}%
\bibitem [{\citenamefont {{Katsuda}}\ \emph {et~al.}(2018)\citenamefont
  {{Katsuda}}, \citenamefont {{Morii}}, \citenamefont {{Janka}}, \citenamefont
  {{Wongwathanarat}}, \citenamefont {{Nakamura}}, \citenamefont {{Kotake}},
  \citenamefont {{Mori}}, \citenamefont {{M{\"u}ller}}, \citenamefont
  {{Takiwaki}}, \citenamefont {{Tanaka}}, \citenamefont {{Tominaga}},\ and\
  \citenamefont {{Tsunemi}}}]{2018ApJ...856...18K}%
  \BibitemOpen
  \bibfield  {author} {\bibinfo {author} {\bibfnamefont {Satoru}\ \bibnamefont
  {{Katsuda}}}, \bibinfo {author} {\bibfnamefont {Mikio}\ \bibnamefont
  {{Morii}}}, \bibinfo {author} {\bibfnamefont {Hans-Thomas}\ \bibnamefont
  {{Janka}}}, \bibinfo {author} {\bibfnamefont {Annop}\ \bibnamefont
  {{Wongwathanarat}}}, \bibinfo {author} {\bibfnamefont {Ko}~\bibnamefont
  {{Nakamura}}}, \bibinfo {author} {\bibfnamefont {Kei}\ \bibnamefont
  {{Kotake}}}, \bibinfo {author} {\bibfnamefont {Koji}\ \bibnamefont {{Mori}}},
  \bibinfo {author} {\bibfnamefont {Ewald}\ \bibnamefont {{M{\"u}ller}}},
  \bibinfo {author} {\bibfnamefont {Tomoya}\ \bibnamefont {{Takiwaki}}},
  \bibinfo {author} {\bibfnamefont {Masaomi}\ \bibnamefont {{Tanaka}}},
  \bibinfo {author} {\bibfnamefont {Nozomu}\ \bibnamefont {{Tominaga}}}, \ and\
  \bibinfo {author} {\bibfnamefont {Hiroshi}\ \bibnamefont {{Tsunemi}}},\
  }\bibfield  {title} {\enquote {\bibinfo {title} {{Intermediate-mass Elements
  in Young Supernova Remnants Reveal Neutron Star Kicks by Asymmetric
  Explosions}},}\ }\href {\doibase 10.3847/1538-4357/aab092} {\bibfield
  {journal} {\bibinfo  {journal} {\apj}\ }\textbf {\bibinfo {volume} {856}},\
  \bibinfo {eid} {18} (\bibinfo {year} {2018})},\ \Eprint
  {http://arxiv.org/abs/1710.10372} {arXiv:1710.10372 [astro-ph.HE]}
  \BibitemShut {NoStop}%
\bibitem [{\citenamefont {{Mayer}}\ and\ \citenamefont
  {{Becker}}(2021)}]{2021A&A...651A..40M}%
  \BibitemOpen
  \bibfield  {author} {\bibinfo {author} {\bibfnamefont {Martin G.~F.}\
  \bibnamefont {{Mayer}}}\ and\ \bibinfo {author} {\bibfnamefont {Werner}\
  \bibnamefont {{Becker}}},\ }\bibfield  {title} {\enquote {\bibinfo {title}
  {{A kinematic study of central compact objects and their host supernova
  remnants}},}\ }\href {\doibase 10.1051/0004-6361/202141119} {\bibfield
  {journal} {\bibinfo  {journal} {\aap}\ }\textbf {\bibinfo {volume} {651}},\
  \bibinfo {eid} {A40} (\bibinfo {year} {2021})},\ \Eprint
  {http://arxiv.org/abs/2106.00700} {arXiv:2106.00700 [astro-ph.HE]}
  \BibitemShut {NoStop}%
\bibitem [{\citenamefont {{Holland-Ashford}}\ \emph {et~al.}(2023)\citenamefont
  {{Holland-Ashford}}, \citenamefont {{Slane}},\ and\ \citenamefont
  {{Long}}}]{2023arXiv231019879H}%
  \BibitemOpen
  \bibfield  {author} {\bibinfo {author} {\bibfnamefont {Tyler}\ \bibnamefont
  {{Holland-Ashford}}}, \bibinfo {author} {\bibfnamefont {Patrick}\
  \bibnamefont {{Slane}}}, \ and\ \bibinfo {author} {\bibfnamefont
  {Xi}~\bibnamefont {{Long}}},\ }\bibfield  {title} {\enquote {\bibinfo {title}
  {{Updated Proper Motion of the Neutron Star in the Supernova Remnant
  Cassiopeia A}},}\ }\href {\doibase 10.48550/arXiv.2310.19879} {\bibfield
  {journal} {\bibinfo  {journal} {arXiv e-prints}\ ,\ \bibinfo {eid}
  {arXiv:2310.19879}} (\bibinfo {year} {2023})},\ \Eprint
  {http://arxiv.org/abs/2310.19879} {arXiv:2310.19879 [astro-ph.HE]}
  \BibitemShut {NoStop}%
\bibitem [{\citenamefont {{Wongwathanarat}}\ \emph {et~al.}(2010)\citenamefont
  {{Wongwathanarat}}, \citenamefont {{Janka}},\ and\ \citenamefont
  {{M{\"u}ller}}}]{2010ApJ...725L.106W}%
  \BibitemOpen
  \bibfield  {author} {\bibinfo {author} {\bibfnamefont {Annop}\ \bibnamefont
  {{Wongwathanarat}}}, \bibinfo {author} {\bibfnamefont {Hans-Thomas}\
  \bibnamefont {{Janka}}}, \ and\ \bibinfo {author} {\bibfnamefont {Ewald}\
  \bibnamefont {{M{\"u}ller}}},\ }\bibfield  {title} {\enquote {\bibinfo
  {title} {{Hydrodynamical Neutron Star Kicks in Three Dimensions}},}\ }\href
  {\doibase 10.1088/2041-8205/725/1/L106} {\bibfield  {journal} {\bibinfo
  {journal} {\apjl}\ }\textbf {\bibinfo {volume} {725}},\ \bibinfo {pages}
  {L106--L110} (\bibinfo {year} {2010})},\ \Eprint
  {http://arxiv.org/abs/1010.0167} {arXiv:1010.0167 [astro-ph.HE]} \BibitemShut
  {NoStop}%
\bibitem [{\citenamefont {{Nordhaus}}\ \emph {et~al.}(2010)\citenamefont
  {{Nordhaus}}, \citenamefont {{Brandt}}, \citenamefont {{Burrows}},
  \citenamefont {{Livne}},\ and\ \citenamefont {{Ott}}}]{2010PhRvD..82j3016N}%
  \BibitemOpen
  \bibfield  {author} {\bibinfo {author} {\bibfnamefont {J.}~\bibnamefont
  {{Nordhaus}}}, \bibinfo {author} {\bibfnamefont {T.~D.}\ \bibnamefont
  {{Brandt}}}, \bibinfo {author} {\bibfnamefont {A.}~\bibnamefont {{Burrows}}},
  \bibinfo {author} {\bibfnamefont {E.}~\bibnamefont {{Livne}}}, \ and\
  \bibinfo {author} {\bibfnamefont {C.~D.}\ \bibnamefont {{Ott}}},\ }\bibfield
  {title} {\enquote {\bibinfo {title} {{Theoretical support for the
  hydrodynamic mechanism of pulsar kicks}},}\ }\href {\doibase
  10.1103/PhysRevD.82.103016} {\bibfield  {journal} {\bibinfo  {journal}
  {\prd}\ }\textbf {\bibinfo {volume} {82}},\ \bibinfo {eid} {103016} (\bibinfo
  {year} {2010})},\ \Eprint {http://arxiv.org/abs/1010.0674} {arXiv:1010.0674
  [astro-ph.HE]} \BibitemShut {NoStop}%
\bibitem [{\citenamefont {{Nordhaus}}\ \emph {et~al.}(2012)\citenamefont
  {{Nordhaus}}, \citenamefont {{Brandt}}, \citenamefont {{Burrows}},\ and\
  \citenamefont {{Almgren}}}]{2012MNRAS.423.1805N}%
  \BibitemOpen
  \bibfield  {author} {\bibinfo {author} {\bibfnamefont {J.}~\bibnamefont
  {{Nordhaus}}}, \bibinfo {author} {\bibfnamefont {T.~D.}\ \bibnamefont
  {{Brandt}}}, \bibinfo {author} {\bibfnamefont {A.}~\bibnamefont {{Burrows}}},
  \ and\ \bibinfo {author} {\bibfnamefont {A.}~\bibnamefont {{Almgren}}},\
  }\bibfield  {title} {\enquote {\bibinfo {title} {{The hydrodynamic origin of
  neutron star kicks}},}\ }\href {\doibase 10.1111/j.1365-2966.2012.21002.x}
  {\bibfield  {journal} {\bibinfo  {journal} {\mnras}\ }\textbf {\bibinfo
  {volume} {423}},\ \bibinfo {pages} {1805--1812} (\bibinfo {year} {2012})},\
  \Eprint {http://arxiv.org/abs/1112.3342} {arXiv:1112.3342 [astro-ph.SR]}
  \BibitemShut {NoStop}%
\bibitem [{\citenamefont {{Wongwathanarat}}\ \emph {et~al.}(2013)\citenamefont
  {{Wongwathanarat}}, \citenamefont {{Janka}},\ and\ \citenamefont
  {{M{\"u}ller}}}]{2013A&A...552A.126W}%
  \BibitemOpen
  \bibfield  {author} {\bibinfo {author} {\bibfnamefont {A.}~\bibnamefont
  {{Wongwathanarat}}}, \bibinfo {author} {\bibfnamefont {H.-T.}\ \bibnamefont
  {{Janka}}}, \ and\ \bibinfo {author} {\bibfnamefont {E.}~\bibnamefont
  {{M{\"u}ller}}},\ }\bibfield  {title} {\enquote {\bibinfo {title}
  {{Three-dimensional neutrino-driven supernovae: Neutron star kicks, spins,
  and asymmetric ejection of nucleosynthesis products}},}\ }\href {\doibase
  10.1051/0004-6361/201220636} {\bibfield  {journal} {\bibinfo  {journal}
  {\aap}\ }\textbf {\bibinfo {volume} {552}},\ \bibinfo {eid} {A126} (\bibinfo
  {year} {2013})},\ \Eprint {http://arxiv.org/abs/1210.8148} {arXiv:1210.8148
  [astro-ph.HE]} \BibitemShut {NoStop}%
\bibitem [{\citenamefont {{Janka}}(2017)}]{2017ApJ...837...84J}%
  \BibitemOpen
  \bibfield  {author} {\bibinfo {author} {\bibfnamefont {Hans-Thomas}\
  \bibnamefont {{Janka}}},\ }\bibfield  {title} {\enquote {\bibinfo {title}
  {{Neutron Star Kicks by the Gravitational Tug-boat Mechanism in Asymmetric
  Supernova Explosions: Progenitor and Explosion Dependence}},}\ }\href
  {\doibase 10.3847/1538-4357/aa618e} {\bibfield  {journal} {\bibinfo
  {journal} {\apj}\ }\textbf {\bibinfo {volume} {837}},\ \bibinfo {eid} {84}
  (\bibinfo {year} {2017})},\ \Eprint {http://arxiv.org/abs/1611.07562}
  {arXiv:1611.07562 [astro-ph.HE]} \BibitemShut {NoStop}%
\bibitem [{\citenamefont {{Burrows}}\ \emph {et~al.}(2023)\citenamefont
  {{Burrows}}, \citenamefont {{Wang}}, \citenamefont {{Vartanyan}},\ and\
  \citenamefont {{Coleman}}}]{2023arXiv231112109B}%
  \BibitemOpen
  \bibfield  {author} {\bibinfo {author} {\bibfnamefont {Adam}\ \bibnamefont
  {{Burrows}}}, \bibinfo {author} {\bibfnamefont {Tianshu}\ \bibnamefont
  {{Wang}}}, \bibinfo {author} {\bibfnamefont {David}\ \bibnamefont
  {{Vartanyan}}}, \ and\ \bibinfo {author} {\bibfnamefont {Matthew S.~B.}\
  \bibnamefont {{Coleman}}},\ }\bibfield  {title} {\enquote {\bibinfo {title}
  {{A Comprehensive Theory for Neutron Star and Black Hole Kicks and Induced
  Spins}},}\ }\href {\doibase 10.48550/arXiv.2311.12109} {\bibfield  {journal}
  {\bibinfo  {journal} {arXiv e-prints}\ ,\ \bibinfo {eid} {arXiv:2311.12109}}
  (\bibinfo {year} {2023})},\ \Eprint {http://arxiv.org/abs/2311.12109}
  {arXiv:2311.12109 [astro-ph.HE]} \BibitemShut {NoStop}%
\bibitem [{\citenamefont {{Woosley}}(1987)}]{1987IAUS..125..255W}%
  \BibitemOpen
  \bibfield  {author} {\bibinfo {author} {\bibfnamefont {S.~E.}\ \bibnamefont
  {{Woosley}}},\ }\bibfield  {title} {\enquote {\bibinfo {title} {{The Birth of
  Neutron Stars}},}\ }in\ \href@noop {} {\emph {\bibinfo {booktitle} {The
  Origin and Evolution of Neutron Stars}}},\ Vol.\ \bibinfo {volume} {125},\
  \bibinfo {editor} {edited by\ \bibinfo {editor} {\bibfnamefont {D.~J.}\
  \bibnamefont {{Helfand}}}\ and\ \bibinfo {editor} {\bibfnamefont {J.~H.}\
  \bibnamefont {{Huang}}}}\ (\bibinfo {year} {1987})\ p.\ \bibinfo {pages}
  {255}\BibitemShut {NoStop}%
\bibitem [{\citenamefont {{Bisnovatyi-Kogan}}(1993)}]{1993A&AT....3..287B}%
  \BibitemOpen
  \bibfield  {author} {\bibinfo {author} {\bibfnamefont {G.~S.}\ \bibnamefont
  {{Bisnovatyi-Kogan}}},\ }\bibfield  {title} {\enquote {\bibinfo {title}
  {{Asymmetric neutrino emission and formation of rapidly moving pulsars}},}\
  }\href {\doibase 10.1080/10556799308230566} {\bibfield  {journal} {\bibinfo
  {journal} {Astronomical and Astrophysical Transactions}\ }\textbf {\bibinfo
  {volume} {3}},\ \bibinfo {pages} {287--294} (\bibinfo {year} {1993})},\
  \Eprint {http://arxiv.org/abs/astro-ph/9707120} {arXiv:astro-ph/9707120
  [astro-ph]} \BibitemShut {NoStop}%
\bibitem [{\citenamefont {{Socrates}}\ \emph {et~al.}(2005)\citenamefont
  {{Socrates}}, \citenamefont {{Blaes}}, \citenamefont {{Hungerford}},\ and\
  \citenamefont {{Fryer}}}]{2005ApJ...632..531S}%
  \BibitemOpen
  \bibfield  {author} {\bibinfo {author} {\bibfnamefont {Aristotle}\
  \bibnamefont {{Socrates}}}, \bibinfo {author} {\bibfnamefont {Omer}\
  \bibnamefont {{Blaes}}}, \bibinfo {author} {\bibfnamefont {Aimee}\
  \bibnamefont {{Hungerford}}}, \ and\ \bibinfo {author} {\bibfnamefont
  {Chris~L.}\ \bibnamefont {{Fryer}}},\ }\bibfield  {title} {\enquote {\bibinfo
  {title} {{The Neutrino Bubble Instability: A Mechanism for Generating Pulsar
  Kicks}},}\ }\href {\doibase 10.1086/431786} {\bibfield  {journal} {\bibinfo
  {journal} {\apj}\ }\textbf {\bibinfo {volume} {632}},\ \bibinfo {pages}
  {531--562} (\bibinfo {year} {2005})},\ \Eprint
  {http://arxiv.org/abs/astro-ph/0412144} {arXiv:astro-ph/0412144 [astro-ph]}
  \BibitemShut {NoStop}%
\bibitem [{\citenamefont {{Fryer}}\ and\ \citenamefont
  {{Kusenko}}(2006)}]{2006ApJS..163..335F}%
  \BibitemOpen
  \bibfield  {author} {\bibinfo {author} {\bibfnamefont {Christopher~L.}\
  \bibnamefont {{Fryer}}}\ and\ \bibinfo {author} {\bibfnamefont {Alexander}\
  \bibnamefont {{Kusenko}}},\ }\bibfield  {title} {\enquote {\bibinfo {title}
  {{Effects of Neutrino-driven Kicks on the Supernova Explosion Mechanism}},}\
  }\href {\doibase 10.1086/500933} {\bibfield  {journal} {\bibinfo  {journal}
  {\apjs}\ }\textbf {\bibinfo {volume} {163}},\ \bibinfo {pages} {335--343}
  (\bibinfo {year} {2006})},\ \Eprint {http://arxiv.org/abs/astro-ph/0512033}
  {arXiv:astro-ph/0512033 [astro-ph]} \BibitemShut {NoStop}%
\bibitem [{\citenamefont {{Nagakura}}\ \emph
  {et~al.}(2019{\natexlab{a}})\citenamefont {{Nagakura}}, \citenamefont
  {{Sumiyoshi}},\ and\ \citenamefont {{Yamada}}}]{2019ApJ...880L..28N}%
  \BibitemOpen
  \bibfield  {author} {\bibinfo {author} {\bibfnamefont {Hiroki}\ \bibnamefont
  {{Nagakura}}}, \bibinfo {author} {\bibfnamefont {Kohsuke}\ \bibnamefont
  {{Sumiyoshi}}}, \ and\ \bibinfo {author} {\bibfnamefont {Shoichi}\
  \bibnamefont {{Yamada}}},\ }\bibfield  {title} {\enquote {\bibinfo {title}
  {{Possible Early Linear Acceleration of Proto-neutron Stars via Asymmetric
  Neutrino Emission in Core-collapse Supernovae}},}\ }\href {\doibase
  10.3847/2041-8213/ab30ca} {\bibfield  {journal} {\bibinfo  {journal} {\apjl}\
  }\textbf {\bibinfo {volume} {880}},\ \bibinfo {eid} {L28} (\bibinfo {year}
  {2019}{\natexlab{a}})},\ \Eprint {http://arxiv.org/abs/1907.04863}
  {arXiv:1907.04863 [astro-ph.HE]} \BibitemShut {NoStop}%
\bibitem [{\citenamefont {{Coleman}}\ and\ \citenamefont
  {{Burrows}}(2022)}]{2022MNRAS.517.3938C}%
  \BibitemOpen
  \bibfield  {author} {\bibinfo {author} {\bibfnamefont {Matthew S.~B.}\
  \bibnamefont {{Coleman}}}\ and\ \bibinfo {author} {\bibfnamefont {Adam}\
  \bibnamefont {{Burrows}}},\ }\bibfield  {title} {\enquote {\bibinfo {title}
  {{Kicks and induced spins of neutron stars at birth}},}\ }\href {\doibase
  10.1093/mnras/stac2573} {\bibfield  {journal} {\bibinfo  {journal} {\mnras}\
  }\textbf {\bibinfo {volume} {517}},\ \bibinfo {pages} {3938--3961} (\bibinfo
  {year} {2022})},\ \Eprint {http://arxiv.org/abs/2209.02711} {arXiv:2209.02711
  [astro-ph.HE]} \BibitemShut {NoStop}%
\bibitem [{\citenamefont {{Fryer}}(2004)}]{2004ApJ...601L.175F}%
  \BibitemOpen
  \bibfield  {author} {\bibinfo {author} {\bibfnamefont {C.~L.}\ \bibnamefont
  {{Fryer}}},\ }\bibfield  {title} {\enquote {\bibinfo {title} {{Neutron Star
  Kicks from Asymmetric Collapse}},}\ }\href {\doibase 10.1086/382044}
  {\bibfield  {journal} {\bibinfo  {journal} {\apjl}\ }\textbf {\bibinfo
  {volume} {601}},\ \bibinfo {pages} {L175--L178} (\bibinfo {year} {2004})},\
  \Eprint {http://arxiv.org/abs/astro-ph/0312265} {astro-ph/0312265}
  \BibitemShut {NoStop}%
\bibitem [{\citenamefont {{Nagakura}}\ \emph
  {et~al.}(2019{\natexlab{b}})\citenamefont {{Nagakura}}, \citenamefont
  {{Sumiyoshi}},\ and\ \citenamefont {{Yamada}}}]{2019ApJ...878..160N}%
  \BibitemOpen
  \bibfield  {author} {\bibinfo {author} {\bibfnamefont {Hiroki}\ \bibnamefont
  {{Nagakura}}}, \bibinfo {author} {\bibfnamefont {Kohsuke}\ \bibnamefont
  {{Sumiyoshi}}}, \ and\ \bibinfo {author} {\bibfnamefont {Shoichi}\
  \bibnamefont {{Yamada}}},\ }\bibfield  {title} {\enquote {\bibinfo {title}
  {{Three-dimensional Boltzmann-hydro Code for Core-collapse in Massive Stars.
  III. A New Method for Momentum Feedback from Neutrino to Matter}},}\ }\href
  {\doibase 10.3847/1538-4357/ab2189} {\bibfield  {journal} {\bibinfo
  {journal} {\apj}\ }\textbf {\bibinfo {volume} {878}},\ \bibinfo {eid} {160}
  (\bibinfo {year} {2019}{\natexlab{b}})},\ \Eprint
  {http://arxiv.org/abs/1906.10143} {arXiv:1906.10143 [astro-ph.HE]}
  \BibitemShut {NoStop}%
\bibitem [{\citenamefont {{Janka}}\ \emph {et~al.}(2022)\citenamefont
  {{Janka}}, \citenamefont {{Wongwathanarat}},\ and\ \citenamefont
  {{Kramer}}}]{2022ApJ...926....9J}%
  \BibitemOpen
  \bibfield  {author} {\bibinfo {author} {\bibfnamefont {Hans-Thomas}\
  \bibnamefont {{Janka}}}, \bibinfo {author} {\bibfnamefont {Annop}\
  \bibnamefont {{Wongwathanarat}}}, \ and\ \bibinfo {author} {\bibfnamefont
  {Michael}\ \bibnamefont {{Kramer}}},\ }\bibfield  {title} {\enquote {\bibinfo
  {title} {{Supernova Fallback as Origin of Neutron Star Spins and Spin-kick
  Alignment}},}\ }\href {\doibase 10.3847/1538-4357/ac403c} {\bibfield
  {journal} {\bibinfo  {journal} {\apj}\ }\textbf {\bibinfo {volume} {926}},\
  \bibinfo {eid} {9} (\bibinfo {year} {2022})},\ \Eprint
  {http://arxiv.org/abs/2104.07493} {arXiv:2104.07493 [astro-ph.HE]}
  \BibitemShut {NoStop}%
\bibitem [{\citenamefont {{M{\"u}ller}}(2023)}]{2023MNRAS.526.2880M}%
  \BibitemOpen
  \bibfield  {author} {\bibinfo {author} {\bibfnamefont {Bernhard}\
  \bibnamefont {{M{\"u}ller}}},\ }\bibfield  {title} {\enquote {\bibinfo
  {title} {{Fallback onto kicked neutron stars and its effect on spin-kick
  alignment}},}\ }\href {\doibase 10.1093/mnras/stad2881} {\bibfield  {journal}
  {\bibinfo  {journal} {\mnras}\ }\textbf {\bibinfo {volume} {526}},\ \bibinfo
  {pages} {2880--2888} (\bibinfo {year} {2023})},\ \Eprint
  {http://arxiv.org/abs/2308.08312} {arXiv:2308.08312 [astro-ph.HE]}
  \BibitemShut {NoStop}%
\bibitem [{\citenamefont {{Harrison}}\ and\ \citenamefont
  {{Tademaru}}(1975)}]{1975ApJ...201..447H}%
  \BibitemOpen
  \bibfield  {author} {\bibinfo {author} {\bibfnamefont {E.~R.}\ \bibnamefont
  {{Harrison}}}\ and\ \bibinfo {author} {\bibfnamefont {E.}~\bibnamefont
  {{Tademaru}}},\ }\bibfield  {title} {\enquote {\bibinfo {title}
  {{Acceleration of pulsars by asymmetric radiation.}}}\ }\href {\doibase
  10.1086/153907} {\bibfield  {journal} {\bibinfo  {journal} {\apj}\ }\textbf
  {\bibinfo {volume} {201}},\ \bibinfo {pages} {447--461} (\bibinfo {year}
  {1975})}\BibitemShut {NoStop}%
\bibitem [{\citenamefont {{Lai}}\ \emph {et~al.}(2001)\citenamefont {{Lai}},
  \citenamefont {{Chernoff}},\ and\ \citenamefont
  {{Cordes}}}]{2001ApJ...549.1111L}%
  \BibitemOpen
  \bibfield  {author} {\bibinfo {author} {\bibfnamefont {Dong}\ \bibnamefont
  {{Lai}}}, \bibinfo {author} {\bibfnamefont {David~F.}\ \bibnamefont
  {{Chernoff}}}, \ and\ \bibinfo {author} {\bibfnamefont {James~M.}\
  \bibnamefont {{Cordes}}},\ }\bibfield  {title} {\enquote {\bibinfo {title}
  {{Pulsar Jets: Implications for Neutron Star Kicks and Initial Spins}},}\
  }\href {\doibase 10.1086/319455} {\bibfield  {journal} {\bibinfo  {journal}
  {\apj}\ }\textbf {\bibinfo {volume} {549}},\ \bibinfo {pages} {1111--1118}
  (\bibinfo {year} {2001})},\ \Eprint {http://arxiv.org/abs/astro-ph/0007272}
  {arXiv:astro-ph/0007272 [astro-ph]} \BibitemShut {NoStop}%
\bibitem [{\citenamefont {{P{\'e}tri}}(2019)}]{2019MNRAS.488.4161P}%
  \BibitemOpen
  \bibfield  {author} {\bibinfo {author} {\bibfnamefont {J.}~\bibnamefont
  {{P{\'e}tri}}},\ }\bibfield  {title} {\enquote {\bibinfo {title} {{Impact of
  an off-centred dipole on neutron star binaries}},}\ }\href {\doibase
  10.1093/mnras/stz2021} {\bibfield  {journal} {\bibinfo  {journal} {\mnras}\
  }\textbf {\bibinfo {volume} {488}},\ \bibinfo {pages} {4161--4168} (\bibinfo
  {year} {2019})},\ \Eprint {http://arxiv.org/abs/1907.07551} {arXiv:1907.07551
  [astro-ph.HE]} \BibitemShut {NoStop}%
\bibitem [{\citenamefont {{Igoshev}}\ \emph {et~al.}(2021)\citenamefont
  {{Igoshev}}, \citenamefont {{Chruslinska}}, \citenamefont {{Dorozsmai}},\
  and\ \citenamefont {{Toonen}}}]{2021MNRAS.508.3345I}%
  \BibitemOpen
  \bibfield  {author} {\bibinfo {author} {\bibfnamefont {Andrei~P.}\
  \bibnamefont {{Igoshev}}}, \bibinfo {author} {\bibfnamefont {Martyna}\
  \bibnamefont {{Chruslinska}}}, \bibinfo {author} {\bibfnamefont {Andris}\
  \bibnamefont {{Dorozsmai}}}, \ and\ \bibinfo {author} {\bibfnamefont
  {Silvia}\ \bibnamefont {{Toonen}}},\ }\bibfield  {title} {\enquote {\bibinfo
  {title} {{Combined analysis of neutron star natal kicks using proper motions
  and parallax measurements for radio pulsars and Be X-ray binaries}},}\ }\href
  {\doibase 10.1093/mnras/stab2734} {\bibfield  {journal} {\bibinfo  {journal}
  {\mnras}\ }\textbf {\bibinfo {volume} {508}},\ \bibinfo {pages} {3345--3364}
  (\bibinfo {year} {2021})},\ \Eprint {http://arxiv.org/abs/2109.10362}
  {arXiv:2109.10362 [astro-ph.HE]} \BibitemShut {NoStop}%
\bibitem [{\citenamefont {{Sawyer}}(2005)}]{2005PhRvD..72d5003S}%
  \BibitemOpen
  \bibfield  {author} {\bibinfo {author} {\bibfnamefont {R.~F.}\ \bibnamefont
  {{Sawyer}}},\ }\bibfield  {title} {\enquote {\bibinfo {title} {{Speed-up of
  neutrino transformations in a supernova environment}},}\ }\href {\doibase
  10.1103/PhysRevD.72.045003} {\bibfield  {journal} {\bibinfo  {journal}
  {\prd}\ }\textbf {\bibinfo {volume} {72}},\ \bibinfo {eid} {045003} (\bibinfo
  {year} {2005})},\ \Eprint {http://arxiv.org/abs/hep-ph/0503013}
  {arXiv:hep-ph/0503013 [astro-ph]} \BibitemShut {NoStop}%
\bibitem [{\citenamefont {{Tamborra}}\ and\ \citenamefont
  {{Shalgar}}(2021)}]{2021ARNPS..71..165T}%
  \BibitemOpen
  \bibfield  {author} {\bibinfo {author} {\bibfnamefont {Irene}\ \bibnamefont
  {{Tamborra}}}\ and\ \bibinfo {author} {\bibfnamefont {Shashank}\ \bibnamefont
  {{Shalgar}}},\ }\bibfield  {title} {\enquote {\bibinfo {title} {{New
  Developments in Flavor Evolution of a Dense Neutrino Gas}},}\ }\href
  {\doibase 10.1146/annurev-nucl-102920-050505} {\bibfield  {journal} {\bibinfo
   {journal} {Annual Review of Nuclear and Particle Science}\ }\textbf
  {\bibinfo {volume} {71}},\ \bibinfo {pages} {165--188} (\bibinfo {year}
  {2021})},\ \Eprint {http://arxiv.org/abs/2011.01948} {arXiv:2011.01948
  [astro-ph.HE]} \BibitemShut {NoStop}%
\bibitem [{\citenamefont {{Capozzi}}\ and\ \citenamefont
  {{Saviano}}(2022)}]{2022Univ....8...94C}%
  \BibitemOpen
  \bibfield  {author} {\bibinfo {author} {\bibfnamefont {Francesco}\
  \bibnamefont {{Capozzi}}}\ and\ \bibinfo {author} {\bibfnamefont {Ninetta}\
  \bibnamefont {{Saviano}}},\ }\bibfield  {title} {\enquote {\bibinfo {title}
  {{Neutrino Flavor Conversions in High-Density Astrophysical and Cosmological
  Environments}},}\ }\href {\doibase 10.3390/universe8020094} {\bibfield
  {journal} {\bibinfo  {journal} {Universe}\ }\textbf {\bibinfo {volume} {8}},\
  \bibinfo {pages} {94} (\bibinfo {year} {2022})},\ \Eprint
  {http://arxiv.org/abs/2202.02494} {arXiv:2202.02494 [hep-ph]} \BibitemShut
  {NoStop}%
\bibitem [{\citenamefont {{Richers}}\ and\ \citenamefont
  {{Sen}}(2022)}]{2022arXiv220703561R}%
  \BibitemOpen
  \bibfield  {author} {\bibinfo {author} {\bibfnamefont {Sherwood}\
  \bibnamefont {{Richers}}}\ and\ \bibinfo {author} {\bibfnamefont {Manibrata}\
  \bibnamefont {{Sen}}},\ }\bibfield  {title} {\enquote {\bibinfo {title}
  {{Fast Flavor Transformations}},}\ }\href@noop {} {\bibfield  {journal}
  {\bibinfo  {journal} {arXiv e-prints}\ ,\ \bibinfo {eid} {arXiv:2207.03561}}
  (\bibinfo {year} {2022})},\ \Eprint {http://arxiv.org/abs/2207.03561}
  {arXiv:2207.03561 [astro-ph.HE]} \BibitemShut {NoStop}%
\bibitem [{\citenamefont {{Volpe}}(2023)}]{2023arXiv230111814V}%
  \BibitemOpen
  \bibfield  {author} {\bibinfo {author} {\bibfnamefont {Maria~Cristina}\
  \bibnamefont {{Volpe}}},\ }\bibfield  {title} {\enquote {\bibinfo {title}
  {{Neutrinos from dense: flavor mechanisms, theoretical approaches,
  observations, new directions}},}\ }\href {\doibase 10.48550/arXiv.2301.11814}
  {\bibfield  {journal} {\bibinfo  {journal} {arXiv e-prints}\ ,\ \bibinfo
  {eid} {arXiv:2301.11814}} (\bibinfo {year} {2023})},\ \Eprint
  {http://arxiv.org/abs/2301.11814} {arXiv:2301.11814 [hep-ph]} \BibitemShut
  {NoStop}%
\bibitem [{\citenamefont {{Fischer}}\ \emph {et~al.}(2023)\citenamefont
  {{Fischer}}, \citenamefont {{Guo}}, \citenamefont {{Langanke}}, \citenamefont
  {{Martinez-Pinedo}}, \citenamefont {{Qian}},\ and\ \citenamefont
  {{Wu}}}]{2023arXiv230803962F}%
  \BibitemOpen
  \bibfield  {author} {\bibinfo {author} {\bibfnamefont {Tobias}\ \bibnamefont
  {{Fischer}}}, \bibinfo {author} {\bibfnamefont {Gang}\ \bibnamefont {{Guo}}},
  \bibinfo {author} {\bibfnamefont {Karlheinz}\ \bibnamefont {{Langanke}}},
  \bibinfo {author} {\bibfnamefont {Gabriel}\ \bibnamefont
  {{Martinez-Pinedo}}}, \bibinfo {author} {\bibfnamefont {Yong-Zhong}\
  \bibnamefont {{Qian}}}, \ and\ \bibinfo {author} {\bibfnamefont {Meng-Ru}\
  \bibnamefont {{Wu}}},\ }\bibfield  {title} {\enquote {\bibinfo {title}
  {{Neutrinos and nucleosynthesis of elements}},}\ }\href {\doibase
  10.48550/arXiv.2308.03962} {\bibfield  {journal} {\bibinfo  {journal} {arXiv
  e-prints}\ ,\ \bibinfo {eid} {arXiv:2308.03962}} (\bibinfo {year} {2023})},\
  \Eprint {http://arxiv.org/abs/2308.03962} {arXiv:2308.03962 [astro-ph.HE]}
  \BibitemShut {NoStop}%
\bibitem [{\citenamefont {{Abbar}}\ \emph {et~al.}(2021)\citenamefont
  {{Abbar}}, \citenamefont {{Capozzi}}, \citenamefont {{Glas}}, \citenamefont
  {{Janka}},\ and\ \citenamefont {{Tamborra}}}]{2021PhRvD.103f3033A}%
  \BibitemOpen
  \bibfield  {author} {\bibinfo {author} {\bibfnamefont {Sajad}\ \bibnamefont
  {{Abbar}}}, \bibinfo {author} {\bibfnamefont {Francesco}\ \bibnamefont
  {{Capozzi}}}, \bibinfo {author} {\bibfnamefont {Robert}\ \bibnamefont
  {{Glas}}}, \bibinfo {author} {\bibfnamefont {H.~Thomas}\ \bibnamefont
  {{Janka}}}, \ and\ \bibinfo {author} {\bibfnamefont {Irene}\ \bibnamefont
  {{Tamborra}}},\ }\bibfield  {title} {\enquote {\bibinfo {title} {{On the
  characteristics of fast neutrino flavor instabilities in three-dimensional
  core-collapse supernova models}},}\ }\href {\doibase
  10.1103/PhysRevD.103.063033} {\bibfield  {journal} {\bibinfo  {journal}
  {\prd}\ }\textbf {\bibinfo {volume} {103}},\ \bibinfo {eid} {063033}
  (\bibinfo {year} {2021})},\ \Eprint {http://arxiv.org/abs/2012.06594}
  {arXiv:2012.06594 [astro-ph.HE]} \BibitemShut {NoStop}%
\bibitem [{\citenamefont {{Nagakura}}\ \emph {et~al.}(2021)\citenamefont
  {{Nagakura}}, \citenamefont {{Burrows}}, \citenamefont {{Johns}},\ and\
  \citenamefont {{Fuller}}}]{2021PhRvD.104h3025N}%
  \BibitemOpen
  \bibfield  {author} {\bibinfo {author} {\bibfnamefont {Hiroki}\ \bibnamefont
  {{Nagakura}}}, \bibinfo {author} {\bibfnamefont {Adam}\ \bibnamefont
  {{Burrows}}}, \bibinfo {author} {\bibfnamefont {Lucas}\ \bibnamefont
  {{Johns}}}, \ and\ \bibinfo {author} {\bibfnamefont {George~M.}\ \bibnamefont
  {{Fuller}}},\ }\bibfield  {title} {\enquote {\bibinfo {title} {{Where, when,
  and why: Occurrence of fast-pairwise collective neutrino oscillation in
  three-dimensional core-collapse supernova models}},}\ }\href {\doibase
  10.1103/PhysRevD.104.083025} {\bibfield  {journal} {\bibinfo  {journal}
  {\prd}\ }\textbf {\bibinfo {volume} {104}},\ \bibinfo {eid} {083025}
  (\bibinfo {year} {2021})},\ \Eprint {http://arxiv.org/abs/2108.07281}
  {arXiv:2108.07281 [astro-ph.HE]} \BibitemShut {NoStop}%
\bibitem [{\citenamefont {{Harada}}\ and\ \citenamefont
  {{Nagakura}}(2022)}]{2022ApJ...924..109H}%
  \BibitemOpen
  \bibfield  {author} {\bibinfo {author} {\bibfnamefont {Akira}\ \bibnamefont
  {{Harada}}}\ and\ \bibinfo {author} {\bibfnamefont {Hiroki}\ \bibnamefont
  {{Nagakura}}},\ }\bibfield  {title} {\enquote {\bibinfo {title} {{Prospects
  of Fast Flavor Neutrino Conversion in Rotating Core-collapse Supernovae}},}\
  }\href {\doibase 10.3847/1538-4357/ac38a0} {\bibfield  {journal} {\bibinfo
  {journal} {\apj}\ }\textbf {\bibinfo {volume} {924}},\ \bibinfo {eid} {109}
  (\bibinfo {year} {2022})},\ \Eprint {http://arxiv.org/abs/2110.08291}
  {arXiv:2110.08291 [astro-ph.HE]} \BibitemShut {NoStop}%
\bibitem [{\citenamefont {{Capozzi}}\ \emph {et~al.}(2021)\citenamefont
  {{Capozzi}}, \citenamefont {{Abbar}}, \citenamefont {{Bollig}},\ and\
  \citenamefont {{Janka}}}]{2021PhRvD.103f3013C}%
  \BibitemOpen
  \bibfield  {author} {\bibinfo {author} {\bibfnamefont {Francesco}\
  \bibnamefont {{Capozzi}}}, \bibinfo {author} {\bibfnamefont {Sajad}\
  \bibnamefont {{Abbar}}}, \bibinfo {author} {\bibfnamefont {Robert}\
  \bibnamefont {{Bollig}}}, \ and\ \bibinfo {author} {\bibfnamefont
  {H.~Thomas}\ \bibnamefont {{Janka}}},\ }\bibfield  {title} {\enquote
  {\bibinfo {title} {{Fast neutrino flavor conversions in one-dimensional
  core-collapse supernova models with and without muon creation}},}\ }\href
  {\doibase 10.1103/PhysRevD.103.063013} {\bibfield  {journal} {\bibinfo
  {journal} {\prd}\ }\textbf {\bibinfo {volume} {103}},\ \bibinfo {eid}
  {063013} (\bibinfo {year} {2021})},\ \Eprint
  {http://arxiv.org/abs/2012.08525} {arXiv:2012.08525 [astro-ph.HE]}
  \BibitemShut {NoStop}%
\bibitem [{\citenamefont {{Morinaga}}(2022)}]{2022PhRvD.105j1301M}%
  \BibitemOpen
  \bibfield  {author} {\bibinfo {author} {\bibfnamefont {Taiki}\ \bibnamefont
  {{Morinaga}}},\ }\bibfield  {title} {\enquote {\bibinfo {title} {{Fast
  neutrino flavor instability and neutrino flavor lepton number crossings}},}\
  }\href {\doibase 10.1103/PhysRevD.105.L101301} {\bibfield  {journal}
  {\bibinfo  {journal} {\prd}\ }\textbf {\bibinfo {volume} {105}},\ \bibinfo
  {eid} {L101301} (\bibinfo {year} {2022})},\ \Eprint
  {http://arxiv.org/abs/2103.15267} {arXiv:2103.15267 [hep-ph]} \BibitemShut
  {NoStop}%
\bibitem [{\citenamefont {{Abbar}}\ \emph {et~al.}(2019)\citenamefont
  {{Abbar}}, \citenamefont {{Duan}}, \citenamefont {{Sumiyoshi}}, \citenamefont
  {{Takiwaki}},\ and\ \citenamefont {{Volpe}}}]{2019PhRvD.100d3004A}%
  \BibitemOpen
  \bibfield  {author} {\bibinfo {author} {\bibfnamefont {Sajad}\ \bibnamefont
  {{Abbar}}}, \bibinfo {author} {\bibfnamefont {Huaiyu}\ \bibnamefont
  {{Duan}}}, \bibinfo {author} {\bibfnamefont {Kohsuke}\ \bibnamefont
  {{Sumiyoshi}}}, \bibinfo {author} {\bibfnamefont {Tomoya}\ \bibnamefont
  {{Takiwaki}}}, \ and\ \bibinfo {author} {\bibfnamefont {Maria~Cristina}\
  \bibnamefont {{Volpe}}},\ }\bibfield  {title} {\enquote {\bibinfo {title}
  {{On the occurrence of fast neutrino flavor conversions in multidimensional
  supernova models}},}\ }\href {\doibase 10.1103/PhysRevD.100.043004}
  {\bibfield  {journal} {\bibinfo  {journal} {\prd}\ }\textbf {\bibinfo
  {volume} {100}},\ \bibinfo {eid} {043004} (\bibinfo {year} {2019})},\ \Eprint
  {http://arxiv.org/abs/1812.06883} {arXiv:1812.06883 [astro-ph.HE]}
  \BibitemShut {NoStop}%
\bibitem [{\citenamefont {{Nagakura}}\ \emph
  {et~al.}(2019{\natexlab{c}})\citenamefont {{Nagakura}}, \citenamefont
  {{Morinaga}}, \citenamefont {{Kato}},\ and\ \citenamefont
  {{Yamada}}}]{2019ApJ...886..139N}%
  \BibitemOpen
  \bibfield  {author} {\bibinfo {author} {\bibfnamefont {Hiroki}\ \bibnamefont
  {{Nagakura}}}, \bibinfo {author} {\bibfnamefont {Taiki}\ \bibnamefont
  {{Morinaga}}}, \bibinfo {author} {\bibfnamefont {Chinami}\ \bibnamefont
  {{Kato}}}, \ and\ \bibinfo {author} {\bibfnamefont {Shoichi}\ \bibnamefont
  {{Yamada}}},\ }\bibfield  {title} {\enquote {\bibinfo {title} {{Fast-pairwise
  Collective Neutrino Oscillations Associated with Asymmetric Neutrino
  Emissions in Core-collapse Supernovae}},}\ }\href {\doibase
  10.3847/1538-4357/ab4cf2} {\bibfield  {journal} {\bibinfo  {journal} {\apj}\
  }\textbf {\bibinfo {volume} {886}},\ \bibinfo {eid} {139} (\bibinfo {year}
  {2019}{\natexlab{c}})},\ \Eprint {http://arxiv.org/abs/1910.04288}
  {arXiv:1910.04288 [astro-ph.HE]} \BibitemShut {NoStop}%
\bibitem [{\citenamefont {{Tamborra}}\ \emph {et~al.}(2017)\citenamefont
  {{Tamborra}}, \citenamefont {{H{\"u}depohl}}, \citenamefont {{Raffelt}},\
  and\ \citenamefont {{Janka}}}]{2017ApJ...839..132T}%
  \BibitemOpen
  \bibfield  {author} {\bibinfo {author} {\bibfnamefont {Irene}\ \bibnamefont
  {{Tamborra}}}, \bibinfo {author} {\bibfnamefont {Lorenz}\ \bibnamefont
  {{H{\"u}depohl}}}, \bibinfo {author} {\bibfnamefont {Georg~G.}\ \bibnamefont
  {{Raffelt}}}, \ and\ \bibinfo {author} {\bibfnamefont {Hans-Thomas}\
  \bibnamefont {{Janka}}},\ }\bibfield  {title} {\enquote {\bibinfo {title}
  {{Flavor-dependent Neutrino Angular Distribution in Core-collapse
  Supernovae}},}\ }\href {\doibase 10.3847/1538-4357/aa6a18} {\bibfield
  {journal} {\bibinfo  {journal} {\apj}\ }\textbf {\bibinfo {volume} {839}},\
  \bibinfo {eid} {132} (\bibinfo {year} {2017})},\ \Eprint
  {http://arxiv.org/abs/1702.00060} {arXiv:1702.00060 [astro-ph.HE]}
  \BibitemShut {NoStop}%
\bibitem [{\citenamefont {{Morinaga}}\ \emph {et~al.}(2020)\citenamefont
  {{Morinaga}}, \citenamefont {{Nagakura}}, \citenamefont {{Kato}},\ and\
  \citenamefont {{Yamada}}}]{2020PhRvR...2a2046M}%
  \BibitemOpen
  \bibfield  {author} {\bibinfo {author} {\bibfnamefont {Taiki}\ \bibnamefont
  {{Morinaga}}}, \bibinfo {author} {\bibfnamefont {Hiroki}\ \bibnamefont
  {{Nagakura}}}, \bibinfo {author} {\bibfnamefont {Chinami}\ \bibnamefont
  {{Kato}}}, \ and\ \bibinfo {author} {\bibfnamefont {Shoichi}\ \bibnamefont
  {{Yamada}}},\ }\bibfield  {title} {\enquote {\bibinfo {title} {{Fast
  neutrino-flavor conversion in the preshock region of core-collapse
  supernovae}},}\ }\href {\doibase 10.1103/PhysRevResearch.2.012046} {\bibfield
   {journal} {\bibinfo  {journal} {Physical Review Research}\ }\textbf
  {\bibinfo {volume} {2}},\ \bibinfo {eid} {012046} (\bibinfo {year} {2020})},\
  \Eprint {http://arxiv.org/abs/1909.13131} {arXiv:1909.13131 [astro-ph.HE]}
  \BibitemShut {NoStop}%
\bibitem [{\citenamefont {{Nagakura}}\ \emph {et~al.}(2020)\citenamefont
  {{Nagakura}}, \citenamefont {{Burrows}}, \citenamefont {{Radice}},\ and\
  \citenamefont {{Vartanyan}}}]{2020MNRAS.492.5764N}%
  \BibitemOpen
  \bibfield  {author} {\bibinfo {author} {\bibfnamefont {Hiroki}\ \bibnamefont
  {{Nagakura}}}, \bibinfo {author} {\bibfnamefont {Adam}\ \bibnamefont
  {{Burrows}}}, \bibinfo {author} {\bibfnamefont {David}\ \bibnamefont
  {{Radice}}}, \ and\ \bibinfo {author} {\bibfnamefont {David}\ \bibnamefont
  {{Vartanyan}}},\ }\bibfield  {title} {\enquote {\bibinfo {title} {{A
  systematic study of proto-neutron star convection in three-dimensional
  core-collapse supernova simulations}},}\ }\href {\doibase
  10.1093/mnras/staa261} {\bibfield  {journal} {\bibinfo  {journal} {\mnras}\
  }\textbf {\bibinfo {volume} {492}},\ \bibinfo {pages} {5764--5779} (\bibinfo
  {year} {2020})},\ \Eprint {http://arxiv.org/abs/1912.07615} {arXiv:1912.07615
  [astro-ph.HE]} \BibitemShut {NoStop}%
\bibitem [{\citenamefont {{Glas}}\ \emph {et~al.}(2020)\citenamefont {{Glas}},
  \citenamefont {{Janka}}, \citenamefont {{Capozzi}}, \citenamefont {{Sen}},
  \citenamefont {{Dasgupta}}, \citenamefont {{Mirizzi}},\ and\ \citenamefont
  {{Sigl}}}]{2020PhRvD.101f3001G}%
  \BibitemOpen
  \bibfield  {author} {\bibinfo {author} {\bibfnamefont {Robert}\ \bibnamefont
  {{Glas}}}, \bibinfo {author} {\bibfnamefont {H.~Thomas}\ \bibnamefont
  {{Janka}}}, \bibinfo {author} {\bibfnamefont {Francesco}\ \bibnamefont
  {{Capozzi}}}, \bibinfo {author} {\bibfnamefont {Manibrata}\ \bibnamefont
  {{Sen}}}, \bibinfo {author} {\bibfnamefont {Basudeb}\ \bibnamefont
  {{Dasgupta}}}, \bibinfo {author} {\bibfnamefont {Alessandro}\ \bibnamefont
  {{Mirizzi}}}, \ and\ \bibinfo {author} {\bibfnamefont {G{\"u}nter}\
  \bibnamefont {{Sigl}}},\ }\bibfield  {title} {\enquote {\bibinfo {title}
  {{Fast neutrino flavor instability in the neutron-star convection layer of
  three-dimensional supernova models}},}\ }\href {\doibase
  10.1103/PhysRevD.101.063001} {\bibfield  {journal} {\bibinfo  {journal}
  {\prd}\ }\textbf {\bibinfo {volume} {101}},\ \bibinfo {eid} {063001}
  (\bibinfo {year} {2020})},\ \Eprint {http://arxiv.org/abs/1912.00274}
  {arXiv:1912.00274 [astro-ph.HE]} \BibitemShut {NoStop}%
\bibitem [{\citenamefont {{Delfan Azari}}\ \emph {et~al.}(2020)\citenamefont
  {{Delfan Azari}}, \citenamefont {{Yamada}}, \citenamefont {{Morinaga}},
  \citenamefont {{Nagakura}}, \citenamefont {{Furusawa}}, \citenamefont
  {{Harada}}, \citenamefont {{Okawa}}, \citenamefont {{Iwakami}},\ and\
  \citenamefont {{Sumiyoshi}}}]{2020PhRvD.101b3018D}%
  \BibitemOpen
  \bibfield  {author} {\bibinfo {author} {\bibfnamefont {Milad}\ \bibnamefont
  {{Delfan Azari}}}, \bibinfo {author} {\bibfnamefont {Shoichi}\ \bibnamefont
  {{Yamada}}}, \bibinfo {author} {\bibfnamefont {Taiki}\ \bibnamefont
  {{Morinaga}}}, \bibinfo {author} {\bibfnamefont {Hiroki}\ \bibnamefont
  {{Nagakura}}}, \bibinfo {author} {\bibfnamefont {Shun}\ \bibnamefont
  {{Furusawa}}}, \bibinfo {author} {\bibfnamefont {Akira}\ \bibnamefont
  {{Harada}}}, \bibinfo {author} {\bibfnamefont {Hirotada}\ \bibnamefont
  {{Okawa}}}, \bibinfo {author} {\bibfnamefont {Wakana}\ \bibnamefont
  {{Iwakami}}}, \ and\ \bibinfo {author} {\bibfnamefont {Kohsuke}\ \bibnamefont
  {{Sumiyoshi}}},\ }\bibfield  {title} {\enquote {\bibinfo {title} {{Fast
  collective neutrino oscillations inside the neutrino sphere in core-collapse
  supernovae}},}\ }\href {\doibase 10.1103/PhysRevD.101.023018} {\bibfield
  {journal} {\bibinfo  {journal} {\prd}\ }\textbf {\bibinfo {volume} {101}},\
  \bibinfo {eid} {023018} (\bibinfo {year} {2020})},\ \Eprint
  {http://arxiv.org/abs/1910.06176} {arXiv:1910.06176 [astro-ph.HE]}
  \BibitemShut {NoStop}%
\bibitem [{\citenamefont {{Tamborra}}\ \emph {et~al.}(2014)\citenamefont
  {{Tamborra}}, \citenamefont {{Hanke}}, \citenamefont {{Janka}}, \citenamefont
  {{M{\"u}ller}}, \citenamefont {{Raffelt}},\ and\ \citenamefont
  {{Marek}}}]{2014ApJ...792...96T}%
  \BibitemOpen
  \bibfield  {author} {\bibinfo {author} {\bibfnamefont {Irene}\ \bibnamefont
  {{Tamborra}}}, \bibinfo {author} {\bibfnamefont {Florian}\ \bibnamefont
  {{Hanke}}}, \bibinfo {author} {\bibfnamefont {Hans-Thomas}\ \bibnamefont
  {{Janka}}}, \bibinfo {author} {\bibfnamefont {Bernhard}\ \bibnamefont
  {{M{\"u}ller}}}, \bibinfo {author} {\bibfnamefont {Georg~G.}\ \bibnamefont
  {{Raffelt}}}, \ and\ \bibinfo {author} {\bibfnamefont {Andreas}\ \bibnamefont
  {{Marek}}},\ }\bibfield  {title} {\enquote {\bibinfo {title} {{Self-sustained
  Asymmetry of Lepton-number Emission: A New Phenomenon during the Supernova
  Shock-accretion Phase in Three Dimensions}},}\ }\href {\doibase
  10.1088/0004-637X/792/2/96} {\bibfield  {journal} {\bibinfo  {journal}
  {\apj}\ }\textbf {\bibinfo {volume} {792}},\ \bibinfo {eid} {96} (\bibinfo
  {year} {2014})},\ \Eprint {http://arxiv.org/abs/1402.5418} {arXiv:1402.5418
  [astro-ph.SR]} \BibitemShut {NoStop}%
\bibitem [{\citenamefont {{Glas}}\ \emph {et~al.}(2019)\citenamefont {{Glas}},
  \citenamefont {{Janka}}, \citenamefont {{Melson}}, \citenamefont
  {{Stockinger}},\ and\ \citenamefont {{Just}}}]{2019ApJ...881...36G}%
  \BibitemOpen
  \bibfield  {author} {\bibinfo {author} {\bibfnamefont {Robert}\ \bibnamefont
  {{Glas}}}, \bibinfo {author} {\bibfnamefont {H.~Thomas}\ \bibnamefont
  {{Janka}}}, \bibinfo {author} {\bibfnamefont {Tobias}\ \bibnamefont
  {{Melson}}}, \bibinfo {author} {\bibfnamefont {Georg}\ \bibnamefont
  {{Stockinger}}}, \ and\ \bibinfo {author} {\bibfnamefont {Oliver}\
  \bibnamefont {{Just}}},\ }\bibfield  {title} {\enquote {\bibinfo {title}
  {{Effects of LESA in Three-dimensional Supernova Simulations with
  Multidimensional and Ray-by-ray-plus Neutrino Transport}},}\ }\href {\doibase
  10.3847/1538-4357/ab275c} {\bibfield  {journal} {\bibinfo  {journal} {\apj}\
  }\textbf {\bibinfo {volume} {881}},\ \bibinfo {eid} {36} (\bibinfo {year}
  {2019})},\ \Eprint {http://arxiv.org/abs/1809.10150} {arXiv:1809.10150
  [astro-ph.HE]} \BibitemShut {NoStop}%
\bibitem [{\citenamefont {{Powell}}\ and\ \citenamefont
  {{M{\"u}ller}}(2019)}]{2019MNRAS.487.1178P}%
  \BibitemOpen
  \bibfield  {author} {\bibinfo {author} {\bibfnamefont {Jade}\ \bibnamefont
  {{Powell}}}\ and\ \bibinfo {author} {\bibfnamefont {Bernhard}\ \bibnamefont
  {{M{\"u}ller}}},\ }\bibfield  {title} {\enquote {\bibinfo {title}
  {{Gravitational wave emission from 3D explosion models of core-collapse
  supernovae with low and normal explosion energies}},}\ }\href {\doibase
  10.1093/mnras/stz1304} {\bibfield  {journal} {\bibinfo  {journal} {\mnras}\
  }\textbf {\bibinfo {volume} {487}},\ \bibinfo {pages} {1178--1190} (\bibinfo
  {year} {2019})},\ \Eprint {http://arxiv.org/abs/1812.05738} {arXiv:1812.05738
  [astro-ph.HE]} \BibitemShut {NoStop}%
\bibitem [{\citenamefont {{Li}}\ and\ \citenamefont
  {{Siegel}}(2021)}]{2021PhRvL.126y1101L}%
  \BibitemOpen
  \bibfield  {author} {\bibinfo {author} {\bibfnamefont {Xinyu}\ \bibnamefont
  {{Li}}}\ and\ \bibinfo {author} {\bibfnamefont {Daniel~M.}\ \bibnamefont
  {{Siegel}}},\ }\bibfield  {title} {\enquote {\bibinfo {title} {{Neutrino Fast
  Flavor Conversions in Neutron-Star Postmerger Accretion Disks}},}\ }\href
  {\doibase 10.1103/PhysRevLett.126.251101} {\bibfield  {journal} {\bibinfo
  {journal} {\prl}\ }\textbf {\bibinfo {volume} {126}},\ \bibinfo {eid}
  {251101} (\bibinfo {year} {2021})},\ \Eprint
  {http://arxiv.org/abs/2103.02616} {arXiv:2103.02616 [astro-ph.HE]}
  \BibitemShut {NoStop}%
\bibitem [{\citenamefont {{Just}}\ \emph {et~al.}(2022)\citenamefont {{Just}},
  \citenamefont {{Abbar}}, \citenamefont {{Wu}}, \citenamefont {{Tamborra}},
  \citenamefont {{Janka}},\ and\ \citenamefont
  {{Capozzi}}}]{2022PhRvD.105h3024J}%
  \BibitemOpen
  \bibfield  {author} {\bibinfo {author} {\bibfnamefont {Oliver}\ \bibnamefont
  {{Just}}}, \bibinfo {author} {\bibfnamefont {Sajad}\ \bibnamefont {{Abbar}}},
  \bibinfo {author} {\bibfnamefont {Meng-Ru}\ \bibnamefont {{Wu}}}, \bibinfo
  {author} {\bibfnamefont {Irene}\ \bibnamefont {{Tamborra}}}, \bibinfo
  {author} {\bibfnamefont {Hans-Thomas}\ \bibnamefont {{Janka}}}, \ and\
  \bibinfo {author} {\bibfnamefont {Francesco}\ \bibnamefont {{Capozzi}}},\
  }\bibfield  {title} {\enquote {\bibinfo {title} {{Fast neutrino conversion in
  hydrodynamic simulations of neutrino-cooled accretion disks}},}\ }\href
  {\doibase 10.1103/PhysRevD.105.083024} {\bibfield  {journal} {\bibinfo
  {journal} {\prd}\ }\textbf {\bibinfo {volume} {105}},\ \bibinfo {eid}
  {083024} (\bibinfo {year} {2022})},\ \Eprint
  {http://arxiv.org/abs/2203.16559} {arXiv:2203.16559 [astro-ph.HE]}
  \BibitemShut {NoStop}%
\bibitem [{\citenamefont {{Fern{\'a}ndez}}\ \emph {et~al.}(2022)\citenamefont
  {{Fern{\'a}ndez}}, \citenamefont {{Richers}}, \citenamefont {{Mulyk}},\ and\
  \citenamefont {{Fahlman}}}]{2022PhRvD.106j3003F}%
  \BibitemOpen
  \bibfield  {author} {\bibinfo {author} {\bibfnamefont {Rodrigo}\ \bibnamefont
  {{Fern{\'a}ndez}}}, \bibinfo {author} {\bibfnamefont {Sherwood}\ \bibnamefont
  {{Richers}}}, \bibinfo {author} {\bibfnamefont {Nicole}\ \bibnamefont
  {{Mulyk}}}, \ and\ \bibinfo {author} {\bibfnamefont {Steven}\ \bibnamefont
  {{Fahlman}}},\ }\bibfield  {title} {\enquote {\bibinfo {title} {{Fast flavor
  instability in hypermassive neutron star disk outflows}},}\ }\href {\doibase
  10.1103/PhysRevD.106.103003} {\bibfield  {journal} {\bibinfo  {journal}
  {\prd}\ }\textbf {\bibinfo {volume} {106}},\ \bibinfo {eid} {103003}
  (\bibinfo {year} {2022})},\ \Eprint {http://arxiv.org/abs/2207.10680}
  {arXiv:2207.10680 [astro-ph.HE]} \BibitemShut {NoStop}%
\bibitem [{\citenamefont {{Ehring}}\ \emph
  {et~al.}(2023{\natexlab{a}})\citenamefont {{Ehring}}, \citenamefont
  {{Abbar}}, \citenamefont {{Janka}}, \citenamefont {{Raffelt}},\ and\
  \citenamefont {{Tamborra}}}]{2023PhRvD.107j3034E}%
  \BibitemOpen
  \bibfield  {author} {\bibinfo {author} {\bibfnamefont {Jakob}\ \bibnamefont
  {{Ehring}}}, \bibinfo {author} {\bibfnamefont {Sajad}\ \bibnamefont
  {{Abbar}}}, \bibinfo {author} {\bibfnamefont {Hans-Thomas}\ \bibnamefont
  {{Janka}}}, \bibinfo {author} {\bibfnamefont {Georg}\ \bibnamefont
  {{Raffelt}}}, \ and\ \bibinfo {author} {\bibfnamefont {Irene}\ \bibnamefont
  {{Tamborra}}},\ }\bibfield  {title} {\enquote {\bibinfo {title} {{Fast
  neutrino flavor conversion in core-collapse supernovae: A parametric study in
  1D models}},}\ }\href {\doibase 10.1103/PhysRevD.107.103034} {\bibfield
  {journal} {\bibinfo  {journal} {\prd}\ }\textbf {\bibinfo {volume} {107}},\
  \bibinfo {eid} {103034} (\bibinfo {year} {2023}{\natexlab{a}})},\ \Eprint
  {http://arxiv.org/abs/2301.11938} {arXiv:2301.11938 [astro-ph.HE]}
  \BibitemShut {NoStop}%
\bibitem [{\citenamefont {{Ehring}}\ \emph
  {et~al.}(2023{\natexlab{b}})\citenamefont {{Ehring}}, \citenamefont
  {{Abbar}}, \citenamefont {{Janka}}, \citenamefont {{Raffelt}},\ and\
  \citenamefont {{Tamborra}}}]{2023PhRvL.131f1401E}%
  \BibitemOpen
  \bibfield  {author} {\bibinfo {author} {\bibfnamefont {Jakob}\ \bibnamefont
  {{Ehring}}}, \bibinfo {author} {\bibfnamefont {Sajad}\ \bibnamefont
  {{Abbar}}}, \bibinfo {author} {\bibfnamefont {Hans-Thomas}\ \bibnamefont
  {{Janka}}}, \bibinfo {author} {\bibfnamefont {Georg}\ \bibnamefont
  {{Raffelt}}}, \ and\ \bibinfo {author} {\bibfnamefont {Irene}\ \bibnamefont
  {{Tamborra}}},\ }\bibfield  {title} {\enquote {\bibinfo {title} {{Fast
  Neutrino Flavor Conversions Can Help and Hinder Neutrino-Driven
  Explosions}},}\ }\href {\doibase 10.1103/PhysRevLett.131.061401} {\bibfield
  {journal} {\bibinfo  {journal} {\prl}\ }\textbf {\bibinfo {volume} {131}},\
  \bibinfo {eid} {061401} (\bibinfo {year} {2023}{\natexlab{b}})},\ \Eprint
  {http://arxiv.org/abs/2305.11207} {arXiv:2305.11207 [astro-ph.HE]}
  \BibitemShut {NoStop}%
\bibitem [{\citenamefont {{Nagakura}}(2023)}]{2023PhRvL.130u1401N}%
  \BibitemOpen
  \bibfield  {author} {\bibinfo {author} {\bibfnamefont {Hiroki}\ \bibnamefont
  {{Nagakura}}},\ }\bibfield  {title} {\enquote {\bibinfo {title} {{Roles of
  Fast Neutrino-Flavor Conversion on the Neutrino-Heating Mechanism of
  Core-Collapse Supernova}},}\ }\href {\doibase 10.1103/PhysRevLett.130.211401}
  {\bibfield  {journal} {\bibinfo  {journal} {\prl}\ }\textbf {\bibinfo
  {volume} {130}},\ \bibinfo {eid} {211401} (\bibinfo {year} {2023})},\ \Eprint
  {http://arxiv.org/abs/2301.10785} {arXiv:2301.10785 [astro-ph.HE]}
  \BibitemShut {NoStop}%
\bibitem [{\citenamefont {{Nagakura}}\ and\ \citenamefont
  {{Zaizen}}(2023{\natexlab{a}})}]{2023PhRvD.108l3003N}%
  \BibitemOpen
  \bibfield  {author} {\bibinfo {author} {\bibfnamefont {Hiroki}\ \bibnamefont
  {{Nagakura}}}\ and\ \bibinfo {author} {\bibfnamefont {Masamichi}\
  \bibnamefont {{Zaizen}}},\ }\bibfield  {title} {\enquote {\bibinfo {title}
  {{Basic characteristics of neutrino flavor conversions in the postshock
  regions of core-collapse supernova}},}\ }\href {\doibase
  10.1103/PhysRevD.108.123003} {\bibfield  {journal} {\bibinfo  {journal}
  {\prd}\ }\textbf {\bibinfo {volume} {108}},\ \bibinfo {eid} {123003}
  (\bibinfo {year} {2023}{\natexlab{a}})},\ \Eprint
  {http://arxiv.org/abs/2308.14800} {arXiv:2308.14800 [astro-ph.HE]}
  \BibitemShut {NoStop}%
\bibitem [{\citenamefont {{Bollig}}\ \emph {et~al.}(2017)\citenamefont
  {{Bollig}}, \citenamefont {{Janka}}, \citenamefont {{Lohs}}, \citenamefont
  {{Mart{\'\i}nez-Pinedo}}, \citenamefont {{Horowitz}},\ and\ \citenamefont
  {{Melson}}}]{2017PhRvL.119x2702B}%
  \BibitemOpen
  \bibfield  {author} {\bibinfo {author} {\bibfnamefont {R.}~\bibnamefont
  {{Bollig}}}, \bibinfo {author} {\bibfnamefont {H.~T.}\ \bibnamefont
  {{Janka}}}, \bibinfo {author} {\bibfnamefont {A.}~\bibnamefont {{Lohs}}},
  \bibinfo {author} {\bibfnamefont {G.}~\bibnamefont {{Mart{\'\i}nez-Pinedo}}},
  \bibinfo {author} {\bibfnamefont {C.~J.}\ \bibnamefont {{Horowitz}}}, \ and\
  \bibinfo {author} {\bibfnamefont {T.}~\bibnamefont {{Melson}}},\ }\bibfield
  {title} {\enquote {\bibinfo {title} {{Muon Creation in Supernova Matter
  Facilitates Neutrino-Driven Explosions}},}\ }\href {\doibase
  10.1103/PhysRevLett.119.242702} {\bibfield  {journal} {\bibinfo  {journal}
  {\prl}\ }\textbf {\bibinfo {volume} {119}},\ \bibinfo {eid} {242702}
  (\bibinfo {year} {2017})},\ \Eprint {http://arxiv.org/abs/1706.04630}
  {arXiv:1706.04630 [astro-ph.HE]} \BibitemShut {NoStop}%
\bibitem [{\citenamefont {{Fischer}}\ \emph {et~al.}(2020)\citenamefont
  {{Fischer}}, \citenamefont {{Guo}}, \citenamefont {{Mart{\'\i}nez-Pinedo}},
  \citenamefont {{Liebend{\"o}rfer}},\ and\ \citenamefont
  {{Mezzacappa}}}]{2020PhRvD.102l3001F}%
  \BibitemOpen
  \bibfield  {author} {\bibinfo {author} {\bibfnamefont {Tobias}\ \bibnamefont
  {{Fischer}}}, \bibinfo {author} {\bibfnamefont {Gang}\ \bibnamefont {{Guo}}},
  \bibinfo {author} {\bibfnamefont {Gabriel}\ \bibnamefont
  {{Mart{\'\i}nez-Pinedo}}}, \bibinfo {author} {\bibfnamefont {Matthias}\
  \bibnamefont {{Liebend{\"o}rfer}}}, \ and\ \bibinfo {author} {\bibfnamefont
  {Anthony}\ \bibnamefont {{Mezzacappa}}},\ }\bibfield  {title} {\enquote
  {\bibinfo {title} {{Muonization of supernova matter}},}\ }\href {\doibase
  10.1103/PhysRevD.102.123001} {\bibfield  {journal} {\bibinfo  {journal}
  {\prd}\ }\textbf {\bibinfo {volume} {102}},\ \bibinfo {eid} {123001}
  (\bibinfo {year} {2020})},\ \Eprint {http://arxiv.org/abs/2008.13628}
  {arXiv:2008.13628 [astro-ph.HE]} \BibitemShut {NoStop}%
\bibitem [{\citenamefont {{Guo}}\ \emph {et~al.}(2020)\citenamefont {{Guo}},
  \citenamefont {{Mart{\'\i}nez-Pinedo}}, \citenamefont {{Lohs}},\ and\
  \citenamefont {{Fischer}}}]{2020PhRvD.102b3037G}%
  \BibitemOpen
  \bibfield  {author} {\bibinfo {author} {\bibfnamefont {Gang}\ \bibnamefont
  {{Guo}}}, \bibinfo {author} {\bibfnamefont {Gabriel}\ \bibnamefont
  {{Mart{\'\i}nez-Pinedo}}}, \bibinfo {author} {\bibfnamefont {A.}~\bibnamefont
  {{Lohs}}}, \ and\ \bibinfo {author} {\bibfnamefont {Tobias}\ \bibnamefont
  {{Fischer}}},\ }\bibfield  {title} {\enquote {\bibinfo {title}
  {{Charged-current muonic reactions in core-collapse supernovae}},}\ }\href
  {\doibase 10.1103/PhysRevD.102.023037} {\bibfield  {journal} {\bibinfo
  {journal} {\prd}\ }\textbf {\bibinfo {volume} {102}},\ \bibinfo {eid}
  {023037} (\bibinfo {year} {2020})},\ \Eprint
  {http://arxiv.org/abs/2006.12051} {arXiv:2006.12051 [hep-ph]} \BibitemShut
  {NoStop}%
\bibitem [{\citenamefont {{Fujimoto}}\ and\ \citenamefont
  {{Nagakura}}(2023)}]{2023MNRAS.519.2623F}%
  \BibitemOpen
  \bibfield  {author} {\bibinfo {author} {\bibfnamefont {Shin-ichiro}\
  \bibnamefont {{Fujimoto}}}\ and\ \bibinfo {author} {\bibfnamefont {Hiroki}\
  \bibnamefont {{Nagakura}}},\ }\bibfield  {title} {\enquote {\bibinfo {title}
  {{Explosive nucleosynthesis with fast neutrino-flavour conversion in
  core-collapse supernovae}},}\ }\href {\doibase 10.1093/mnras/stac3763}
  {\bibfield  {journal} {\bibinfo  {journal} {\mnras}\ }\textbf {\bibinfo
  {volume} {519}},\ \bibinfo {pages} {2623--2629} (\bibinfo {year} {2023})},\
  \Eprint {http://arxiv.org/abs/2210.02106} {arXiv:2210.02106 [astro-ph.HE]}
  \BibitemShut {NoStop}%
\bibitem [{\citenamefont {{Holland-Ashford}}\ \emph {et~al.}(2020)\citenamefont
  {{Holland-Ashford}}, \citenamefont {{Lopez}},\ and\ \citenamefont
  {{Auchettl}}}]{2020ApJ...889..144H}%
  \BibitemOpen
  \bibfield  {author} {\bibinfo {author} {\bibfnamefont {Tyler}\ \bibnamefont
  {{Holland-Ashford}}}, \bibinfo {author} {\bibfnamefont {Laura~A.}\
  \bibnamefont {{Lopez}}}, \ and\ \bibinfo {author} {\bibfnamefont {Katie}\
  \bibnamefont {{Auchettl}}},\ }\bibfield  {title} {\enquote {\bibinfo {title}
  {{Asymmetries of Heavy Elements in the Young Supernova Remnant Cassiopeia
  A}},}\ }\href {\doibase 10.3847/1538-4357/ab64e4} {\bibfield  {journal}
  {\bibinfo  {journal} {\apj}\ }\textbf {\bibinfo {volume} {889}},\ \bibinfo
  {eid} {144} (\bibinfo {year} {2020})},\ \Eprint
  {http://arxiv.org/abs/1904.06357} {arXiv:1904.06357 [astro-ph.HE]}
  \BibitemShut {NoStop}%
\bibitem [{\citenamefont {{XRISM Science Team}}(2020)}]{2020arXiv200304962X}%
  \BibitemOpen
  \bibfield  {author} {\bibinfo {author} {\bibnamefont {{XRISM Science
  Team}}},\ }\bibfield  {title} {\enquote {\bibinfo {title} {{Science with the
  X-ray Imaging and Spectroscopy Mission (XRISM)}},}\ }\href@noop {} {\bibfield
   {journal} {\bibinfo  {journal} {arXiv e-prints}\ ,\ \bibinfo {eid}
  {arXiv:2003.04962}} (\bibinfo {year} {2020})},\ \Eprint
  {http://arxiv.org/abs/2003.04962} {arXiv:2003.04962 [astro-ph.HE]}
  \BibitemShut {NoStop}%
\bibitem [{\citenamefont {{Nagakura}}\ \emph {et~al.}(2018)\citenamefont
  {{Nagakura}}, \citenamefont {{Iwakami}}, \citenamefont {{Furusawa}},
  \citenamefont {{Okawa}}, \citenamefont {{Harada}}, \citenamefont
  {{Sumiyoshi}}, \citenamefont {{Yamada}}, \citenamefont {{Matsufuru}},\ and\
  \citenamefont {{Imakura}}}]{2018ApJ...854..136N}%
  \BibitemOpen
  \bibfield  {author} {\bibinfo {author} {\bibfnamefont {H.}~\bibnamefont
  {{Nagakura}}}, \bibinfo {author} {\bibfnamefont {W.}~\bibnamefont
  {{Iwakami}}}, \bibinfo {author} {\bibfnamefont {S.}~\bibnamefont
  {{Furusawa}}}, \bibinfo {author} {\bibfnamefont {H.}~\bibnamefont {{Okawa}}},
  \bibinfo {author} {\bibfnamefont {A.}~\bibnamefont {{Harada}}}, \bibinfo
  {author} {\bibfnamefont {K.}~\bibnamefont {{Sumiyoshi}}}, \bibinfo {author}
  {\bibfnamefont {S.}~\bibnamefont {{Yamada}}}, \bibinfo {author}
  {\bibfnamefont {H.}~\bibnamefont {{Matsufuru}}}, \ and\ \bibinfo {author}
  {\bibfnamefont {A.}~\bibnamefont {{Imakura}}},\ }\bibfield  {title} {\enquote
  {\bibinfo {title} {{Simulations of Core-collapse Supernovae in Spatial
  Axisymmetry with Full Boltzmann Neutrino Transport}},}\ }\href {\doibase
  10.3847/1538-4357/aaac29} {\bibfield  {journal} {\bibinfo  {journal} {\apj}\
  }\textbf {\bibinfo {volume} {854}},\ \bibinfo {eid} {136} (\bibinfo {year}
  {2018})},\ \Eprint {http://arxiv.org/abs/1702.01752} {arXiv:1702.01752
  [astro-ph.HE]} \BibitemShut {NoStop}%
\bibitem [{\citenamefont {{Woosley}}\ \emph {et~al.}(2002)\citenamefont
  {{Woosley}}, \citenamefont {{Heger}},\ and\ \citenamefont
  {{Weaver}}}]{2002RvMP...74.1015W}%
  \BibitemOpen
  \bibfield  {author} {\bibinfo {author} {\bibfnamefont {S.~E.}\ \bibnamefont
  {{Woosley}}}, \bibinfo {author} {\bibfnamefont {A.}~\bibnamefont {{Heger}}},
  \ and\ \bibinfo {author} {\bibfnamefont {T.~A.}\ \bibnamefont {{Weaver}}},\
  }\bibfield  {title} {\enquote {\bibinfo {title} {{The evolution and explosion
  of massive stars}},}\ }\href {\doibase 10.1103/RevModPhys.74.1015} {\bibfield
   {journal} {\bibinfo  {journal} {Reviews of Modern Physics}\ }\textbf
  {\bibinfo {volume} {74}},\ \bibinfo {pages} {1015--1071} (\bibinfo {year}
  {2002})}\BibitemShut {NoStop}%
\bibitem [{\citenamefont {{Furusawa}}\ \emph {et~al.}(2017)\citenamefont
  {{Furusawa}}, \citenamefont {{Togashi}}, \citenamefont {{Nagakura}},
  \citenamefont {{Sumiyoshi}}, \citenamefont {{Yamada}}, \citenamefont
  {{Suzuki}},\ and\ \citenamefont {{Takano}}}]{2017JPhG...44i4001F}%
  \BibitemOpen
  \bibfield  {author} {\bibinfo {author} {\bibfnamefont {S.}~\bibnamefont
  {{Furusawa}}}, \bibinfo {author} {\bibfnamefont {H.}~\bibnamefont
  {{Togashi}}}, \bibinfo {author} {\bibfnamefont {H.}~\bibnamefont
  {{Nagakura}}}, \bibinfo {author} {\bibfnamefont {K.}~\bibnamefont
  {{Sumiyoshi}}}, \bibinfo {author} {\bibfnamefont {S.}~\bibnamefont
  {{Yamada}}}, \bibinfo {author} {\bibfnamefont {H.}~\bibnamefont {{Suzuki}}},
  \ and\ \bibinfo {author} {\bibfnamefont {M.}~\bibnamefont {{Takano}}},\
  }\bibfield  {title} {\enquote {\bibinfo {title} {{A new equation of state for
  core-collapse supernovae based on realistic nuclear forces and including a
  full nuclear ensemble}},}\ }\href {\doibase 10.1088/1361-6471/aa7f35}
  {\bibfield  {journal} {\bibinfo  {journal} {Journal of Physics G Nuclear
  Physics}\ }\textbf {\bibinfo {volume} {44}},\ \bibinfo {pages} {094001}
  (\bibinfo {year} {2017})},\ \Eprint {http://arxiv.org/abs/1707.06410}
  {arXiv:1707.06410 [astro-ph.HE]} \BibitemShut {NoStop}%
\bibitem [{\citenamefont {{Sumiyoshi}}\ and\ \citenamefont
  {{Yamada}}(2012)}]{2012ApJS..199...17S}%
  \BibitemOpen
  \bibfield  {author} {\bibinfo {author} {\bibfnamefont {K.}~\bibnamefont
  {{Sumiyoshi}}}\ and\ \bibinfo {author} {\bibfnamefont {S.}~\bibnamefont
  {{Yamada}}},\ }\bibfield  {title} {\enquote {\bibinfo {title} {{Neutrino
  Transfer in Three Dimensions for Core-collapse Supernovae. I. Static
  Configurations}},}\ }\href {\doibase 10.1088/0067-0049/199/1/17} {\bibfield
  {journal} {\bibinfo  {journal} {\apjs}\ }\textbf {\bibinfo {volume} {199}},\
  \bibinfo {eid} {17} (\bibinfo {year} {2012})},\ \Eprint
  {http://arxiv.org/abs/1201.2244} {arXiv:1201.2244 [astro-ph.HE]} \BibitemShut
  {NoStop}%
\bibitem [{\citenamefont {{Sumiyoshi}}\ \emph {et~al.}(2015)\citenamefont
  {{Sumiyoshi}}, \citenamefont {{Takiwaki}}, \citenamefont {{Matsufuru}},\ and\
  \citenamefont {{Yamada}}}]{2015ApJS..216....5S}%
  \BibitemOpen
  \bibfield  {author} {\bibinfo {author} {\bibfnamefont {K.}~\bibnamefont
  {{Sumiyoshi}}}, \bibinfo {author} {\bibfnamefont {T.}~\bibnamefont
  {{Takiwaki}}}, \bibinfo {author} {\bibfnamefont {H.}~\bibnamefont
  {{Matsufuru}}}, \ and\ \bibinfo {author} {\bibfnamefont {S.}~\bibnamefont
  {{Yamada}}},\ }\bibfield  {title} {\enquote {\bibinfo {title}
  {{Multi-dimensional Features of Neutrino Transfer in Core-collapse
  Supernovae}},}\ }\href {\doibase 10.1088/0067-0049/216/1/5} {\bibfield
  {journal} {\bibinfo  {journal} {\apjs}\ }\textbf {\bibinfo {volume} {216}},\
  \bibinfo {eid} {5} (\bibinfo {year} {2015})},\ \Eprint
  {http://arxiv.org/abs/1403.4476} {arXiv:1403.4476 [astro-ph.HE]} \BibitemShut
  {NoStop}%
\bibitem [{\citenamefont {{Sumiyoshi}}\ \emph {et~al.}(2021)\citenamefont
  {{Sumiyoshi}}, \citenamefont {{Fujibayashi}}, \citenamefont {{Sekiguchi}},\
  and\ \citenamefont {{Shibata}}}]{2021ApJ...907...92S}%
  \BibitemOpen
  \bibfield  {author} {\bibinfo {author} {\bibfnamefont {Kohsuke}\ \bibnamefont
  {{Sumiyoshi}}}, \bibinfo {author} {\bibfnamefont {Sho}\ \bibnamefont
  {{Fujibayashi}}}, \bibinfo {author} {\bibfnamefont {Yuichiro}\ \bibnamefont
  {{Sekiguchi}}}, \ and\ \bibinfo {author} {\bibfnamefont {Masaru}\
  \bibnamefont {{Shibata}}},\ }\bibfield  {title} {\enquote {\bibinfo {title}
  {{Properties of Neutrino Transfer in a Deformed Remnant of a Neutron Star
  Merger}},}\ }\href {\doibase 10.3847/1538-4357/abce63} {\bibfield  {journal}
  {\bibinfo  {journal} {\apj}\ }\textbf {\bibinfo {volume} {907}},\ \bibinfo
  {eid} {92} (\bibinfo {year} {2021})},\ \Eprint
  {http://arxiv.org/abs/2010.10865} {arXiv:2010.10865 [astro-ph.HE]}
  \BibitemShut {NoStop}%
\bibitem [{\citenamefont {{Nagakura}}\ \emph
  {et~al.}(2019{\natexlab{d}})\citenamefont {{Nagakura}}, \citenamefont
  {{Furusawa}}, \citenamefont {{Togashi}}, \citenamefont {{Richers}},
  \citenamefont {{Sumiyoshi}},\ and\ \citenamefont
  {{Yamada}}}]{2019ApJS..240...38N}%
  \BibitemOpen
  \bibfield  {author} {\bibinfo {author} {\bibfnamefont {Hiroki}\ \bibnamefont
  {{Nagakura}}}, \bibinfo {author} {\bibfnamefont {Shun}\ \bibnamefont
  {{Furusawa}}}, \bibinfo {author} {\bibfnamefont {Hajime}\ \bibnamefont
  {{Togashi}}}, \bibinfo {author} {\bibfnamefont {Sherwood}\ \bibnamefont
  {{Richers}}}, \bibinfo {author} {\bibfnamefont {Kohsuke}\ \bibnamefont
  {{Sumiyoshi}}}, \ and\ \bibinfo {author} {\bibfnamefont {Shoichi}\
  \bibnamefont {{Yamada}}},\ }\bibfield  {title} {\enquote {\bibinfo {title}
  {{Comparing Treatments of Weak Reactions with Nuclei in Simulations of
  Core-collapse Supernovae}},}\ }\href {\doibase 10.3847/1538-4365/aafac9}
  {\bibfield  {journal} {\bibinfo  {journal} {\apjs}\ }\textbf {\bibinfo
  {volume} {240}},\ \bibinfo {eid} {38} (\bibinfo {year}
  {2019}{\natexlab{d}})},\ \Eprint {http://arxiv.org/abs/1812.09811}
  {arXiv:1812.09811 [astro-ph.HE]} \BibitemShut {NoStop}%
\bibitem [{\citenamefont {{Harada}}\ \emph {et~al.}(2020)\citenamefont
  {{Harada}}, \citenamefont {{Nagakura}}, \citenamefont {{Iwakami}},
  \citenamefont {{Okawa}}, \citenamefont {{Furusawa}}, \citenamefont
  {{Sumiyoshi}}, \citenamefont {{Matsufuru}},\ and\ \citenamefont
  {{Yamada}}}]{2020ApJ...902..150H}%
  \BibitemOpen
  \bibfield  {author} {\bibinfo {author} {\bibfnamefont {Akira}\ \bibnamefont
  {{Harada}}}, \bibinfo {author} {\bibfnamefont {Hiroki}\ \bibnamefont
  {{Nagakura}}}, \bibinfo {author} {\bibfnamefont {Wakana}\ \bibnamefont
  {{Iwakami}}}, \bibinfo {author} {\bibfnamefont {Hirotada}\ \bibnamefont
  {{Okawa}}}, \bibinfo {author} {\bibfnamefont {Shun}\ \bibnamefont
  {{Furusawa}}}, \bibinfo {author} {\bibfnamefont {Kohsuke}\ \bibnamefont
  {{Sumiyoshi}}}, \bibinfo {author} {\bibfnamefont {Hideo}\ \bibnamefont
  {{Matsufuru}}}, \ and\ \bibinfo {author} {\bibfnamefont {Shoichi}\
  \bibnamefont {{Yamada}}},\ }\bibfield  {title} {\enquote {\bibinfo {title}
  {{The Boltzmann-radiation-hydrodynamics Simulations of Core-collapse
  Supernovae with Different Equations of State: The Role of Nuclear Composition
  and the Behavior of Neutrinos}},}\ }\href {\doibase 10.3847/1538-4357/abb5a9}
  {\bibfield  {journal} {\bibinfo  {journal} {\apj}\ }\textbf {\bibinfo
  {volume} {902}},\ \bibinfo {eid} {150} (\bibinfo {year} {2020})},\ \Eprint
  {http://arxiv.org/abs/2003.08630} {arXiv:2003.08630 [astro-ph.HE]}
  \BibitemShut {NoStop}%
\bibitem [{\citenamefont {{Iwakami}}\ \emph {et~al.}(2020)\citenamefont
  {{Iwakami}}, \citenamefont {{Okawa}}, \citenamefont {{Nagakura}},
  \citenamefont {{Harada}}, \citenamefont {{Furusawa}}, \citenamefont
  {{Sumiyoshi}}, \citenamefont {{Matsufuru}},\ and\ \citenamefont
  {{Yamada}}}]{2020ApJ...903...82I}%
  \BibitemOpen
  \bibfield  {author} {\bibinfo {author} {\bibfnamefont {Wakana}\ \bibnamefont
  {{Iwakami}}}, \bibinfo {author} {\bibfnamefont {Hirotada}\ \bibnamefont
  {{Okawa}}}, \bibinfo {author} {\bibfnamefont {Hiroki}\ \bibnamefont
  {{Nagakura}}}, \bibinfo {author} {\bibfnamefont {Akira}\ \bibnamefont
  {{Harada}}}, \bibinfo {author} {\bibfnamefont {Shun}\ \bibnamefont
  {{Furusawa}}}, \bibinfo {author} {\bibfnamefont {Kosuke}\ \bibnamefont
  {{Sumiyoshi}}}, \bibinfo {author} {\bibfnamefont {Hideo}\ \bibnamefont
  {{Matsufuru}}}, \ and\ \bibinfo {author} {\bibfnamefont {Shoichi}\
  \bibnamefont {{Yamada}}},\ }\bibfield  {title} {\enquote {\bibinfo {title}
  {{Simulations of the Early Postbounce Phase of Core-collapse Supernovae in
  Three-dimensional Space with Full Boltzmann Neutrino Transport}},}\ }\href
  {\doibase 10.3847/1538-4357/abb8cf} {\bibfield  {journal} {\bibinfo
  {journal} {\apj}\ }\textbf {\bibinfo {volume} {903}},\ \bibinfo {eid} {82}
  (\bibinfo {year} {2020})},\ \Eprint {http://arxiv.org/abs/2004.02091}
  {arXiv:2004.02091 [astro-ph.HE]} \BibitemShut {NoStop}%
\bibitem [{\citenamefont {{Nagakura}}\ \emph {et~al.}(2014)\citenamefont
  {{Nagakura}}, \citenamefont {{Sumiyoshi}},\ and\ \citenamefont
  {{Yamada}}}]{2014ApJS..214...16N}%
  \BibitemOpen
  \bibfield  {author} {\bibinfo {author} {\bibfnamefont {H.}~\bibnamefont
  {{Nagakura}}}, \bibinfo {author} {\bibfnamefont {K.}~\bibnamefont
  {{Sumiyoshi}}}, \ and\ \bibinfo {author} {\bibfnamefont {S.}~\bibnamefont
  {{Yamada}}},\ }\bibfield  {title} {\enquote {\bibinfo {title}
  {{Three-dimensional Boltzmann Hydro Code for Core Collapse in Massive Stars.
  I. Special Relativistic Treatments}},}\ }\href {\doibase
  10.1088/0067-0049/214/2/16} {\bibfield  {journal} {\bibinfo  {journal}
  {\apjs}\ }\textbf {\bibinfo {volume} {214}},\ \bibinfo {eid} {16} (\bibinfo
  {year} {2014})},\ \Eprint {http://arxiv.org/abs/1407.5632} {arXiv:1407.5632
  [astro-ph.HE]} \BibitemShut {NoStop}%
\bibitem [{\citenamefont {{Richers}}\ \emph {et~al.}(2017)\citenamefont
  {{Richers}}, \citenamefont {{Nagakura}}, \citenamefont {{Ott}}, \citenamefont
  {{Dolence}}, \citenamefont {{Sumiyoshi}},\ and\ \citenamefont
  {{Yamada}}}]{2017ApJ...847..133R}%
  \BibitemOpen
  \bibfield  {author} {\bibinfo {author} {\bibfnamefont {S.}~\bibnamefont
  {{Richers}}}, \bibinfo {author} {\bibfnamefont {H.}~\bibnamefont
  {{Nagakura}}}, \bibinfo {author} {\bibfnamefont {C.~D.}\ \bibnamefont
  {{Ott}}}, \bibinfo {author} {\bibfnamefont {J.}~\bibnamefont {{Dolence}}},
  \bibinfo {author} {\bibfnamefont {K.}~\bibnamefont {{Sumiyoshi}}}, \ and\
  \bibinfo {author} {\bibfnamefont {S.}~\bibnamefont {{Yamada}}},\ }\bibfield
  {title} {\enquote {\bibinfo {title} {{A Detailed Comparison of
  Multidimensional Boltzmann Neutrino Transport Methods in Core-collapse
  Supernovae}},}\ }\href {\doibase 10.3847/1538-4357/aa8bb2} {\bibfield
  {journal} {\bibinfo  {journal} {\apj}\ }\textbf {\bibinfo {volume} {847}},\
  \bibinfo {eid} {133} (\bibinfo {year} {2017})},\ \Eprint
  {http://arxiv.org/abs/1706.06187} {arXiv:1706.06187 [astro-ph.HE]}
  \BibitemShut {NoStop}%
\bibitem [{\citenamefont {{Nagakura}}(2021)}]{2021MNRAS.500..319N}%
  \BibitemOpen
  \bibfield  {author} {\bibinfo {author} {\bibfnamefont {Hiroki}\ \bibnamefont
  {{Nagakura}}},\ }\bibfield  {title} {\enquote {\bibinfo {title} {{Retrieval
  of energy spectra for all flavours of neutrinos from core-collapse supernova
  with multiple detectors}},}\ }\href {\doibase 10.1093/mnras/staa3287}
  {\bibfield  {journal} {\bibinfo  {journal} {\mnras}\ }\textbf {\bibinfo
  {volume} {500}},\ \bibinfo {pages} {319--332} (\bibinfo {year} {2021})},\
  \Eprint {http://arxiv.org/abs/2008.10082} {arXiv:2008.10082 [astro-ph.HE]}
  \BibitemShut {NoStop}%
\bibitem [{\citenamefont {{Johns}}(2024)}]{2024arXiv240115247J}%
  \BibitemOpen
  \bibfield  {author} {\bibinfo {author} {\bibfnamefont {Lucas}\ \bibnamefont
  {{Johns}}},\ }\bibfield  {title} {\enquote {\bibinfo {title} {{Subgrid
  modeling of neutrino oscillations in astrophysics}},}\ }\href {\doibase
  10.48550/arXiv.2401.15247} {\bibfield  {journal} {\bibinfo  {journal} {arXiv
  e-prints}\ ,\ \bibinfo {eid} {arXiv:2401.15247}} (\bibinfo {year} {2024})},\
  \Eprint {http://arxiv.org/abs/2401.15247} {arXiv:2401.15247 [astro-ph.HE]}
  \BibitemShut {NoStop}%
\bibitem [{\citenamefont {{Johns}}(2023)}]{2023arXiv230614982J}%
  \BibitemOpen
  \bibfield  {author} {\bibinfo {author} {\bibfnamefont {Lucas}\ \bibnamefont
  {{Johns}}},\ }\bibfield  {title} {\enquote {\bibinfo {title} {{Thermodynamics
  of oscillating neutrinos}},}\ }\href {\doibase 10.48550/arXiv.2306.14982}
  {\bibfield  {journal} {\bibinfo  {journal} {arXiv e-prints}\ ,\ \bibinfo
  {eid} {arXiv:2306.14982}} (\bibinfo {year} {2023})},\ \Eprint
  {http://arxiv.org/abs/2306.14982} {arXiv:2306.14982 [hep-ph]} \BibitemShut
  {NoStop}%
\bibitem [{\citenamefont {{Nagakura}}\ \emph {et~al.}(2023)\citenamefont
  {{Nagakura}}, \citenamefont {{Johns}},\ and\ \citenamefont
  {{Zaizen}}}]{2023arXiv231216285N}%
  \BibitemOpen
  \bibfield  {author} {\bibinfo {author} {\bibfnamefont {Hiroki}\ \bibnamefont
  {{Nagakura}}}, \bibinfo {author} {\bibfnamefont {Lucas}\ \bibnamefont
  {{Johns}}}, \ and\ \bibinfo {author} {\bibfnamefont {Masamichi}\ \bibnamefont
  {{Zaizen}}},\ }\bibfield  {title} {\enquote {\bibinfo {title} {{BGK subgrid
  model for neutrino quantum kinetics}},}\ }\href {\doibase
  10.48550/arXiv.2312.16285} {\bibfield  {journal} {\bibinfo  {journal} {arXiv
  e-prints}\ ,\ \bibinfo {eid} {arXiv:2312.16285}} (\bibinfo {year} {2023})},\
  \Eprint {http://arxiv.org/abs/2312.16285} {arXiv:2312.16285 [astro-ph.HE]}
  \BibitemShut {NoStop}%
\bibitem [{\citenamefont {{Xiong}}\ \emph
  {et~al.}(2024{\natexlab{a}})\citenamefont {{Xiong}}, \citenamefont {{Wu}},
  \citenamefont {{George}},\ and\ \citenamefont {{Lin}}}]{2024arXiv240317269X}%
  \BibitemOpen
  \bibfield  {author} {\bibinfo {author} {\bibfnamefont {Zewei}\ \bibnamefont
  {{Xiong}}}, \bibinfo {author} {\bibfnamefont {Meng-Ru}\ \bibnamefont {{Wu}}},
  \bibinfo {author} {\bibfnamefont {Manu}\ \bibnamefont {{George}}}, \ and\
  \bibinfo {author} {\bibfnamefont {Chun-Yu}\ \bibnamefont {{Lin}}},\
  }\bibfield  {title} {\enquote {\bibinfo {title} {{Robust integration of fast
  flavor conversions in classical neutrino transport}},}\ }\href {\doibase
  10.48550/arXiv.2403.17269} {\bibfield  {journal} {\bibinfo  {journal} {arXiv
  e-prints}\ ,\ \bibinfo {eid} {arXiv:2403.17269}} (\bibinfo {year}
  {2024}{\natexlab{a}})},\ \Eprint {http://arxiv.org/abs/2403.17269}
  {arXiv:2403.17269 [astro-ph.HE]} \BibitemShut {NoStop}%
\bibitem [{\citenamefont {{Nagakura}}\ and\ \citenamefont
  {{Zaizen}}(2023{\natexlab{b}})}]{2023PhRvD.107f3033N}%
  \BibitemOpen
  \bibfield  {author} {\bibinfo {author} {\bibfnamefont {Hiroki}\ \bibnamefont
  {{Nagakura}}}\ and\ \bibinfo {author} {\bibfnamefont {Masamichi}\
  \bibnamefont {{Zaizen}}},\ }\bibfield  {title} {\enquote {\bibinfo {title}
  {{Connecting small-scale to large-scale structures of fast neutrino-flavor
  conversion}},}\ }\href {\doibase 10.1103/PhysRevD.107.063033} {\bibfield
  {journal} {\bibinfo  {journal} {\prd}\ }\textbf {\bibinfo {volume} {107}},\
  \bibinfo {eid} {063033} (\bibinfo {year} {2023}{\natexlab{b}})},\ \Eprint
  {http://arxiv.org/abs/2211.01398} {arXiv:2211.01398 [astro-ph.HE]}
  \BibitemShut {NoStop}%
\bibitem [{\citenamefont {{Zaizen}}\ and\ \citenamefont
  {{Nagakura}}(2023{\natexlab{a}})}]{2023PhRvD.107j3022Z}%
  \BibitemOpen
  \bibfield  {author} {\bibinfo {author} {\bibfnamefont {Masamichi}\
  \bibnamefont {{Zaizen}}}\ and\ \bibinfo {author} {\bibfnamefont {Hiroki}\
  \bibnamefont {{Nagakura}}},\ }\bibfield  {title} {\enquote {\bibinfo {title}
  {{Simple method for determining asymptotic states of fast neutrino-flavor
  conversion}},}\ }\href {\doibase 10.1103/PhysRevD.107.103022} {\bibfield
  {journal} {\bibinfo  {journal} {\prd}\ }\textbf {\bibinfo {volume} {107}},\
  \bibinfo {eid} {103022} (\bibinfo {year} {2023}{\natexlab{a}})},\ \Eprint
  {http://arxiv.org/abs/2211.09343} {arXiv:2211.09343 [astro-ph.HE]}
  \BibitemShut {NoStop}%
\bibitem [{\citenamefont {{Wu}}\ \emph {et~al.}(2021)\citenamefont {{Wu}},
  \citenamefont {{George}}, \citenamefont {{Lin}},\ and\ \citenamefont
  {{Xiong}}}]{2021PhRvD.104j3003W}%
  \BibitemOpen
  \bibfield  {author} {\bibinfo {author} {\bibfnamefont {Meng-Ru}\ \bibnamefont
  {{Wu}}}, \bibinfo {author} {\bibfnamefont {Manu}\ \bibnamefont {{George}}},
  \bibinfo {author} {\bibfnamefont {Chun-Yu}\ \bibnamefont {{Lin}}}, \ and\
  \bibinfo {author} {\bibfnamefont {Zewei}\ \bibnamefont {{Xiong}}},\
  }\bibfield  {title} {\enquote {\bibinfo {title} {{Collective fast neutrino
  flavor conversions in a 1D box: Initial conditions and long-term
  evolution}},}\ }\href {\doibase 10.1103/PhysRevD.104.103003} {\bibfield
  {journal} {\bibinfo  {journal} {\prd}\ }\textbf {\bibinfo {volume} {104}},\
  \bibinfo {eid} {103003} (\bibinfo {year} {2021})},\ \Eprint
  {http://arxiv.org/abs/2108.09886} {arXiv:2108.09886 [hep-ph]} \BibitemShut
  {NoStop}%
\bibitem [{\citenamefont {{Xiong}}\ \emph
  {et~al.}(2024{\natexlab{b}})\citenamefont {{Xiong}}, \citenamefont {{Wu}},
  \citenamefont {{George}}, \citenamefont {{Lin}}, \citenamefont {{Khosravi
  Largani}}, \citenamefont {{Fischer}},\ and\ \citenamefont
  {{Mart{\'\i}nez-Pinedo}}}]{2024arXiv240219252X}%
  \BibitemOpen
  \bibfield  {author} {\bibinfo {author} {\bibfnamefont {Zewei}\ \bibnamefont
  {{Xiong}}}, \bibinfo {author} {\bibfnamefont {Meng-Ru}\ \bibnamefont {{Wu}}},
  \bibinfo {author} {\bibfnamefont {Manu}\ \bibnamefont {{George}}}, \bibinfo
  {author} {\bibfnamefont {Chun-Yu}\ \bibnamefont {{Lin}}}, \bibinfo {author}
  {\bibfnamefont {Noshad}\ \bibnamefont {{Khosravi Largani}}}, \bibinfo
  {author} {\bibfnamefont {Tobias}\ \bibnamefont {{Fischer}}}, \ and\ \bibinfo
  {author} {\bibfnamefont {Gabriel}\ \bibnamefont {{Mart{\'\i}nez-Pinedo}}},\
  }\bibfield  {title} {\enquote {\bibinfo {title} {{Fast neutrino flavor
  conversions in a supernova: emergence, evolution, and effects}},}\ }\href
  {\doibase 10.48550/arXiv.2402.19252} {\bibfield  {journal} {\bibinfo
  {journal} {arXiv e-prints}\ ,\ \bibinfo {eid} {arXiv:2402.19252}} (\bibinfo
  {year} {2024}{\natexlab{b}})},\ \Eprint {http://arxiv.org/abs/2402.19252}
  {arXiv:2402.19252 [astro-ph.HE]} \BibitemShut {NoStop}%
\bibitem [{\citenamefont {{Zaizen}}\ and\ \citenamefont
  {{Nagakura}}(2023{\natexlab{b}})}]{2023PhRvD.107l3021Z}%
  \BibitemOpen
  \bibfield  {author} {\bibinfo {author} {\bibfnamefont {Masamichi}\
  \bibnamefont {{Zaizen}}}\ and\ \bibinfo {author} {\bibfnamefont {Hiroki}\
  \bibnamefont {{Nagakura}}},\ }\bibfield  {title} {\enquote {\bibinfo {title}
  {{Characterizing quasisteady states of fast neutrino-flavor conversion by
  stability and conservation laws}},}\ }\href {\doibase
  10.1103/PhysRevD.107.123021} {\bibfield  {journal} {\bibinfo  {journal}
  {\prd}\ }\textbf {\bibinfo {volume} {107}},\ \bibinfo {eid} {123021}
  (\bibinfo {year} {2023}{\natexlab{b}})},\ \Eprint
  {http://arxiv.org/abs/2304.05044} {arXiv:2304.05044 [astro-ph.HE]}
  \BibitemShut {NoStop}%
\bibitem [{\citenamefont {{Xiong}}\ \emph {et~al.}(2023)\citenamefont
  {{Xiong}}, \citenamefont {{Wu}}, \citenamefont {{Abbar}}, \citenamefont
  {{Bhattacharyya}}, \citenamefont {{George}},\ and\ \citenamefont
  {{Lin}}}]{2023PhRvD.108f3003X}%
  \BibitemOpen
  \bibfield  {author} {\bibinfo {author} {\bibfnamefont {Zewei}\ \bibnamefont
  {{Xiong}}}, \bibinfo {author} {\bibfnamefont {Meng-Ru}\ \bibnamefont {{Wu}}},
  \bibinfo {author} {\bibfnamefont {Sajad}\ \bibnamefont {{Abbar}}}, \bibinfo
  {author} {\bibfnamefont {Soumya}\ \bibnamefont {{Bhattacharyya}}}, \bibinfo
  {author} {\bibfnamefont {Manu}\ \bibnamefont {{George}}}, \ and\ \bibinfo
  {author} {\bibfnamefont {Chun-Yu}\ \bibnamefont {{Lin}}},\ }\bibfield
  {title} {\enquote {\bibinfo {title} {{Evaluating approximate asymptotic
  distributions for fast neutrino flavor conversions in a periodic 1D box}},}\
  }\href {\doibase 10.1103/PhysRevD.108.063003} {\bibfield  {journal} {\bibinfo
   {journal} {\prd}\ }\textbf {\bibinfo {volume} {108}},\ \bibinfo {eid}
  {063003} (\bibinfo {year} {2023})},\ \Eprint
  {http://arxiv.org/abs/2307.11129} {arXiv:2307.11129 [astro-ph.HE]}
  \BibitemShut {NoStop}%
\end{thebibliography}%

\end{document}